\documentclass[]{aastex}

\usepackage{emulateapj5}
\usepackage{onecolfloat}
\usepackage{graphicx} 
\usepackage{fancyheadings} 
\usepackage{ulem}
\usepackage{rotating}
\usepackage{lscape}

\newcommand{\XFe}[1]{<[{\rm #1/Fe}]>}

\newcommand{\nstrs}{10~}
\newcommand{\msol}{M_\odot}

\newcommand{\lya}{Ly$\alpha$ }

\newcommand{\kms}{km~s$^{-1}$ }
\newcommand{\cm}[1]{\, {\rm cm^{#1}}}
\newcommand{\N}[1]{{N({\rm #1})}}
\newcommand{\e}[1]{{\epsilon({\rm #1})}}

\newcommand{\rAA}{{\AA \enskip}}
\newcommand{\sci}[1]{{\rm \; \times \; 10^{#1}}}
\newcommand{\ltk}{\left [ \,}

\newcommand{\rtk}{\, \right  ] }

\newcommand{\ohf}{{1 \over 2}}

\newcommand{\perd}{\;\;\; .}
\newcommand{\cmma}{\;\;\; ,}

\newcommand{\mkms}{{\rm \; km\;s^{-1}}}
\newcommand{\ew}{W_\lambda}

\begin{document}

\twocolumn[%
\submitted{Accepted to the Astronomical Journal July 21, 2000}

\title{THE GALACTIC THICK DISK STELLAR ABUNDANCES}

\author{ JASON X. PROCHASKA\altaffilmark{1,2},
SERGEI O. NAUMOV\altaffilmark{3,4}, BRUCE W. CARNEY\altaffilmark{3},
ANDREW McWILLIAM\altaffilmark{2}, \& ARTHUR M. WOLFE\altaffilmark{1,5}}

\begin{abstract} 

We present first results from a program to 
measure the chemical abundances of a large
($N>30$) sample of thick disk stars with the principal goal of investigating
the formation history of the Galactic
thick disk.  We have obtained high resolution, high signal-to-noise spectra
of 10 thick disk stars with the HIRES spectrograph on the 10m Keck~I telescope.
Our analysis confirms
previous studies of O and Mg in the thick disk stars which
reported enhancements in excess of the thin disk population.  Furthermore,
the observations of Si, Ca, Ti, Mn, Co, V, Zn, Al, and Eu
all argue that the thick disk population has a distinct chemical history
from the thin disk. With the exception of V and Co, the thick disk abundance
patterns match or tend towards the values observed for halo stars with
$\lbrack$Fe/H$\rbrack$~$\approx -1$.  
This suggests that the thick disk stars had a chemical enrichment
history similar to the metal-rich halo stars.  With the
possible exception of Si, {\it the thick disk abundance patterns are 
in excellent agreement with the chemical abundances observed 
in the metal-poor bulge stars} suggesting the two populations
formed from the same gas reservoir at a common epoch.

The principal results of our analysis are as follows.  
(i) All 10 stars exhibit enhanced 
$\alpha$/Fe ratios with O, Si, and Ca showing tentative trends of 
decreasing overabundances with increasing $\lbrack$Fe/H$\rbrack$.  In contrast, the Mg and
Ti enhancements are constant.
(ii)  The light elements Na and Al are enhanced in these stars.
(iii) With the exception of Ni, Cr and possibly Cu, the iron-peak elements
show significant departures from the solar abundances.
The stars are deficient in Mn, but overabundant in V, Co, Sc, and Zn.
(iv) The heavy elements Ba and Y are consistent with solar abundances
but Eu is significantly enhanced.

If the trends of decreasing O, Si, and Ca with 
increasing $\lbrack$Fe/H$\rbrack$ are explained 
by the onset of Type~Ia SN, then the thick disk stars formed
over the course of $\gtrsim 1$~Gyr.  We argue that this formation
time-scale would rule out
most dissipational collapse scenarios for the formation of the thick disk.
Models which consider the heating of an initial thin disk -- either
through 'gradual' heating mechanisms or a sudden merger event -- are favored.

These observations provide new tests of theories of nucleosynthesis 
in the early universe.  In particular, the enhancements of 
Sc, V, Co, and Zn may imply overproduction during an enhanced
$\alpha$-rich freeze out fueled by neutrino-driven winds.  Meanwhile,
the conflicting trends for Mg, Ti, Ca, Si, and O
pose a difficult challenge to our current understanding
of nucleosynthesis in Type~Ia and Type~II SN.
The Ba/Eu ratios favor r-process dominated enrichment
for the heavy elements, consistent with the ages $(t_{age} > 10$~Gyr)
expected for these stars.

Finally, we discuss the impact of the thick disk abundances on 
interpretations of the abundance patterns of the damped \lya systems.
The observations of mildly enhanced Zn/Fe imply an interpretation for
the damped systems which includes a dust depletion pattern on top of
a Type~II SN enrichment pattern.  We also argue that the S/Zn 
ratio is not a good indicator of nucleosynthetic processes.


\end{abstract}

]
\altaffiltext{1}{Visiting Astronomer, W.M. Keck Telescope.
The Keck Observatory is a joint facility of the University
of California and the California Institute of Technology.}
\altaffiltext{2}{The Observatories of the Carnegie Institute of Washington,
813 Santa Barbara St., Pasadena, CA 91101}
\altaffiltext{3}{Department of Physics and Astronomy, 
University of North Carolina, Chapel Hill, North Carolina 27599-3255}
\altaffiltext{4}{Present address: Rostov State University, Russia}
\altaffiltext{5}{Department of Physics, and Center for Astrophysics and 
Space Sciences, University of California, San Diego, C--0424, La Jolla, 
CA 92093-0424}

\pagestyle{fancyplain}
\lhead[\fancyplain{}{\thepage}]{\fancyplain{}{PROCHASKA ET AL.}}
\rhead[\fancyplain{}{THE GALACTIC THICK DISK STELLAR ABUNDANCES}]{\fancyplain{}{\thepage}}
\setlength{\headrulewidth=0pt}
\cfoot{}

\section{INTRODUCTION}

The history of our Galaxy may be read through the long-lived
stellar relics of its past. In their landmark study, \cite{egg62}
employed dynamical and chemical data to argue that the
dynamically hot and metal-poor halo stellar population was
the precursor of the dynamically cool and metal-rich disk
population. While this conclusion has been subjected to
considerable debate, the comparative study of the halo 
and the disk populations is certainly the primary means by
which we learn of the Galaxy's earliest history.

\cite{gilm83} offered the best evidence for the
existence of another Galactic stellar population, the
thick disk. Their data consolidated earlier but less
well-formed views of the ``intermediate population II"
class described in the 1957 Vatican Conference \citep{oconn58}.
The reality of the thick disk population was in its turn
hotly debated \citep{bahc84}, but it is now generally regarded as a separate
population. The key historical question is whether the
thick disk is related to any or all of the other Galactic
stellar populations: the halo, the bulge, and the disk.
The first steps have been to determine the basic parameters
of the thick disk, including its age, its chemical composition,
and its dynamics/distribution. The thick disk appears to
be very old, based on the abrupt cut-off in the numbers
of stars bluer than the main sequence turn-offs of similar
metallicity globular clusters \citep{gilm87,carn89,gilm89}.
The mean metallicity of the thick disk population, $<$[Fe/H]$>$,
lies between $-0.5$ and $-0.7$ 
\citep{gilm85,carn89,gilmore95,layden95a,layden95b}. The spread
in metallicities of thick disk stars ranges from
near solar to [Fe/H] $\approx$ $-1$, although claims for much
lower metallicities have been made 
\citep[cf.][]{norris85,morrison90,allen91,ryan95,
beers95,twarog96,martin98,chiba00}.
The ``asymmetric drift" of the thick disk (the amount by which
it lags the circular orbit motion at the solar Galactocentric
distance) has been estimated to vary between 20 and 50 \kms
\citep{carn89,morrison90,mjk92,beers95,ojha96,chiba00},
although \cite{mjk92}, \cite{chen97}, and \cite{chiba00} 
have argued that the value varies
with distance from the Galactic plane. Values for the
vertical velocity dispersion, $\sigma$(W), are almost all near 40 \kms 
\citep{norris86,carn89,beers95,ojha96,chiba00}, which implies a vertical
scale height of order 1~kpc or less. This may be compared to the
older stars of the thin disk, which obey a density distribution
consistent with a vertical scale height of about 300~pc. 
Although the thin disk is $\approx 10$ times more massive than the thick
disk, at distances of 1 to 2 kpc above the plane, the thick disk population
dominates.

The properties of the thick disk thus place it between those of
the halo and the thin disk. In turn, one questions whether
it is closely related to either of them
in terms of the Galaxy's chemical and dynamical evolution, or if it might
be the result of a merger event (see Gilmore et al.\ 1989
and Majewski 1993 for excellent reviews of the various models).
Most evolutionary models predict that there should be
continuity in the thick disk and disk dynamical and chemical
histories, and that thick disks should be found in other galaxies.
A merger scenario, conversely, would require
some degree of discontinuity
in the chemical and dynamical parameters of the thick disk and
the thin disk, or observations that indicate not 
all disk galaxies have thick disks.
It is interesting in this regard that very deep surface photometry
of edge-on spirals reveals thick disks in some cases (e.g., NGC~891:
van der Kruit \& Searle 1981; Morrison et al.\ 1997)
but not in all cases (e.g., NGC~5907: Morrison, Boroson, \& Harding 1994;
NGC~4244: Fry et al.\ 1999).

In this paper we study the problem using Galactic stars whose
motions are consistent with thick disk membership. Our 
goal here is to compare their abundance patterns, [X/Fe] vs.\ [Fe/H], 
with those of the other major Galactic stellar populations: the halo,
thin disk and bulge.  If the histories of the thick disk and the disk
are closely related, for example, so should be the derived chemical abundances
patterns vs.\ metallicity. It is well established that
very metal-poor (halo) stars show enhanced levels of the light
``$\alpha$"-rich nuclei elements oxygen, magnesium, silicon, calcium,
and even titanium, but at a metallicity of [Fe/H] $\approx$ $-1$
the [$\alpha$/Fe] values begin to decline from +0.4 dex or so
to solar values at [Fe/H] = 0 (see Wheeler, Sneden, \& Truran 1989;
McWilliam 1997).  A fundamental comparison then is whether
thick disk stars show similar [$\alpha$/Fe]
and other element abundance ratios at the same [Fe/H] values as
the thin disk stars. Similarities would favor the ``evolutionary"
history of the thick disk; discontinuities would support a merger origin.

Large stellar samples with high-precision abundance
analyses have appeared over the last several years, which may,
in principle, answer this question. \cite{edv93}
studied a large sample of F and G dwarfs and found
that lower metallicity stars had, in general, enhanced [$\alpha$/Fe]
values, but they did not 
compare the thick disk and thin disk stellar abundance patterns in detail.
\cite{grtt96,grtt00} were the first to directly compare the abundance
ratios of a sample ($\approx 15)$ of thick disk stars with halo and
thin disk populations.  Their measurements of Fe/O and Fe/Mg ratios
indicate a stark difference in the Fe/O and Fe/Mg abundance
of the thick and thin disk populations with the thick disk stars exhibiting
halo-like ratios.  These authors argue for an early, rapid 
formation of the Galactic thick disk, prior to the thin disk and 
perhaps due to an early accretion event. 
\cite{fuhr98} also compared Mg/Fe ratios for a sample of thick and thin
disk stars taken from both \cite{edv93} and his own smaller kinematic
sample.  Although
the Edvardsson et al.\ sample does not provide compelling evidence for
a discontinuity, the Fuhrmann sample 
shows strong evidence for a disjunction, which further supports the assertion
that the thick disk and thin disk have not shared the same chemical enrichment
history. Most recently, \cite{chen00} presented results
from a sample of 90 F and G dwarfs, which show no significant
scatter in $\alpha$-element ratios as a function of [Fe/H],
contrary to the results obtained by \cite{fuhr98}. 
We contend, however, that the sample selection employed by 
\cite{chen00} was flawed for a
program aimed at the study of the thick disk. They 
chose to study only stars with effective temperatures
between 5800 and 6400~K, so that few of their stars
have life expectancies as great as the thick disk's age.
As an example, consider the disk, probably thick disk,
globular cluster 47~Tuc. Using the \cite{alonso96} 
temperature scale (employed by Chen et al.\ 2000), the metallicity
of [Fe/H] = $-0.70$ \citep{carr97}, and the
photometry of \cite{hesser87}, we find that the temperatures
of main sequence turn-off stars in the cluster is near 5970~K,
only slightly hotter than the lower limit cut-off for the
\cite{chen00} sample. And for more metal-rich clusters
or stars whose ages are as great as 47~Tuc, the turn-off
temperature is even cooler.
Thus if the thick disk is composed almost exclusively of ancient stars,
the \cite{chen00} sample cannot contain many thick disk stars.
Fuhrmann's (1998) claimed thick disk stars, however, are cool
enough to have long enough life expectancies to be considered
part of the thick disk. We believe that \cite{chen00}
have studied, primarily, the detailed chemical evolution of the
thin disk, which no doubt reaches to quite low metallicity
levels itself (see in particular the study regarding the
overlap in abundances of the thick disk and the thin disk
by Wyse \& Gilmore 1995). We emphasize that comparative
studies of the thick disk and the thin disk must employ stars
with life expectancies as great as the thick disk's age
lest the trace but important population, the thick disk,
be overlooked. We do so here. In a future paper \citep{carn00},
we will study the relation between kinematics
and mean metallicity for long-lived dwarf stars in the
mid-plane, finding further evidence for two distinct populations.

\begin{table*}[ht]
\begin{center}
\caption{{\sc JOURNAL OF OBSERVATIONS} \label{tab:obs}}
\begin{tabular}{lllllccc}
\tableline
\tableline 
Star & Alt Name & HIP ID\tablenotemark{a} & RA (2000)
& DEC (2000) & $V$ & Exp(blue)\tablenotemark{b}  & Exp(red)\\
\tableline 
G66-51    &            &        & 15:00:50.0 & +02:07:37 & 10.63 &  380 &  600$^d$ \\
G84-37    & HD 241253  &  24030 & 05:09:56.9 & +05:33:26 &  9.72 &  350 &  500$^d$ \\
G88-13    & B+17 1524  &  34902 & 07:13:17.4 & +17:26:01 & 10.10 &  800 &  800$^d$ \\
G92-19    & B-02 5072  &  96673 & 19:39:14.7 & $-$02:36:44 & 10.31 &  500 &  600$^d$ \\
G97-45    & HD 36283   &  25860 & 05:31:13.7 & +15:46:24 &  8.64 &  430 &  500$^d$ \\
G114-19   & HD 75530   &  43393 & 08:50:21.0 & $-$05:32:09 &  9.19 &  500 &  600$^c$ \\
G144-52   & B+19 4601  & 103812 & 21:02:12.1 & +19;54:03 &  9.07 &  600 &  300$^d$ \\
G181-46   & B+31 3025  &  85373 & 17:26:41.4 & +31:03:34 &  9.68 &  400 &  600$^d$ \\
G211-5    & B+33 4117  & 103691 & 21:00:43.2 & +33:53:20 &  9.62 &  600 &  600$^d$ \\
G247-32   & B+66 343   &  21921 & 04:42:50.2 & +66:44:08 &  8.28 &  350 &  400$^c$ \\
\tableline
\end{tabular}
\end{center}
\tablenotetext{a}{Hipparcos Identifier, ESA (1997)}
\tablenotetext{b}{$\lambda_{obs} = 4325 - 6760$\AA}
\tablenotetext{c}{$\lambda_{obs} = 6380 - 8750$\AA}
\tablenotetext{d}{$\lambda_{obs} = 6810 - 9200$\AA}
\end{table*}

We have initiated a program to measure the chemical abundances of a large
$(N \sim 50)$ sample of thick disk stars at very high resolution 
$(R \approx 50,000)$ with high signal-to-noise ratio ($S/N > 100$), and a
nearly continuous wavelength coverage from $\lambda \approx 4400 - 9000$\AA.
The stars were selected from the surveys by \cite{carn94,carn96} as
exhibiting disk-like kinematics with large maximum distance from the Galactic
plane.  The current sample is
comprised of \nstrs stars, all brighter than $V = 10.5$.  
We present a detailed description of our stellar abundance analysis and
give first results on a small but meaningful sample of stars.  In part,
our goal is to build the analysis framework for future observations. 
This initial sample, however, suggests a number of exciting results which
we will test through a larger survey. 
In $\S$~\ref{sec-obs} we describe the observations and
present a summary of the stellar parameters of the sample.  
The following section presents a thorough explanation of the techniques
employed to measure the chemical abundances of the program stars.  In
general, we follow standard stellar analysis procedures.  A solar analysis
is discussed in $\S$~4 and $\S$~5 gives an element-by-element account
of the results.  Finally, $\S$~6 compares the thick disk results 
against other stellar populations and discusses the implications for the
formation history of the Galaxy, the damped \lya systems, and 
nucleosynthesis in the early universe.

\section{OBSERVATIONS AND DATA REDUCTION}
\label{sec-obs}

All of the observations presented here
were carried out in twilight time during an
ongoing program to study high redshift damped \lya systems 
with the high resolution echelle spectrograph (HIRES; Vogt et al.\ 1994) on
the 10m Keck~I telescope.  Table~\ref{tab:obs} summarizes the current sample
of program stars and presents our journal of observations.  For each
star we took multiple exposures at two settings to achieve nearly continuous
wavelength coverage from $\lambda \approx 4400 - 9000$\rAA with the exception
of the inter-order gaps longward of 5250\AA.
The blue setup consisted of the C1 decker which affords FWHM~$\approx 6 \mkms$
resolution ($\approx 2 \mkms$ per pixel)
and the kv380 filter to block second order light.
For the red settings, we implemented the longer C2 decker for improved
sky subtraction and the og530 filter to eliminate second order light.
The typical signal-to-noise is in excess of 100 per pixel for all
of the spectra and $> 200$ for most of the stars.
Standard ThAr arc calibrations and quartz flats were taken for reduction
and calibration of the spectra.

\begin{figure}[hb]
\includegraphics[height=3.8in, width=2.8in, angle=-90]{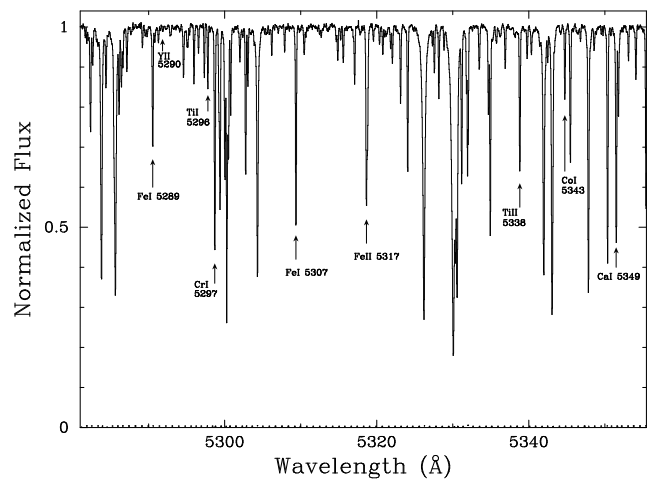}
\caption{Example of a single echelle order from the star G114-19.  The
data is straight sum of two exposures taken with HIRES on the Keck~I 
telescope and normalized to unit flux.  The dotted line in the
figure marks the $1\sigma$ error array.}
\label{exmpspec}
\end{figure}

\begin{table*}[ht]
\begin{center}
\caption{
{\sc STELLAR PHYSICAL PARAMETERS} \label{tab:pparm}}
\begin{tabular}{lcrrrcccccccc}
\tableline
\tableline
Star & $V$ &$\tilde U$& $\tilde V$ & $\tilde W$ & $Z_{max}$ & $R_{apo}$ 
& $R_{per}$ & $d$ & $T_{phot}$&[M/H] & $\log g$ \\
& & (km/s) & (km/s) & (km/s) & (kpc) & (kpc) &(kpc) & (pc) & (K) & & \\
\tableline 
G66-51  & 10.63 & $+102$ & $ -72$ & $ -54$ & 0.79 & 9.39 & 3.69 &    &  5196 & $-1.09$ &        \\
G84-37  &  9.72 & $ -19$ & $ -61$ & $ +88$ & 1.77 & 8.15 & 5.01 &  97 &  5898 & $-0.92$ &  4.47  \\
G88-13  & 10.10 & $ -20$ & $ -41$ & $ -51$ & 0.70 & 8.17 & 5.44 &  89 &  5069 & $-0.44$ &  4.32  \\
G92-19  & 10.31 & $ +88$ & $ -70$ & $ -67$ & 1.08 & 9.10 & 3.80 & 121 &  5433 & $-0.81$ &  4.33  \\
G97-45  &  8.64 & $ +20$ & $ -52$ & $ -49$ & 0.68 & 8.10 & 4.94 &  53 &  5429 & $-0.53$ &  4.37  \\
G114-19 &  9.19 & $ -27$ & $ -85$ & $ -69$ & 1.10 & 8.15 & 3.67 &  54 &  5218 & $-0.52$ &  4.43  \\
G144-52 &  9.07 & $ +54$ & $ -11$ & $ +56$ & 0.87 & 9.13 & 6.36 &  58 &  5497 & $-0.67$ &  4.46  \\
G181-46 &  9.68 & $ +50$ & $ -80$ & $ +54$ & 0.72 & 8.32 & 3.65 &  71 &  5277 & $-0.64$ &  4.43  \\
G211-5  &  9.62 & $ -69$ & $ -21$ & $ -46$ & 0.63 & 9.24 & 5.78 &  67 &  5196 & $-0.50$ &  4.42  \\
G247-32 &  8.28 & $ -49$ & $ -52$ & $ +47$ & 0.61 & 8.45 & 4.75 &  36 &  5270 & $-0.43$ &  4.45  \\
\tableline
\end{tabular}
\end{center}
\end{table*}

The data were extracted and wavelength calibrated with the {\it makee}
package developed by T. Barlow specifically for HIRES observations.
The algorithm performs an optimal extraction using the observed star to
trace the profile, and it solves for a wavelength calibration solution by 
cross-correlating the extracted ThAr spectra with an extensive database
compiled by Dr.\ Barlow.  The pairs of exposures were rebinned to the
same wavelength scale and coadded conserving flux.  Finally, we 
continuum fit the summed spectra with a routine similar to IRAF,
using the Arcturus spectrum \citep{arct68} as a guide in the bluest orders
where the flux rarely recovers to the continuum.  An example of a typical
spectral order is presented in Figure~\ref{exmpspec} and we identify several
representative absorption line features.

The sample of stars were kinematically selected 
from the study of \cite{carn94} to be members of the thick
disk according to the following criteria. 
Our large initial list excluded
stars with any uncertain observational parameters, known subgiants,
stars whose reddenings might exceed 0.05 mag, and all stars known to
be multiple-lined or double-lined. To avoid stars whose lifetimes
are shorter than the age of the thick disk, we avoided stars with
the ``TO" flag (meaning their colors place them near the main
sequence turn-off for globular clusters of similar metallicity).
We further restricted the list to stars with 

\noindent $-$1.1~$\leq$~[M/H]~$\leq -0.4$
to probe the thick disk metallicity regime, and likewise eliminated
stars whose orbits did not carry them farther than 600 pc from the plane.
To further maximize the probability of observing thick disk stars
within this sample, we restricted the $\tilde V$ velocities to lie between
$-20$ and $-100$~\kms.   While these criteria help minimize the contamination
of the thick disk sample from metal-rich halo stars and metal-poor thin
disk stars, these stellar populations do overlap in both metallicity and
kinematic properties and the possibility of contamination exists.
In general, the overlap between the thick and thin disk populations is
small (as determined from proper motion studies; e.g. Carney et al.\ 1989)
but the problem deserves further observational attention. 
Table~\ref{tab:pparm} summarizes values of
the observed stars' photometric temperatures, high-resolution and
low signal-to-noise spectroscopic metallicity, stellar gravities determined
with Hipparcos measurements \citep{hipp97}, and stellar
kinematics and galactic orbital parameters from \cite{carn94}.
All of the stars are G dwarfs found in the solar neighborhood
with distances of $d_{pc} = 50 - 100$~pc.
In those cases where there are Hipparcos parallax measurements,
we calculated the stellar gravity according to the equations presented
in Appendix~\ref{app:grav}.
The uncertainties in the parallax and photometry
imply an error in $\log (g/g_\odot)$ of $\approx 0.1$~dex.  
In the following section, we will compute spectroscopic physical parameters
for each star and compare with the photometric values presented here.

\section{ABUNDANCE ANALYSIS}
\label{sec-abund}

In this section we outline the prescription to measure elemental
abundances for our sample of thick disk stars.  In short, we measured 
equivalent widths with the package {\it getjob}, implement Kurucz model
stellar atmospheres, culled $\log gf$ values from the literature, and
used the stellar analysis package MOOG to constrain the spectroscopic 
physical parameters and determine the elemental abundances.

\subsection{Equivalent Widths}
\label{subsec-EW}

We first compiled a list of nearly 1000 reasonably unblended lines  
from the solar spectrum \citep{moore66} and an extensive
literature search.  The equivalent width, $\ew$, for each line
was then measured with the {\it getjob} package developed by A. McWilliam
for stellar spectroscopic analysis \citep{mcw95a}.  
In the majority of cases, we fit
the absorption lines with single Gaussian profiles which provide an
excellent match to the majority of observations.  
When necessary we fit regions with multiple Gaussians or 
calculated an integrated equivalent width using Simpson's Rule.  
The latter approach was particularly important for strong Ca~I and Mg~I
lines. The {\it getjob} program
yields an error estimate for each $\ew$ value based on the goodness-of-fit
and signal-to-noise of the spectra.  The typical $1\sigma$
error for a single component
fit is $\approx 2$~m\AA.  With the exception of a few special cases,
those lines with errors exceeding $20 \%$ 
were eliminated from the subsequent abundance analysis.  
We also focused on unsaturated lines, specifically lines with 
$\ew / \lambda < (100 {\rm m\AA})/ (5000 {\rm \AA})$.
Table~\ref{tab:ew} lists the $\ew$ values for the measured absorption lines.
We have flagged those absorption lines which we believe are blended or
have incorrect $gf$ values.

\begin{table*}[ht] \footnotesize
\begin{center}
\caption{{\sc EW MEASUREMENTS} \label{tab:ew}}
\begin{tabular}{lccccccccccccccc}
\tableline
\tableline
Ion &$\lambda$& EP & log $gf$ & Ref& Sun & G66-51 & 
G84-37 & G88-13 & G92-19 & G97-45 & 
G114-19 & G144-52 & G181-46 & G211-5 & G247-32 \\
& (\AA) & (eV) & & & (m\AA) & (m\AA) & (m\AA) & (m\AA) & (m\AA) & (m\AA) 
& (m\AA) & (m\AA) & (m\AA) & (m\AA) & (m\AA) \\
\tableline
O I   & 7771.954& 9.140&$ 0.360  $& 62&  71.5&  20.4&  48.6&  30.3&  42.6&  60.7&  36.5&  50.5&  39.8&  33.3&  46.0 \\            
O I   & 7774.177& 9.140&$ 0.210  $& 62&  63.7&  17.2&  37.6&  23.0&  35.7&  52.8&  32.8&  41.8&  33.8&      &  44.2 \\            
O I   & 7775.395& 9.140&$-0.010  $& 62&  53.2&  11.8&  26.8&  17.0&  26.0&  42.5&  23.2&      &  27.9&  21.2&  34.3 \\            
\\                                                                                                                                
Na I  & 5682.647& 2.100&$-0.890  $& 99&  90.0&  51.1&  22.6& 108.4&  66.2&  93.4&  87.5&  74.0&      &  91.1&  89.0 \\            
Na I  & 5688.210& 2.100&$-0.580  $& 99& 119.1&  72.5&  37.2&$>$120&  86.1& 113.2&      &  95.5& 102.8& 115.2& 115.0 \\            
\tableline
\end{tabular}
\end{center}
\tablerefs{Key to References -- 1: \cite{obrian91}; 2: \cite{fuhr88};
3: \cite{black79a}; 4: \cite{black79b}; 5: \cite{black80}; 6: \cite{black82a};
8: \cite{black82e}; 9: \cite{black82b}; 10: \cite{black82c}; 
11: \cite{black83}; 12: \cite{black84}; 13: \cite{black86a}; 
14: \cite{black86b}; 15: \cite{black86c}; 17: \cite{martin88};
18: \cite{fry97}; 19: \cite{bard91}; 21: \cite{bard94}; 25: \cite{mcw94};
26: \cite{cardon82}; 27: \cite{wiese80}; 28: \cite{garz73}; 29: \cite{bizz93};
30: \cite{smith81}; 31: \cite{meylan93}; 32: \cite{biem91};
33: \cite{mcw95}; 36: \cite{hann83}; 37: \cite{moity83};
38: \cite{buur86}; 42: \cite{kock68}; 45: \cite{lawler89};
48: \cite{lamb78}; 54: \cite{fuhrm95}; 55: \cite{wick97};
56: \cite{francois88}; 57: \cite{biem80}; 58: \cite{black76};
60: \cite{schna99}; 61: \cite{kroll87}; 62: \cite{butler91};
66: \cite{booth84a}; 67: \cite{savanov90}; 99: Solar gf (this work);
104: \cite{edv93}; 106: \cite{bev94}}
\tablecomments{The complete version of this table is in the 
electronic edition of the Journal.  The printed edition contains only
a sample}
\end{table*}

\subsection{Model Stellar Atmospheres}
\label{subsec-Atmo}

Throughout the abundance
analysis we adopt Kurucz stellar atmospheres \citep{kur88}
with convection on and 72 layers with optical depth steps, 
$\Delta \tau = 0.125$, ending at $\tau = 100$.
Depending on the application, we either interpolated between the stellar
atmosphere grids kindly provided by R. Kurucz or implemented the Kurucz package
{\it atlas9} to calculate specific models.
The former approach has the advantage that the interpolation can be performed
with minimal human intervention and at minimal computational cost.  In
particular, we relied upon the grids to narrow in on the spectroscopic
physical parameters of each star ($\S$~\ref{subsec-pparm}).
When applicable we constructed model atmospheres with enhanced 
$\alpha$-elements using the $+0.4$~dex enhanced Rosseland opacities and 
the appropriate opacity distribution functions.

\subsection{$gf$ Values}
\label{subsec-loggf}

Columns 4 and 5 of Table~\ref{tab:ew} list the adopted $gf$
values and their references for our sample of measured absorption lines.
In general, we selected the most accurate and recent laboratory measurements 
available, avoiding solar $gf$ values where possible.  
Even with these accurate laboratory values, however, 
the $gf$ values pose a major source of uncertainty in the analysis 
particularly with respect to obtaining measurements relative to the
solar meteoritic abundances which will serve as our abundance reference frame.
We address this issue in $\S$~\ref{sec-solar} by performing
an analysis of the solar spectrum.  In the following, 
we discuss the
criteria established to select the Fe~I and Fe~II $gf$ values which are
critical in determining the spectroscopic atmospheric parameters of
each star.  We reserve comments on the remaining elements to 
$\S$~\ref{sec-elem}.

\begin{figure}
\includegraphics[height=4.3in, width=3.5in]{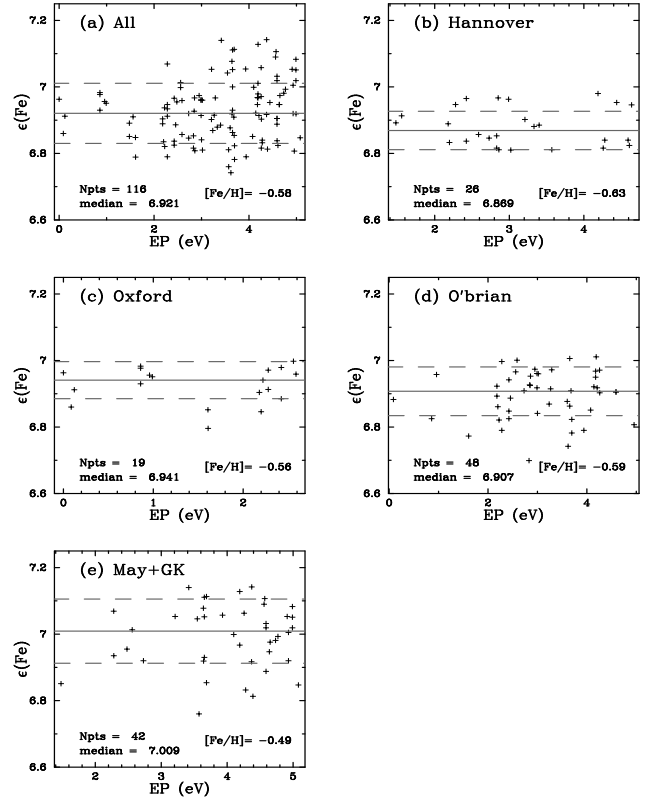}
\caption{Plots of $\e{Fe}$ values versus Excitation Potential (EP) for
six different sets of Fe~I $gf$ values: (a) all samples with the priority given
in the text, (b) Hannover measurements, (c) Oxford measurements, 
(d) O'Brian values, and (e) May74+GK81.  The various data sets yield 
systematically different $\e{Fe}$ results.  In particular, note the offset
between the Hannover and May74+GK81 values which cover nearly the same 
range in EP space.  In our analysis, we have chosen to discount the 
May74+GK81 $gf$ measurements.
}
\label{FeIfig}
\end{figure}

To minimize the uncertainties and systematic errors associated with solar
$gf$ values, we restricted the Fe~I 
analysis to laboratory $gf$ measurements.
The principal sources that we considered are:
(1) the Hannover measurements \citep{bard91,bard94},
(2) the Oxford $gf$ values \citep{black95a}, 
(3) the O'Brian values \citep{obrian91}, and
(4) the May74+GK81\footnote{The GK81 values are solar-$gf$ values normalized
to the May74 measurements.} sample \citep{may74,gk81}.
At present, these are the most accurate measurements for a sizeable
number of Fe~I lines in our wavelength range. To
investigate systematics associated with the various Fe~I
$gf$ data sets,  we performed a thorough Fe abundance analysis
of the star G114$-$19.
In the process, we identified several lines that yielded highly 
discrepant $\e{Fe}$ values even though there was no obvious blend.  
These lines are flagged in Table~\ref{tab:ew} and were removed from
any subsequent analysis.  
In Figure~\ref{FeIfig} we present the $\e{FeI}$ values versus 
excitation potential (EP) for the four sets of Fe~I $gf$ values as well as the
entire Fe~I $gf$ sample for
the G114$-$19 star assuming a Kurucz model atmosphere with
$T_{eff} = 5310$K, $\log g = 4.57$, [M/H]~$= -0.60$, and 
$\xi = 0.80 \mkms$.
Panel (a) presents the complete sample of lines where for lines with 
multiple $\log gf$ measurements we prioritized the values in this
order: Hannover, Oxford, O'Brian, and May74+GK81.
The remaining panels present the $gf$ subsets for the 
(b) Hannover 
(c) Oxford, 
(d) O'Brian, and
(e) May74+GK81 sources.  
We observe the following trends.
First, the Oxford set of lines which all
have EP $< 2.5$~eV yield systematically
higher $\e{Fe}$ values than the Hannover group whose typical EP $> 2$~eV.
The discrepancy in $\e{Fe}$ derived from these two $gf$ data sets
has been discussed at length in the literature 
\citep{black95b,holw95}. 
Both groups are confident in the accuracy of their laboratory measurements
and have argued that the other's
solar equivalent width measurements or spectral analysis techniques are
to blame.  The fact that this offset is also apparent in our solar-like stars
favors the assertion put forth by \cite{grvss99} that this EP dependent
disagreement
indicates an error in the model atmospheres.
\cite{grvss99} suggest an {\it ad hoc} modification to the solar atmosphere
$T-\tau$ relation
which could be applied to our sample of stars but 
is beyond the scope of the paper.  Instead we chose to adopt 
the Hannover and Oxford
$gf$ values with the caveat that the observed offset could result in an 
underestimate of the true effective temperature. 
As described in $\S$~\ref{sec-solar}, we derive a $T_{eff}$ value for the
Sun based on these $gf$ values and the Kurucz atmospheres which is in 
good agreement with the known value.  Furthermore, we perform an abundance
analysis relative to a solar analysis with the same model atmospheres and
stellar abundance techniques which should minimize any of the systematic
effects that the \cite{grvss99} atmosphere addresses.
Contrary to the Oxford/Hannover discrepancy, 
the O'Brian lines (Panel d) cover a larger range of EP
and yield $\e{Fe}$ values with a median in between the Oxford and Hannover
groups.  Furthermore, the few lines with EP $< 2$~eV do not show 
systematically higher $\e{Fe}$ values, although 
this could be the result of small
number statistics or inaccuracies in the O'Brian $gf$ values.
In any case, we include this large sample of reasonably accurate values
which comprise almost $50\%$ of our total Fe~I line-list.
Finally, consider the May74+GK81 lines.  These lines 
cover nearly the same EP range as the Hannover data set yet yield even 
higher $\e{Fe}$ values than the Oxford sample.
Given the greater accuracy of the Hannover measurements, we have decided
to discount the May74+GK81 sample altogether.

In contrast to the Fe~I lines, there are 
few accurate Fe~II $gf$ measurements and a careful intracomparison
is not warranted.  The lack of Fe~II $gf$ values is particularly unfortunate
because Fe$^+$ is the dominant ionization state of Fe in these thick disk
stars and therefore the Fe~II measurements
are less sensitive to non-LTE conditions or model atmosphere inaccuracies.
All of the adopted
Fe~II $gf$ values are laboratory measurements taken
from the following sources 
listed in decreasing priority: 
\cite{schna99,biem91,heise90,mcw95,kroll87,moity83,fuhr88}.

\begin{figure*}[ht]
\includegraphics[height=4.8in, width=3.6in]{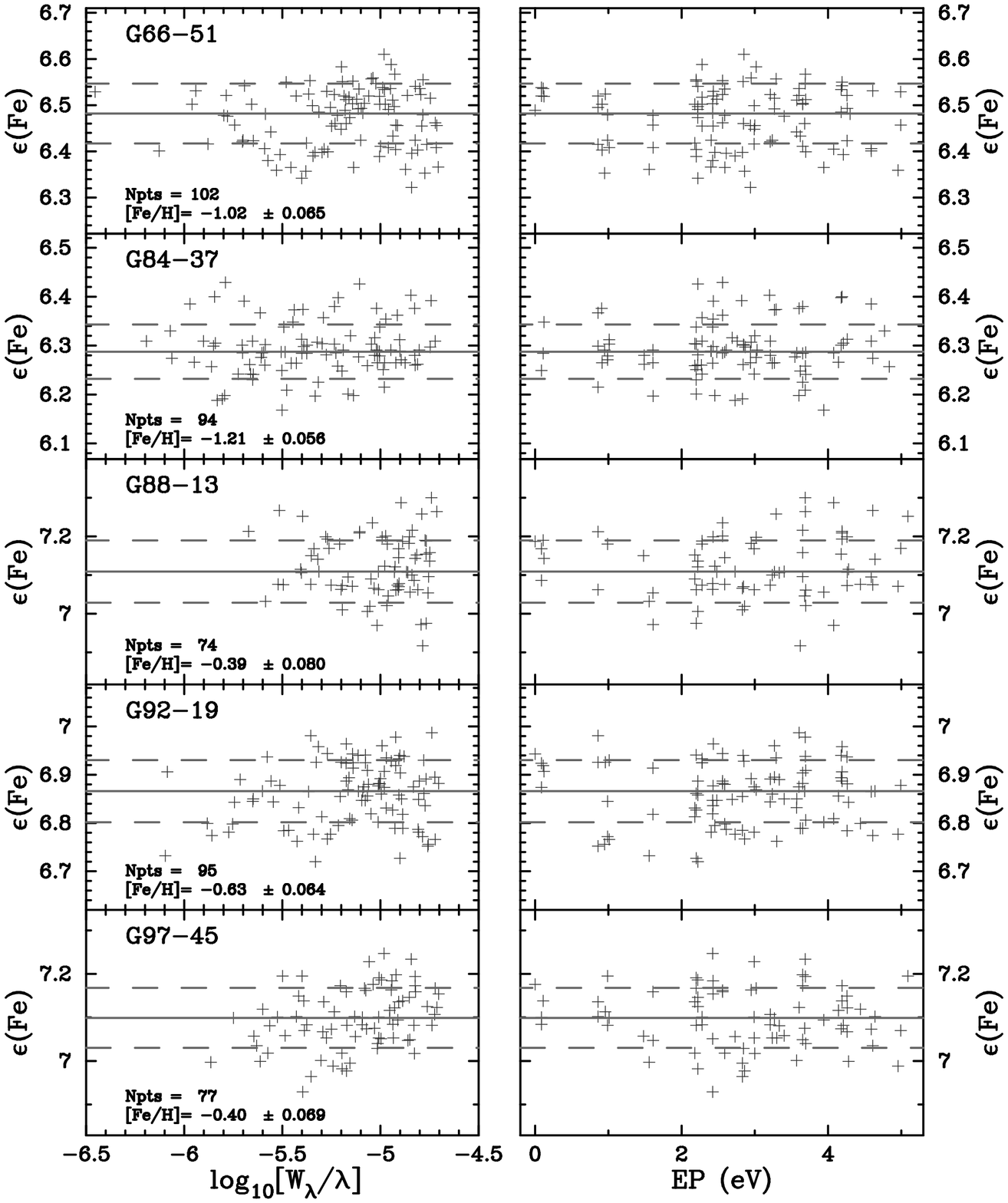}
\includegraphics[height=4.8in, width=3.6in]{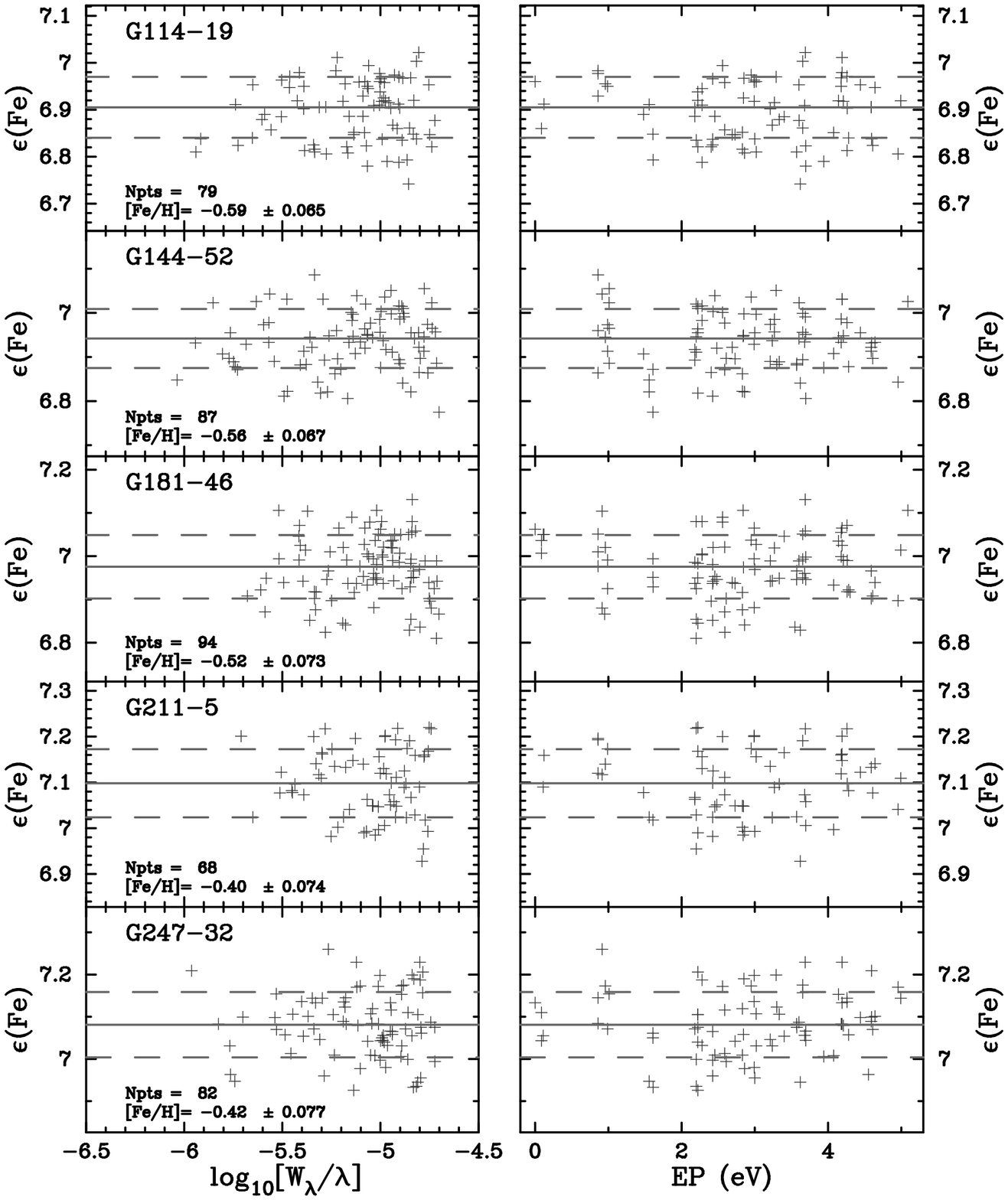}
\caption{Plots of $\e{Fe}$ vs.\ the reduced equivalent width
($\log \ew / \lambda$) and Excitation
Potential (EP) derived from the Fe~I lines
for the 10 program stars.  The $\e{FeI}$ values correspond
to standard model atmospheres with physical parameters listed in 
Table~\ref{tab:sparm}.
These parameters were determined by minimizing the trends of $\e{Fe}$ vs.\
$\log \ew / \lambda$ and EP simultaneously
and by requiring that the median $\e{Fe}$
value from Fe~I match that for Fe~II.}
\label{fig:pparm}
\end{figure*}

\subsection{Spectroscopic Atmospheric Parameters}
\label{subsec-pparm}

We now proceed to determine the spectroscopic atmospheric 
parameters -- temperature
$T_{eff}$, gravity $\log g_{spec}$, microturbulence $\xi$, and
metallicity [Fe/H] -- for each star.  Following
standard practice, we assume
local thermodynamic equilibrium (LTE) holds throughout the stellar atmosphere
which is a good assumption for G dwarf stars.  To perform the analysis,
we have used the stellar line analysis
software package MOOG (v.\ 1997) kindly provided
by C. Sneden.  In the mode {\it abfind}, the MOOG package inputs a model
stellar atmosphere, and a list of data on each absorption line
($\ew$, $\lambda$, EP, $\log gf$) and then matches the observed $\ew$ values
with a computed $\ew$ value by adjusting the elemental abundance.
For damping, we assumed the Unsold approximation with no enhancement.
As a check, 
we experimented with other assumptions for the damping, in particular
the Blackwell correction and a factor of two enhancement to the Unsold
approximation.  To our surprise, under neither of these latter assumptions were
we able to derive a model atmosphere which was physically reasonable.
Specifically, the 
enhancements suggest a lower microturbulence which in turn require
a lower temperature which then implies a lower microturbulence.  
In any case, we have adopted the same damping approximation (no enhancement)
for our solar analysis ($\S$~\ref{sec-solar}) and hope to have minimized the
effects on our final results.

To measure the 
spectroscopic atmospheric parameters, one modifies the model
atmosphere to satisfy three constraints:
(1) minimize the slope of $\e{FeI}$ vs.\ $W_\lambda / \lambda$; 
(2) minimize the slope of $\e{FeI}$ vs.\ EP; and
(3) require that the median $\e{FeI}$ value
equal the median $\e{FeII}$ value. 
The first constraint determines $\xi$ because the adopted microturbulence
value has a significant effect on the abundances derived from large
equivalent width lines, i.e.\ lines which suffer from
saturation.  Therefore, 
requiring that the $\e{FeI}$ values exhibit no trend with $\ew$ sets
the microturbulence of the model atmosphere.
Similarly, the slope of $\e{FeI}$ values vs.\ EP is sensitive
to the effective temperature because the predicted population of various
EP levels is a function of the temperature of the stellar atmosphere.
Finally, $\log g$ is constrained by requiring that the Fe
abundance derived from the Fe~II lines -- which are sensitive to $\log g$ -- 
match the Fe abundance from Fe~I.  This parameter is perhaps the most
uncertain as it is dependent on systematic errors in both the Fe~I and
Fe~II $gf$ values.
In practice, the constraints are mildly
degenerate in the atmospheric parameters and one iteratively solves for the
model atmosphere.
Our approach was to guess values of $\log g$ and [Fe/H]
and then use a $\chi^2$
minimization routine {\it auto\_ab4} developed by A. McWilliam
to find the $\xi$ and 
$T_{eff}$ values which minimize the slope of $\e{FeI}$ vs.\ $\log \ew/\lambda$
and EP.  This notion placed typical error estimates for $T_{eff}$ and
$\xi$ at $\approx 50$~K and 0.05~\kms respectively.
We then adjusted the $\log g$ and [Fe/H] 
values and reran {\it auto\_ab4} until Fe~I and Fe~II were brought
into agreement with one another. 
To minimize our effort, we calculated the stellar atmospheric
models by interpolating the grids provided on R. Kurucz's web 
site\footnote{http://cfaku5.harvard.edu}.
Once we determined reasonable values for $T_{eff}$, $\log g$,
$\xi$, and [Fe/H], we derived a final stellar atmosphere with {\it atlas9}.
This final model atmosphere was then adopted in the abundance analysis
of all of the elements.  As we shall see, nearly every star observed in this
sample has enhanced $\alpha$-elements (Mg, Si, O, etc.).  
For the stars with [Fe/H]~$\approx -0.5$, 
Mg is a principal source of electrons and therefore
an $\alpha$-enhancement can significantly modify the model atmosphere.
In particular, ignoring the higher electron density associated with 
an enhanced magnesium abundance leads to a systematic overestimate of the
stellar gravity.  Therefore, we also derived
an $\alpha$-enhanced model atmosphere with {\it atlas9} (assuming
[$\alpha$/Fe]=+0.4~dex and using the
appropriate opacities) and performed a complete
abundance analysis with this enhanced atmosphere including a reanalysis
of the atmospheric parameters.  We have found, however, that
these atmospheres imply small differences from the 
results of the standard atmospheres.  

\begin{table*}[ht]
\begin{center}
\caption{{\sc ATMOSPHERIC SPECTROSCOPIC PARAMETERS} \label{tab:sparm}}
\begin{tabular}{lcccccccccc}
\tableline
\tableline 
Star & $T_{spec}$ & [M/H] & $\log g$ & $\xi$ && $T_{spec}^\alpha$
& [M/H]$^\alpha$ & $\log g^\alpha$ & $\xi^\alpha$ \\ 
& (K) & & & (km/s) && (K) & & & (km/s) \\
\tableline 
G66-51  &  5220 & $-1.00$ & 
 4.55 & 0.64 &&  5255 & $-1.00$ &  4.48 & 0.90  \\
G84-37  &  5700 & $-1.20$ & 
 4.20 & 1.11 &&  5765 & $-1.20$ &  4.15 & 1.20  \\
G88-13  &  5220 & $-0.40$ & 
 4.60 & 0.86 &&  5270 & $-0.30$ &  4.55 & 1.05  \\
G92-19  &  5530 & $-0.60$ & 
 4.45 & 1.05 &&  5545 & $-0.60$ &  4.40 & 1.15  \\
G97-45  &  5550 & $-0.45$ & 
 4.50 & 0.90 &&  5580 & $-0.40$ &  4.40 & 1.10  \\
G114-19 &  5310 & $-0.60$ & 
 4.57 & 0.80 &&  5350 & $-0.60$ &  4.52 & 0.95  \\
G144-52 &  5575 & $-0.55$ & 
 4.58 & 0.90 &&  5610 & $-0.55$ &  4.45 & 1.13  \\
G181-46 &  5380 & $-0.50$ & 
 4.53 & 0.82 &&  5400 & $-0.50$ &  4.45 & 0.95  \\
G211-5  &  5320 & $-0.40$ & 
 4.55 & 1.07 &&  5360 & $-0.40$ &  4.45 & 1.25  \\
G247-32 &  5360 & $-0.40$ & 
 4.45 & 0.88 &&  5400 & $-0.40$ &  4.35 & 1.05  \\
\tableline 
\end{tabular}
\end{center}
\end{table*}

 Figure~\ref{fig:pparm} presents (a) $\e{FeI}$ vs.\ $\ew/\lambda$ and
(b) $\e{FeI}$ vs.\ EP plots for every thick disk star in the sample assuming
the standard model atmospheres.
The physical parameters of the final model atmospheres (with and without
$\alpha$-enhancement) are listed in Table~\ref{tab:sparm}.
As expected, the $\alpha$-enhanced models tend to have lower
$\log g$ values by $\approx 0.05 - 0.1$~dex.  
We will show that the typical 
$\alpha$-enhancement is $0.2-0.3$~dex so the most accurate spectroscopic
gravity is more likely the average of the two values.  
Furthermore, the $\alpha$-enhanced models require slightly higher $T_{eff}$
values to compensate the larger opacity implied by the increase in electron 
density.
With the notable
exception of G84-37, our spectroscopic $T_{eff}$, $\log g$, and [Fe/H]
values are systematically higher than the photometric values 
based on the \cite{carn94} observations
listed in Table~\ref{tab:pparm}.  
Excluding G84-37, the offset between the spectroscopic 
temperatures and the \cite{carn94} values is 
$<T_{spec} - T_{Carn94}> \; = +87 \pm 15$K for the
standard atmospheres. 
On the other hand, comparing the spectroscopic values 
with the \cite{alonso95} color-temperature
relations based on the Infrared Flux Method (IRFM), we find excellent
agreement: $<T_{spec} - T_{Alonso96}> \; = -1 \pm 12$K 
for the 4 stars with $uvby$ photometry.
The agreement is possibly the result of small number statistics,
however, particularly given the excellent overall agreement between
the two photometric techniques for low mass main
sequence stars \citep{alonso96b}.
As we increase the sample of thick disk stars, it will be important to further
compare the various temperature scales.
Finally, as we will ultimately be interested in a
comparison of the \cite{edv93} and \cite{chen00}
results with our analysis, it is
important to examine their temperature scales.  As an indirect
test, we compared the 33 overlapping stars from the \cite{alonso96}
and \cite{edv93} samples which are nearby (minimally affected by dust)
and in the appropriate temperature range $(T_{eff} < 5900$K).  We find
$<T_{Alonso96} - T_{Edvard}> \; = -95 \pm 6$K, 
so we will assume that our stars are $\approx 100$K cooler than the 
\cite{edv93} temperature scale.  Similarly \cite{chen00}, whose
temperatures are based on the IRFM scale,  report a 
systematically lower temperature $(\approx 70$K) than 
\cite{edv93}.  Therefore, we will assume no temperature offset between
our analysis and that of \cite{chen00}.

While a difference between the spectroscopic and the
color-temperature scales may not be surprising,
the stellar gravity offset is more difficult to explain.
Taking the average spectroscopic gravity from the standard and
$\alpha$-enhanced atmospheres (and again ignoring G84-37 for now), 
the average offset is 
$<\log g_{Hipp} - \log g_{spec}> \; = -0.08 \pm 0.01$~dex.  
We might suggest that a systematic error in the Fe~II $gf$ values
has biased the spectroscopic gravity, but our $\log g$ measurement
for the Sun is in excellent agreement with the known value.
A fraction of the offset (0.03$-$0.05~dex) can be explained by
the difference in assumed effective temperature, but it can not
account for the entire discrepancy.
Nonetheless, the effect of even a 0.1~dex error in $\log g$ has
a minimal impact on the abundances we derive, particularly since we
rely primarily on the Fe~I lines to determine [Fe/H].
Finally, note that the metallicity we compute from the Fe lines 
is systematically
$\approx 0.1$~dex higher than the [M/H] values derived by 
\cite{carn94} from low $S/N$ spectra.  In this case, the difference
is largely explained by the offset in temperature and gravity
as both imply a higher metallicity.  
While a 0.1~dex systematic error in the spectroscopic metallicity
measurements will not significantly affect our conclusions on the abundance
trends of the thick disk stars, a systematic error in the \cite{carn94}
[M/H] measurements
could have important implications for the metallicity distribution
function of the thick disk.  Therefore, we will carefully reassess 
this discrepancy once we have a larger, statistically significant sample
of accurate spectroscopic measurements.


\begin{table*}[ht] \footnotesize
\begin{center}
\caption{ {\sc ERROR ANALYSIS FOR G84-37} \label{tab:erra}}
\begin{tabular}{lcccccccccc}
\tableline
\tableline 
Ion & $N$ & $\Delta T_{eff}$ & $\Delta T_{eff}$ & $\Delta \log g$ & 
$\Delta \log g$& $\Delta \xi$ & $\Delta \xi$ &
$\Delta \ltk \frac{\rm M}{\rm H}\rtk$ & $\Delta \ltk \frac{\rm M}{\rm H}\rtk$ 
& $\alpha$ \\
& & +50K & --50K & +0.05 & --0.05& +0.05 & --0.05& +0.05 & --0.05 \\
\tableline 
 Fe~I/H& 94&$+0.050$&$-0.042$&$-0.002$&$+0.006$&$-0.006$&$+0.011$&$+0.000$&$+0.000$&$+0.025$ \\
 Fe~II/H& 17&$-0.003$&$+0.001$&$+0.018$&$-0.017$&$-0.011$&$+0.013$&$+0.006$&$-0.005$&$-0.009$ \\
 O~I/Fe&  3&$-0.102$&$+0.090$&$+0.017$&$-0.023$&$+0.004$&$-0.009$&$-0.003$&$+0.002$&$-0.080$ \\
 Na~I/Fe&  3&$-0.023$&$+0.015$&$-0.002$&$-0.001$&$+0.005$&$-0.010$&$+0.000$&$+0.000$&$-0.011$ \\
 Mg~I/Fe&  5&$-0.037$&$+0.030$&$+0.002$&$-0.006$&$+0.006$&$-0.011$&$+0.001$&$+0.001$&$-0.017$ \\
 Si~I/Fe& 12&$-0.032$&$+0.025$&$+0.004$&$-0.007$&$+0.004$&$-0.008$&$+0.000$&$+0.000$&$-0.018$ \\
 S~I/Fe&  1&$-0.089$&$+0.077$&$+0.017$&$-0.024$&$+0.005$&$-0.011$&$-0.002$&$+0.001$&$-0.067$ \\
 Ca~I/Fe& 17&$-0.017$&$+0.008$&$-0.003$&$+0.000$&$+0.000$&$-0.005$&$+0.000$&$+0.001$&$-0.014$ \\
 Sc~II/Fe&  4&$-0.040$&$+0.031$&$+0.020$&$-0.025$&$+0.003$&$-0.009$&$+0.005$&$-0.006$&$+0.001$ \\
 Ti~I/Fe& 17&$+0.003$&$-0.012$&$+0.002$&$-0.003$&$+0.002$&$-0.005$&$+0.000$&$+0.001$&$+0.000$ \\
 Ti~II/Fe& 16&$-0.038$&$+0.030$&$+0.021$&$-0.024$&$+0.000$&$-0.008$&$+0.006$&$-0.005$&$-0.009$ \\
 Cr~I/Fe& 12&$+0.003$&$-0.004$&$+0.001$&$-0.003$&$+0.003$&$-0.002$&$+0.001$&$+0.000$&$+0.004$ \\
 Cr~II/Fe&  4&$-0.060$&$+0.049$&$+0.020$&$-0.025$&$+0.001$&$-0.007$&$+0.002$&$-0.003$&$-0.037$ \\
 Mn~I/Fe& 10&$-0.013$&$+0.002$&$+0.002$&$-0.006$&$+0.005$&$-0.011$&$+0.000$&$+0.000$&$-0.005$ \\
 Co~I/Fe&  1&$+0.000$&$-0.009$&$+0.003$&$-0.007$&$+0.005$&$-0.011$&$+0.001$&$+0.001$&$+0.009$ \\
 Ni~I/Fe& 21&$-0.021$&$+0.012$&$+0.002$&$-0.006$&$+0.003$&$-0.009$&$+0.000$&$-0.001$&$-0.013$ \\
 Cu~I/Fe&  2&$-0.010$&$+0.002$&$+0.005$&$-0.006$&$+0.006$&$-0.010$&$+0.000$&$+0.000$&$+0.000$ \\
 Zn~I/Fe&  2&$-0.037$&$+0.029$&$+0.010$&$-0.014$&$-0.001$&$-0.004$&$+0.002$&$-0.002$&$-0.022$ \\
 Y~II/Fe&  5&$-0.035$&$+0.027$&$+0.020$&$-0.026$&$-0.002$&$-0.002$&$+0.006$&$-0.006$&$-0.003$ \\
 Ba~II/Fe&  3&$-0.023$&$+0.017$&$+0.013$&$-0.013$&$-0.023$&$+0.024$&$+0.006$&$+0.001$&$-0.002$ \\
\tableline 
\end{tabular}
\end{center}
\end{table*}
\begin{table*}[ht] \footnotesize
\begin{center}
\caption{ {\sc ERROR ANALYSIS FOR G114-19} \label{tab:errb}}
\begin{tabular}{lcccccccccc}
\tableline
\tableline 
Ion & $N$ & $\Delta T_{eff}$ & $\Delta T_{eff}$ & $\Delta \log g$ & 
$\Delta \log g$& $\Delta \xi$ & $\Delta \xi$ &
$\Delta \ltk \frac{\rm M}{\rm H}\rtk$ & $\Delta \ltk \frac{\rm M}{\rm H}\rtk$ 
& $\alpha$ \\
& & +50K & --50K & +0.05 & --0.05& +0.05 & --0.05& +0.05 & --0.05 \\
\tableline 
 Fe~I/H& 79&$+0.031$&$-0.042$&$-0.001$&$+0.004$&$-0.007$&$+0.006$&$+0.008$&$-0.007$&$+0.006$ \\
 Fe~II/H& 20&$-0.021$&$+0.019$&$+0.026$&$-0.028$&$-0.005$&$+0.005$&$+0.011$&$-0.012$&$+0.031$ \\
 O~I/Fe&  3&$-0.079$&$+0.084$&$+0.025$&$-0.031$&$+0.006$&$-0.005$&$-0.014$&$+0.011$&$-0.015$ \\
 Na~I/Fe&  3&$+0.002$&$+0.007$&$-0.007$&$+0.005$&$+0.006$&$-0.005$&$-0.006$&$+0.005$&$+0.011$ \\
 Mg~I/Fe&  6&$-0.016$&$+0.025$&$-0.005$&$+0.003$&$+0.007$&$-0.005$&$-0.005$&$+0.004$&$+0.023$ \\
 Al~I/Fe&  1&$-0.001$&$+0.011$&$-0.003$&$+0.001$&$+0.006$&$-0.005$&$-0.007$&$+0.007$&$+0.001$ \\
 Si~I/Fe& 14&$-0.031$&$+0.041$&$+0.005$&$-0.008$&$+0.005$&$-0.004$&$+0.001$&$-0.002$&$+0.035$ \\
 S~I/Fe&  1&$-0.068$&$+0.074$&$+0.025$&$-0.030$&$+0.007$&$-0.005$&$-0.012$&$+0.010$&$-0.012$ \\
 Ca~I/Fe& 16&$+0.014$&$-0.003$&$-0.017$&$+0.020$&$+0.000$&$+0.001$&$+0.002$&$+0.001$&$+0.031$ \\
 Sc~II/Fe&  7&$-0.030$&$+0.039$&$+0.023$&$-0.026$&$+0.006$&$-0.002$&$+0.007$&$-0.007$&$+0.049$ \\
 Ti~I/Fe& 41&$+0.030$&$-0.020$&$-0.002$&$+0.002$&$+0.006$&$-0.002$&$-0.006$&$+0.008$&$+0.003$ \\
 Ti~II/Fe& 12&$-0.028$&$+0.038$&$+0.021$&$-0.024$&$-0.003$&$+0.006$&$+0.008$&$-0.007$&$+0.031$ \\
 V~I/Fe& 15&$+0.037$&$-0.028$&$-0.002$&$+0.001$&$+0.006$&$-0.004$&$-0.006$&$+0.007$&$+0.020$ \\
 Cr~I/Fe& 13&$+0.012$&$-0.009$&$-0.015$&$+0.008$&$-0.006$&$+0.000$&$-0.003$&$-0.002$&$+0.001$ \\
 Cr~II/Fe&  5&$-0.044$&$+0.053$&$+0.023$&$-0.027$&$+0.003$&$-0.002$&$+0.000$&$-0.001$&$+0.002$ \\
 Mn~I/Fe& 10&$+0.024$&$-0.027$&$-0.001$&$-0.001$&$+0.005$&$-0.004$&$-0.005$&$+0.005$&$+0.013$ \\
 Co~I/Fe&  9&$+0.003$&$+0.008$&$+0.007$&$-0.010$&$+0.006$&$-0.005$&$-0.002$&$+0.002$&$+0.030$ \\
 Ni~I/Fe& 28&$-0.008$&$+0.014$&$-0.008$&$+0.000$&$-0.003$&$+0.001$&$+0.004$&$-0.005$&$+0.013$ \\
 Cu~I/Fe&  1&$-0.015$&$+0.028$&$+0.008$&$-0.010$&$+0.005$&$-0.003$&$+0.000$&$+0.001$&$+0.021$ \\
 Zn~I/Fe&  2&$-0.031$&$+0.041$&$+0.007$&$-0.010$&$-0.005$&$+0.006$&$+0.005$&$-0.006$&$+0.002$ \\
 Y~II/Fe&  1&$-0.023$&$+0.030$&$+0.016$&$-0.023$&$-0.010$&$+0.007$&$+0.007$&$-0.011$&$+0.000$ \\
 Ba~II/Fe&  2&$-0.016$&$+0.024$&$+0.012$&$-0.016$&$-0.016$&$+0.015$&$+0.011$&$-0.013$&$+0.003$ \\
 Eu~II/Fe&  1&$-0.029$&$+0.039$&$+0.023$&$-0.026$&$+0.006$&$-0.004$&$+0.007$&$-0.008$&$+0.055$ \\
\tableline 
\end{tabular}
\end{center}
\end{table*}

\subsection{Error Analysis}
\label{subsec-Err}

To assess the systematic effects of the model atmospheric parameters on
the elemental abundance ratios,
we have performed a standard abundance error analysis. 
We calculated the elemental abundances for 10 atmospheric models
for two stars (G114-19 and G84-37) using the Kurucz atmosphere grids:
(1) the best fit model; (2,3) 
$T' = T \pm 50$K; (4,5) $\xi' = \xi \pm 0.05 \mkms$; (6,7) 
$\log g' = \log g \pm 0.05$; (8,9) 
[M/H]$'$ = [M/H] $\pm 0.05$~dex; and (10) a +0.4 $\alpha$-enhanced atmosphere. 
The two stars were chosen to have significantly
different stellar atmospheres and the range of parameters roughly corresponds
to our estimated $1 \sigma$ statistical uncertainty.
Tables~\ref{tab:erra} and \ref{tab:errb} summarize the results of the
error analysis for the two stars.
For those elements with $N > 5$ absorption
lines, we expect the uncertainties in the abundances to be dominated by
errors in the atmospheric parameters.  For the remaining elements, the
errors in $\ew$ (i.e. Poissonian noise, blends, continuum error) are 
significant.  In $\S$~\ref{sec-elem} we remark on those elements for which
errors in the $\ew$ measurements are a particular problem.

\subsection{Hyperfine Splitting}
\label{subsec-hfs}

Isotopes with an odd number of protons and/or
neutrons experience hyperfine interactions between
the nucleus and electrons.  These interactions split the lines 
into multiple components with typical separations of $1 - 10$m\AA.  
For strong lines with large equivalent
width the effect is to de-saturate the absorption line, a
phenomenon which must
be taken into account in order to accurately measure the elemental 
abundance.  In the case of several Cu~I lines, for example, hyperfine splitting
leads to a correction of over 0.5~dex.  For our abundance analysis,
we have included hfs corrections 
for Mn, Ba, Sc, Co, and Cu\footnote{The effects of hfs are
negligible for the very weak Eu~II 6645 line and insignificant for the
Y~II lines}.  
With the exception of Ba where we have implemented the results from 
\cite{mcw98}, we adopt the wavelengths of the hfs transitions
from Kurucz's hyperfine tables \citep{kur99}
and calculated the relative $gf$ strengths according to the equations
in Appendix~\ref{app-hfs}.
Table~\ref{tab:hfs} lists all of the hfs transitions considered
here.  For these lines we implemented the MOOG package in the synthesis
mode {\it blends} which matches the observed equivalent widths to that
calculated from a synthesis of the blended hyperfine lines.

\begin{figure}
\includegraphics[height=4.3in, width=3.5in]{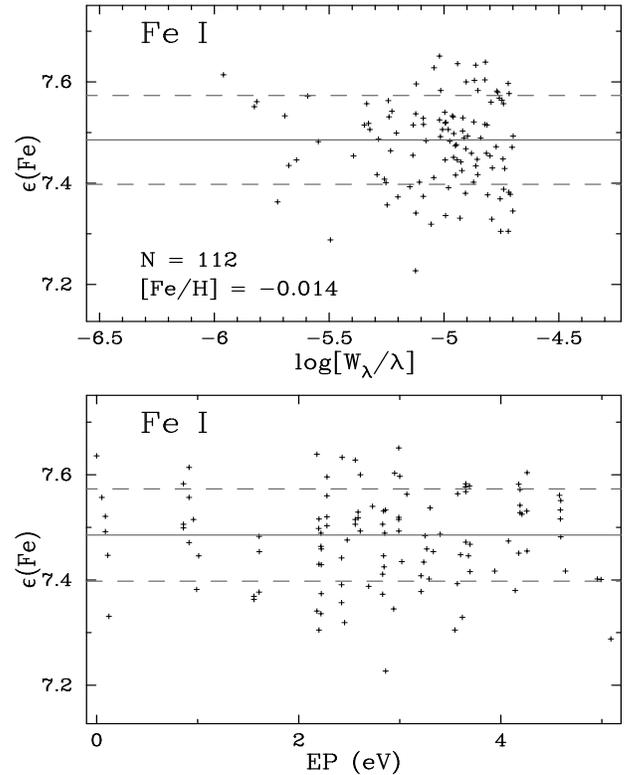}
\caption{$\e{Fe}$ values vs.\ 
$\log \ew / \lambda$ and EP derived from 112 Fe~I measured in the
Kurucz solar spectrum.  By minimizing the slope of $\e{Fe}$ with
$\log \ew / \lambda$ and EP, we derived the 'best' atmospheric parameters
for the Sun ($T_{eff}$, $\log g$, [M/H], $\xi$).  
To our surprise, these parameters agree very well with the
known parameters. }
\label{fig:solar-pparm}
\end{figure}

\section{SOLAR ANALYSIS}
\label{sec-solar}

In order to facilitate abundance comparisons between our thick disk sample
and other stellar populations or galactic systems (e.g.\ the damped \lya
systems), it is crucial to compare our results with a spectroscopic solar
analysis.  
In this fashion, we can report our abundances relative to 
the solar meteoritic abundances \citep{grvss96}
by comparing the solar analysis with the meteoritic values\footnote{Implied in this exercise
is the assumption that the solar spectroscopic
abundances must equal the meteoritic.  While this is supported by the
excellent agreement between the two for many elements, there are notable
exceptions and we warn the reader that this assumption need not hold}.
This exercise also accentuates 
systematic errors associated with the $\log gf$ values or blending
of individual ions.  More ambitiously, by making a line by line comparison
we might eliminate errors in the model atmospheres, damping, and the
stellar analysis package, particularly given that these thick disk stars have
similar spectral types to the Sun.
Therefore, we performed an elemental abundance
analysis of the Sun applying the exact same techniques utilized for the thick
disk stars.  
The equivalent widths were
measured with the {\it getjob} package, we adopted Kurucz model atmospheres,
and constrained the atmospheric parameters with {\it auto\_ab4}.
The only differences lie in the solar spectrum itself; we have
analyzed the Kurucz solar spectrum \citep{kur84} obtained with resolving
power 522,000 and signal-to-noise in excess of 2000.
Column~6 of Table~\ref{tab:ew} lists the $\ew$ 
measurements for the Sun which were included
in the abundance analysis.  We estimate that the typical error of each 
measurement is $1-2$m\AA, the dominant sources of error being line blends
and poor continuum determination.
Figure~\ref{fig:solar-pparm} presents the $\e{Fe}$ vs.\ $\ew/\lambda$
and EP plots for 112 Fe~I lines measured from the solar spectrum.  
Somewhat to our surprise, the physical parameters
that we derive are in excellent agreement with the known values:
$T_{eff} = 5750 \pm 50$~K, 
$\log g = 4.44 \pm 0.03$, $\xi = 1.00 \pm 0.03 \mkms$, and [M/H] = 0.0~dex.
Perhaps most astonishing, 
the Fe abundance matches the meteoritic value
to within 0.015~dex. While the nearly exact agreement is
probably fortuitous, our analysis indicates no significant disagreement
between the photometric and meteoritic solar Fe abundance.

\begin{table}[ht]
\begin{center}
\caption{ {\sc SOLAR ABUNDANCES RELATIVE TO METEORITIC} \label{tab:solar}}
\begin{tabular}{lccrrc}
\tableline
\tableline
Ion & $\e{X}\tablenotemark{a}$ & $N$ & 
[X/H]$_{d}$ & [X/H]$_{n}$ & $\sigma$ \\
\tableline
C I   &  8.55 &   4 & $ 0.23$ & $ 0.19$ &  0.18 \\
O I   &  8.87 &   3 & $ 0.12$ & $ 0.12$ &  0.02 \\
Na I  &  6.32 &   3 & $ 0.00$ & $-0.01$ &  0.02 \\
Mg I  &  7.58 &   3 & $ 0.03$ & $ 0.04$ &  0.02 \\
Al I  &  6.49 &   2 & $-0.09$ & $-0.09$ &  0.05 \\
Si I  &  7.56 &  16 & $ 0.00$ & $ 0.01$ &  0.04 \\
S I   &  7.20 &   2 & $ 0.35$ & $ 0.35$ &  0.00 \\
Ca I  &  6.35 &  20 & $-0.07$ & $-0.05$ &  0.10 \\
Sc II &  3.10 &   9 & $ 0.17$ & $ 0.17$ &  0.08 \\
Ti I  &  4.94 &  47 & $-0.05$ & $-0.04$ &  0.07 \\
Ti II &  4.94 &  19 & $ 0.11$ & $ 0.14$ &  0.13 \\
V I   &  4.02 &  17 & $-0.12$ & $-0.12$ &  0.06 \\
Cr I  &  5.67 &  14 & $ 0.01$ & $ 0.02$ &  0.07 \\
Cr II &  5.67 &   6 & $ 0.09$ & $ 0.14$ &  0.12 \\
Mn I  &  5.53 &   9 & $-0.21$ & $-0.20$ &  0.07 \\
Fe I  &  7.50 & 112 & $-0.01$ & $-0.02$ &  0.09 \\
Fe II &  7.50 &  33 & $ 0.00$ & $ 0.01$ &  0.10 \\
Co I  &  4.91 &  10 & $ 0.03$ & $ 0.03$ &  0.11 \\
Ni I  &  6.25 &  33 & $ 0.00$ & $-0.01$ &  0.10 \\
Cu I  &  4.29 &   3 & $-0.01$ & $-0.05$ &  0.08 \\
Zn I  &  4.67 &   2 & $ 0.00$ & $ 0.00$ &  0.06 \\
Y II  &  2.23 &   3 & $ 0.02$ & $ 0.01$ &  0.02 \\
Ba II &  2.22 &   2 & $ 0.11$ & $ 0.11$ &  0.12 \\
Eu II &  0.54 &   1 & $ 0.10$ & $ 0.10$ \\
\tableline
\end{tabular}
\end{center}
\tablenotetext{a}{Meteoritic Solar abundances from \cite{grvss96}}
\end{table}

With the atmospheric parameters determined, we
constructed a final model atmosphere with {\it atlas9} and measured
the elemental abundances of the remaining absorption lines.
Table~\ref{tab:solar} lists the ion, the number of absorption lines
analyzed, the median and mean abundance relative to the meteoritic value,
and the standard deviation of these measurements.  While the majority of
ions are consistent with the meteoritic values, there are notable 
exceptions: Ti~II (+0.11~dex), S~I (+0.35~dex), C~I (+0.23~dex),
Mn~I ($-$0.21~dex), Cu~I ($-$0.13~dex), Cr~II (+0.09~dex),
and Sc~II (+0.19~dex).  In each of these cases, either the absolute scale
of the $gf$ values is poorly determined or the abundances are very sensitive
to the model atmospheres. 
For example, the S~I measurement is based on a single line with very high
EP and an uncertain solar $gf$ value from the lunar analysis by
\cite{francois88}.
Therefore, we expect the differences between the photometric and 
meteoritic abundance are entirely due to systematic errors in the
$gf$ measurements or errors in the details of the solar model atmosphere.
We also considered a solar analysis with the Holweger-M$\ddot{\rm u}$ller
solar atmosphere \citep{holw74} with $\xi = 1.15 \mkms$ as is commonly
adopted in stellar abundance studies.  With the exception of vanadium,
all of the derived abundances are within 0.1~dex of the values listed in
Table~\ref{tab:solar}.  There is a noticeable increase in $\e{Fe}$
derived from the Fe~I lines of +0.07~dex which would tend toward slightly
higher solar-corrected $\alpha$/Fe values, but by less than +0.05 in 
most cases.

In order to report abundances relative to the solar meteoritic values,
we must apply
any offsets between the solar photospheric values we derived and the
meteoritic values.
The zeroth order correction is to 
modify our final results by the median value computed for each element
as listed in Table~\ref{tab:solar}.  With the exception of a few elements, 
these median corrections are robust and should allow for a
reasonable estimate of the absolute abundance.  Another approach
is to measure a correction for every measured solar absorption line,
$\delta \equiv \e{X}_{obs} - \e{X}_{meteor}$, and subtract this offset
from the abundances calculated for each line in
the program stars. This is akin to adopting solar $gf$ values.
It has the advantage that the results are nearly
independent of errors in the $gf$ values and that systematic errors in
the model atmospheres and stellar analysis package are minimized.
Unfortunately, there are two significant drawbacks: (1) the error in our
solar equivalent width measurements are comparable to the expected error
in the $gf$ values, and (2) saturated or blended solar lines are excluded
limiting one to a smaller sample of absorption lines.  In fact, for Fe~I this
approach tends to increase the scatter in the $\e{Fe}$ values.
Nevertheless, we will consider this line-by-line solar correction
with all of our stars.

\section{ELEMENTAL ABUNDANCES}
\label{sec-elem}

We have computed the elemental abundances for each
absorption line with three different approaches (i) standard model atmospheres,
(ii) $\alpha$-enhanced model atmospheres, and (iii) standard model atmospheres
corrected by 
the solar abundance analysis to the solar meteoritic
abundances as performed in the previous section.
Tables~\ref{abndG66-51}-\ref{abndG247-32} present
[X/Fe], the logarithmic abundances of ion X relative
to Fe normalized to solar meteoritic abundances \citep{grvss96},
for the \nstrs thick disk stars comprising our
current sample.  Column 2 indicates the number of absorption lines
analyzed with the standard and $\alpha$-enhanced atmospheres.  Columns 3$-$5
present the median, mean, and standard deviation of the $\e{X}$ values
for the standard models while columns 6$-$8 present the same quantities
for the $\alpha$-enhanced stellar atmospheres.  
Other than the few exceptions noted below, the $\alpha$-enhanced models have
minimal effect on the [X/Fe] ratios.
Finally, column 9 lists the number of lines from the solar-corrected 
analysis and columns 10$-$12 present the corrected mean and median
values and the resultant standard deviation.

We turn now to comment on the analysis and results for individual elements.
Unless otherwise noted, we consider the median measurements in our
discussions and the error bars in the figures
refer to the unreduced standard deviation
which in the majority of cases
are {\it conservative} estimates of the true error.  In the few
cases where the standard deviation is very small $(< 0.05$), we plot
a minimum error bar of 0.05~dex because we feel this is a lower limit
to the statistical error associated with the measurements.
We also plot in the upper-right hand corner an estimated systematic error
for each abundance measurement derived by adding in quadrature the values
listed in Tables~\ref{tab:erra} and \ref{tab:errb}.
For most of the plots we present the solar-corrected ratios and in
a few cases where the corrections are very large
we also present the uncorrected values.
We also discuss the sensitivity of the results to uncertainties in the
atmospheric models, paying particular attention to the possibility 
that we have 
overestimated $T_{eff}$ for the majority of stars ($\S$~\ref{subsec-pparm}).
In the following section, we will compare the observed trends of these
thick disk stars with the halo \citep{mcw95}, thin disk
\citep{edv93,chen00}, and bulge \citep{mcw94} stellar populations.

\begin{table}[ht]
\caption{ {\sc ABUNDANCES FOR G66-51} \label{abndG66-51}}
\tablecomments{Tables~\ref{abndG66-51}-17 can be obtained electronically at\\
http://www.ociw.edu/$\sim$xavier/Science/Stars/index.html}
\end{table}

\begin{table}[ht]
\tablenum{17}
\caption{ {\sc ABUNDANCES FOR G247-32} \label{abndG247-32}}
\tablecomments{Tables~\ref{abndG66-51}-17 can be obtained electronically at\\
http://www.ociw.edu/$\sim$xavier/Science/Stars/index.html}
\end{table}

\subsection{Iron}

We computed the [Fe/H] values for the program stars from the Fe~I and
Fe~II measurements under the constraint that the adopted stellar gravity
gives a median $\e{Fe}$ value from the Fe~II lines
within 0.03~dex of the median $\e{Fe}$ value from the Fe~I lines.
In this section and all further analysis, we 
take [Fe/H] from the median $\e{Fe}$ value of the Fe~I lines.
While significant uncertainties exist for the 
Fe~I and Fe~II $gf$ values, the excellent agreement between our solar
spectroscopic Fe abundance and the meteoritic abundance raises our
confidence in the absolute value of our [Fe/H] measurements.
Furthermore, with the exception of one star, the [Fe/H] values
are essentially independent of $\alpha$-enhancement.  
This one exception, G88-13, has the highest metallicity of all of our
program stars but is otherwise unpeculiar.  While the difference in 
[Fe/H] is significant for this star, the majority of elemental abundances
relative to Fe are insensitive to the $\alpha$-enhancement and we will
generally not include this approach in our discussion.

\subsection{Alpha Elements -- O, Mg, Si, S, Ca, Ti}
\label{subsec-alpha}

Our observations include measurements on a number of $\alpha$-elements.
In this subsection, we describe the results.  We have grouped the elements
into two subsets, one with relatively robust results and the other with
poorly constrained measurements. 

\subsubsection{Silicon, Calcium, Titanium}

For the abundances of Si, Ca, and Ti, we have measured 
over 15 absorption lines and have reasonably accurate laboratory 
$gf$ values.  In the case of silicon, we have adopted the
$gf$ values from 
\cite{garz73} adjusted by $+0.1$~dex as recommended by \cite{becker80} 
as well as the solar $gf$ values from \cite{fry97} and \cite{mcw94} which give 
$\e{Si}$ values in good agreement with the adjusted \cite{garz73} abundances.
For calcium, we rely solely on the laboratory measurements by \cite{smith81}
while the titanium $gf$ values were gleaned from a number of sources with the
Oxford measurements given highest priority. 

\begin{figure}[ht]
\includegraphics[height=3.8in, width=2.8in,angle=-90]{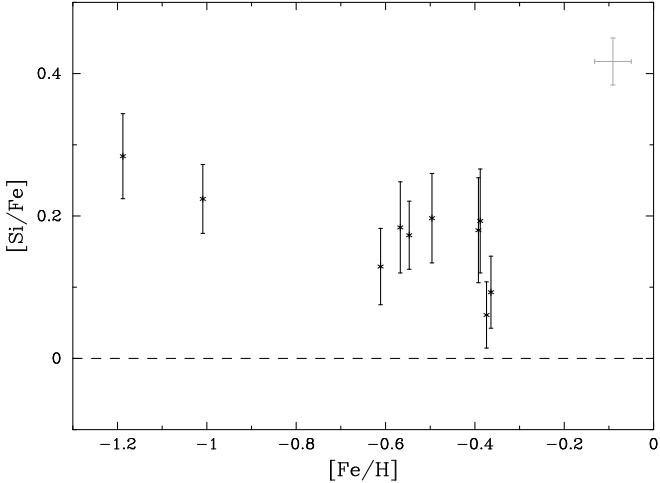}
\caption{Solar-corrected [Si/Fe] abundance ratios 
vs.\ [Fe/H] metallicity for the 10 thick disk stars.  The observations
show a clear enhancement and hint at a trend for increasing [Si/Fe]
with decreasing [Fe/H].  The dashed line at [Si/Fe] = 0 indicates the
solar meteoritic Si/Fe ratio and the gray error bars reflect the 
$1\sigma$ systematic error from the atmospheric uncertainties.}
\label{fig:Si}
\end{figure}

Silicon is a prototypical $\alpha$-element.  In addition to exhibiting
an enhancement in metal-poor stars \citep{mcw97}, theoretically
it is expected to be synthesized predominantly in moderate mass
$(\approx 20 \msol$) Type~II SN \citep{ww95}.  
In Figure~\ref{fig:Si}, we plot the solar-corrected silicon abundances 
vs.\ [Fe/H] for the thick disk stars.  The majority are significantly
enhanced and there is an indication of higher [Si/Fe] at lower metallicity
as found in most metal-poor stellar abundance studies.
In terms of the uncertainty in the atmospheric parameters, 
a decrease in $T_{eff}$ of 100K would further enhance the Si/Fe ratio
by 0.06$-$0.08~dex and the ratio is largely insensitive to the other
parameters.
As discussed in the following section, we contend that the majority of
stars are even more enhanced than their thin disk counterparts at the
same metallicity.
The obvious exceptions are 
the two highest metallicity stars (G88-13, G211-5) which exhibit 
relatively low [Si/Fe] values.
We shall note, however, that these two stars also show lower values
of calcium and oxygen yet significantly enhanced Ti and Mg.
These trends might be indicative of an overestimate of temperature for 
these two stars because Si, Ca, and O are most sensitive to $T_{eff}$ 
in G dwarf stars.  

\begin{figure}[ht]
\includegraphics[height=3.8in, width=2.8in,angle=-90]{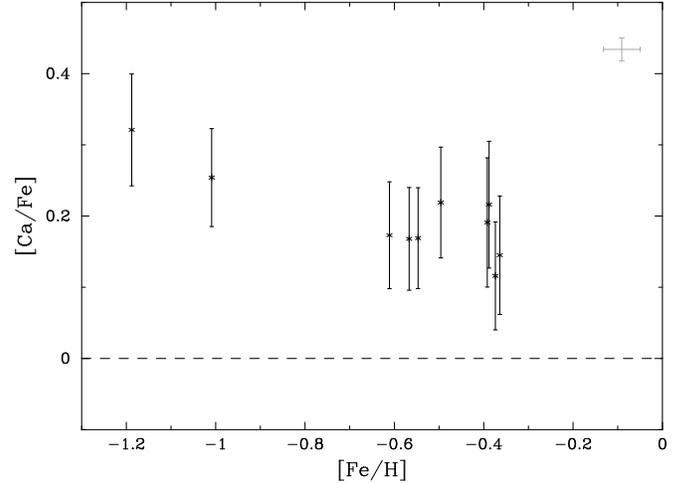}
\caption{Solar-corrected [Ca/Fe] abundance ratios 
vs.\ [Fe/H] metallicity for the 10 thick disk stars.  The observations
show a clear enhancement and hint at a trend for increasing [Ca/Fe]
with decreasing [Fe/H].  
The dashed line at [Ca/Fe] = 0 indicates the solar meteoritic Ca/Fe ratio.}
\label{fig:Ca}
\end{figure}

Like silicon, Ca is enhanced in metal-poor stars \citep{mcw97} and
is predicted to be produced in intermediate mass Type~II SN along with
silicon \citep{ww95}.
Not surprisingly, then, 
the abundance trends that we observe for silicon are well matched by calcium.  
Figure~\ref{fig:Ca} presents the solar-corrected Ca abundances relative to Fe
vs.\ [Fe/H].  All of the stars exhibit enhanced Ca and,
similar to Si,  there is a mild trend
to higher [Ca/Fe] at lower [Fe/H].  Also similar to the silicon 
results, the two highest metallicity stars show somewhat lower [Ca/Fe].
In contrast to most of the other well measured elements, the Ca measurements
for a given star exhibit a fairly large scatter.  We expect this is due
to a greater uncertainty in the equivalent width measurements for 
the Ca~I lines which often have significant damping wings.

\begin{figure}[ht]
\includegraphics[height=3.8in, width=2.8in,angle=-90]{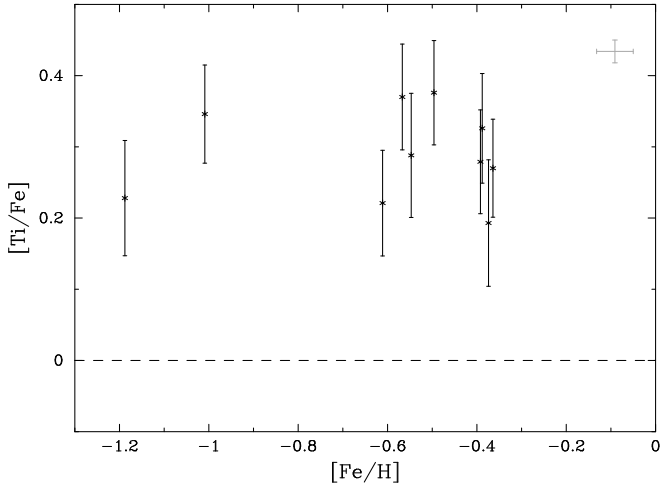}
\caption{Solar-corrected [Ti/Fe] abundance ratios 
vs.\ [Fe/H] metallicity for the 10 thick disk stars.  The observations
show a clear enhancement yet no trend with metallicity.
The dashed line at [Ti/Fe] = 0 indicates the solar meteoritic Ti/Fe ratio.}
\label{fig:Ti}
\end{figure}

Titanium is traditionally referred to as an $\alpha$-element 
because it exhibits enhanced
abundances in metal-poor stars \citep{gratton91}, but
it is unclear if the nucleosynthesis of Ti
is related to the other $\alpha$-elements \citep{ww95}.  
Therefore, it would not be surprising if
the Ti abundances differ from the results for Si and Ca.  
In almost every star (G114-19 is an exception), the $\e{Ti}$ abundance
derived from Ti~II exceeds that from Ti~I by $\approx 0.10-0.15$~dex.
This offset has been discussed in the literature (e.g.\ Luck \& Bond 1985)
and has been attributed to non-LTE effects and other possible systematic
errors.
We find a similar offset in our solar analysis such that the solar-corrected
$\e{Ti}$ values from Ti~I and Ti~II abundances are in 
good agreement for the majority
of stars.  Because the Ti~I results are statistically more robust,
we restrict our further analysis to the Ti~I results.
In Figure~\ref{fig:Ti} we present the solar-corrected
[Ti/Fe] measurements as a function of metallicity for the 10 thick disk 
stars.  All of the observations are consistent with a single-valued
enhancement, $\XFe{Ti} = 0.29 \pm 0.02$,
and there is no indication of a trend with
metallicity.  The latter observation contradicts 
the general picture described by Si and Ca.

\subsubsection{Oxygen, Magnesium, Sulfur}

\begin{figure}[ht]
\includegraphics[height=3.8in, width=2.8in,angle=-90]{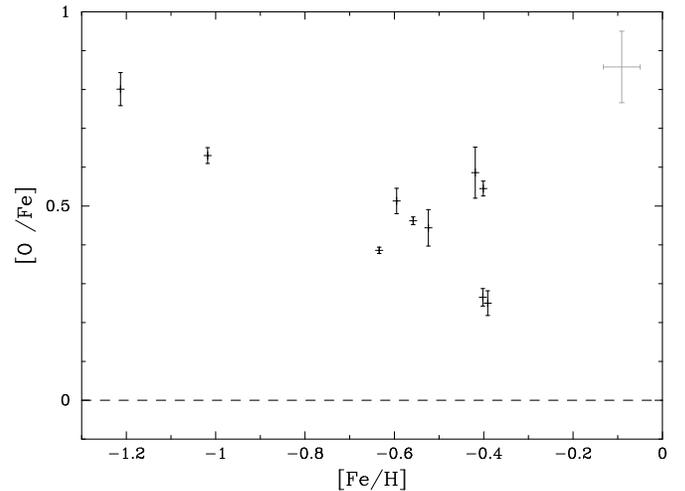}
\caption{Standard (uncorrected) [O/Fe] abundance ratios 
vs.\ [Fe/H] metallicity for the 10 thick disk stars.  The observations
show a clear enhancement and a likely trend with metallicity.
Note the error bars only reflect the scatter in the $\e{O}$ values
derived from the three O~I lines at $\lambda \approx 7775$\rAA for which
the systematic uncertainties are very significant $(> 0.1$~dex).
The dashed line at [O/Fe] = 0 indicates the solar meteoritic O/Fe ratio.}
\label{fig:O}
\end{figure}

For our choice of observational setup the forbidden
O~I~$\lambda 6300$ line lies within the inter-order gaps of HIRES.
Therefore, we rely on the triplet of O~I lines at 
$\lambda \approx 7775$~\rAA,
which have very high excitation potential and are very sensitive
to the effective temperature (Tables~\ref{tab:erra} and \ref{tab:errb})
and non-LTE effects.  
We account for CO, CH, and OH molecule formation in deriving our final oxygen
abundances which leads to an enhancement of $\approx +0.025$~dex over
the abundance derived without molecules. 
For the Sun, adopting the laboratory $gf$ values from
\cite{bev94} and \cite{butler91} we find a median $\e{O}$ value of
[O/H]~=~$+0.10$~dex with very small scatter.  Given the large uncertainty
associated with using the O~I triplet for an oxygen 
abundance analysis, the agreement
between our solar analysis and the meteoritic value is surprisingly good.
Figure~\ref{fig:O} plots the [O/Fe] values for the thick disk stars against the
stellar metallicity without the solar-correction.  
We do not apply the solar-correction here because
we believe the uncertainty in the solar measurement from the O~I triplet
is at least as large
as 0.1~dex and therefore we would be more likely to introduce an error
in the final abundances.
The error bars reflect the scatter in the individual
measurements for each star which in each case is significantly smaller
than the uncertainties from the atmospheric parameters.  In particular,
even a 50K error in $T_{eff}$ results in nearly a 0.1~dex 
uncertainty in [O/Fe].
We point out, however, that with the exception of the most metal-poor star,
the spectroscopic $T_{eff}$ values are at least as high as the photometric
values, implying if anything that we have 
underestimated the [O/Fe] ratios.
Examining the figure, one notes that the majority of stars exhibit enhanced
[O/Fe] abundances with tentative evidence for an increasing oxygen abundance
at lower metallicity.  The two stars with the highest metallicity, however,
have nearly solar oxygen abundance.   
It will be imperative in the future for us to improve on the oxygen 
measurements, either through the forbidden O~I [6300] line or perhaps
the near-IR OH lines.

\begin{figure}[ht]
\includegraphics[height=3.8in, width=2.8in,angle=-90]{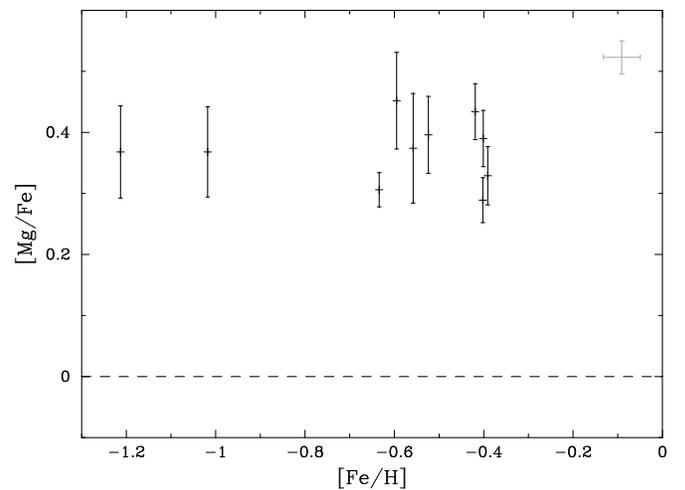}
\caption{Standard (uncorrected) [Mg/Fe] abundance ratios 
vs.\ [Fe/H] metallicity for the 10 thick disk stars.  The observations
show a clear enhancement and no trend with metallicity.
The dashed line at [Mg/Fe] = 0 indicates the solar meteoritic Mg/Fe ratio.}
\label{fig:Mg}
\end{figure}

As noted in $\S$~\ref{subsec-pparm}, measurements of magnesium 
are particularly important because Mg is a significant contributor of
electrons in the stellar atmospheres of our stars.  
Unfortunately, there are very few Mg~I lines with reported
$gf$ values that we do not find saturated.  Therefore, we are
compelled to include a few lines with solar $gf$ values taken from the
literature \citep{edv93,mcw95}.
In general, we find good agreement between the
various lines and have reasonable confidence in our results.  
Figure~\ref{fig:Mg} plots the [Mg/Fe] values versus [Fe/H]
for the 10 program stars.
We observe enhanced Mg in every case, $\XFe{Mg} = +0.37 \pm .02$,
with no suggestion of a trend with metallicity.  
It should be noted, however, that the Mg results for the two most metal-poor
stars are derived from a different set of Mg~I lines than the other stars.
Given the uncertainty in the $gf$ values we adopted and the fact that we
could not perform a solar analysis it is possible that there is a systematic
error in comparing against the two metal-poor stars although there is
no evidence of any offset.
The Mg/Fe ratio is insensitive
to uncertainties in the atmospheric parameters and we believe the observed
enhancement is robust aside from possible errors in the $gf$ values.
Because Mg is a principal source of electrons in the stellar atmospheres
of our stars and we observe an enhancement of Mg/Fe in every
case, it is important to consider $\alpha$-enhanced stellar atmospheres
as we have done throughout our analysis.  

\begin{figure}[ht]
\includegraphics[height=3.8in, width=2.8in,angle=-90]{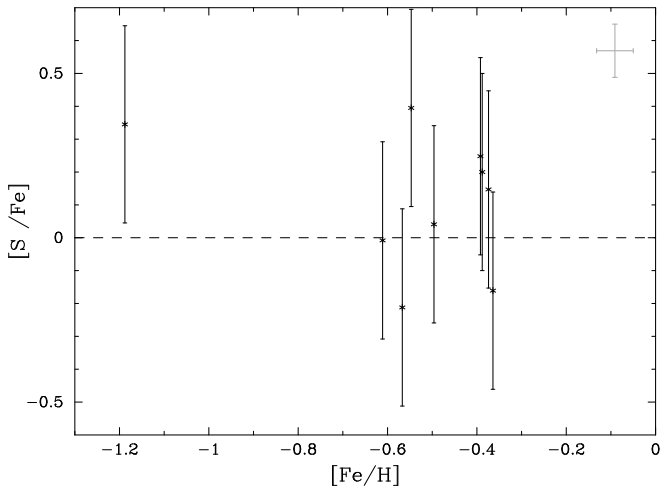}
\caption{Solar-corrected [S/Fe] abundance ratios 
vs.\ [Fe/H] metallicity for 9 of the 10 thick disk stars.  
For these measurements, the error bars reflect an estimated 0.3~dex 
uncertainty in the measurements due to the measurement of a single, weak
S~I line which has a very large EP and therefore a high sensitivity to
$T_{eff}$ and non-LTE effects.
There is no indication of a significant enhancement
and no trend with metallicity.
The dashed line at [S/Fe] = 0 indicates the solar meteoritic S/Fe ratio.}
\label{fig:S}
\end{figure}

The difficulties associated with sulfur are even more dire than the problems
associated with oxygen and magnesium:  there is only one useful transition 
(S~I $\lambda 8694$); it lies toward the red end of the spectrum
where the sensitivity of HIRES is markedly reduced; it has a high 
excitation potential with a correspondingly large temperature sensitivity; 
it is very weak (only 30m\rAA in the Sun); and there
is no reliable laboratory $gf$ value so a solar analysis is required.
We first adopted the $gf$ value from \cite{francois88} for the standard
and $\alpha$-enhanced values reported
in Tables~\ref{tab:solar}$-$\ref{abndG247-32}, but our solar analysis shows
[S/Fe]~$\approx +0.2$~dex indicating a significant correction to the $gf$
value.  Figure~\ref{fig:S} plots the [S/Fe] abundances for the 9 stars
with a measured sulfur equivalent width with the 
solar-corrected values plotted.  In a few cases, we have included
$\ew$ values less than 10 m\AA.  For these stars, the measurement error
even exceeds the uncertainties due to errors in
the atmospheric parameters.  We estimate the total 1$\sigma$ uncertainty to
be 0.3~dex and have plotted the error bars accordingly.  Given the large
errors associated with the [S/Fe] measurements, it is difficult to make
any meaningful statements about the sulfur abundance.  
It is somewhat surprising, however, that the mean ratio
$<$[S/Fe]$> \, = 0.11 \pm 0.08$ is consistent with the solar abundance.  
Given the importance of sulfur in quasar absorption line studies
($\S$~\ref{sec:dla}), a more careful and extensive stellar abundance
analysis of sulfur in metal-poor stars is warranted.

\begin{figure}[ht]
\includegraphics[height=3.8in, width=2.8in,angle=-90]{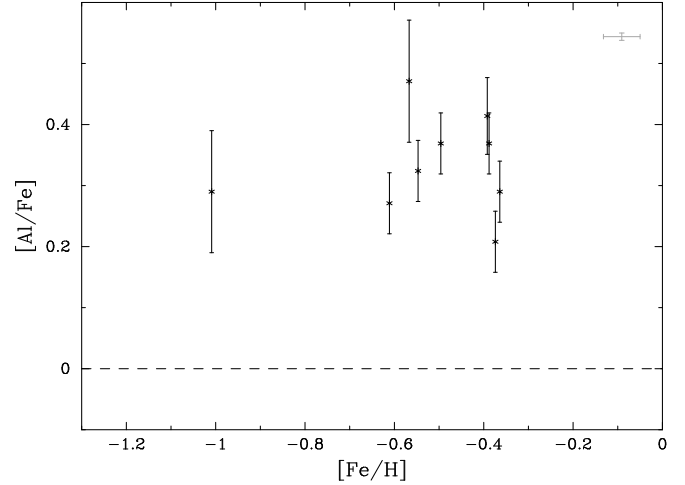}
\caption{Solar-corrected [Al/Fe] abundance ratios vs.\ [Fe/H] 
metallicity for 9 of the 10 thick disk stars.  
All of the stars show enhanced Al/Fe and there is no trend with
metallicity.
The dashed line at [Zn/Fe] = 0 indicates the solar meteoritic Zn/Fe ratio.}
\label{fig:Al}
\end{figure}

\subsection{Light Elements -- Al, Na}

With the exception of the extremely metal-poor stars 
\citep{gratton88,mcw95,shetr96}, Al is mildly
enhanced in metal-poor stars \citep{tomkin85,edv93}. As such, Al is sometimes
classified as an $\alpha$-element.  Contrary to the majority of Al 
studies, however, \cite{chen00} found an enhancement in Al in
disk stars with [Fe/H]~$\sim 0$ and no enhancement in their metal-poor
F and G dwarfs.  The \cite{chen00} 
analysis primarily focuses on the pair of Al~I lines
at 7830\AA, while \cite{edv93} examined the pair near 8773\AA.  Unfortunately,
all of these transitions fall into the inter-order gasps of our echelle
setup.  Instead our observations cover three other Al~I
absorption lines, all with measured laboratory $gf$ values \citep{buur86}:
$\lambda = 5577,6696,6698$\AA.
In general, we find good agreement for the Al abundances derived
from the two lines at $\lambda \approx 6700$\AA,
but the $\lambda 5577$ line yields systematically
lower values.  Unfortunately, this line is too badly blended in the Sun to 
determine a solar-correction. As such, we have decided
to remove the $\lambda 5577$ line from our analysis.  For the Sun, the two
remaining lines yield a solar Al abundance $\approx 0.1$~dex lower than the 
meteoritic value therefore we have applied a 0.1~dex offset in the
solar-corrected values.  
In Figure~\ref{fig:Al} we plot the solar-corrected [Al/Fe] values 
for the 9 stars where we measured at least one of the lines at 
$\lambda \approx 6700$\AA; the Al~I lines are too weak in the most
metal-poor star.
All of the stars exhibit significant Al 
enhancements, $<$[Al/Fe]$> = +0.334 \pm 0.028$ and there is 
no trend with metallicity.
Our error analysis indicates the Al/Fe ratio is insensitive to the
atmospheric parameters; the principal uncertainty lies in 
the paucity of Al~I lines.
There is a further systematic error associated
with these Al~I lines, however. A non-LTE analysis by \cite{baum97}
suggests a further enhancement to [Al/Fe] of $\approx 0.15$~dex for
stars with metallicity [Fe/H]~$\approx -0.5$~dex.
We have chosen not to include this non-LTE correction in our analysis,
but warn the reader that the reported Al/Fe values may be a lower limit
to the true ratio.

\begin{figure}[ht]
\includegraphics[height=3.8in, width=2.8in,angle=-90]{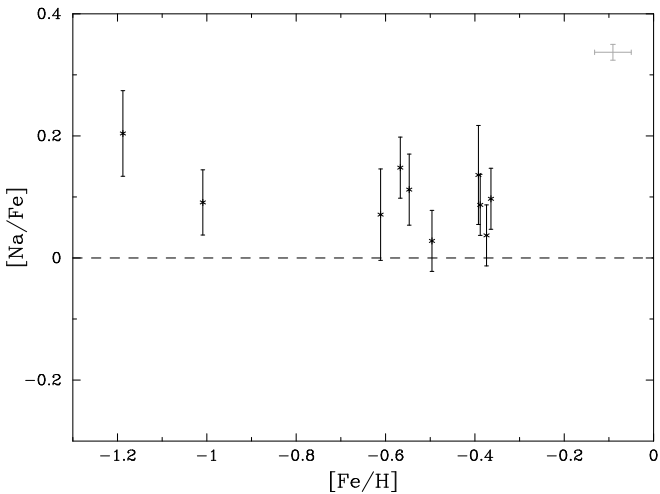}
\caption{Solar-corrected [Na/Fe] abundance ratios vs.\ [Fe/H] 
metallicity for the 10 thick disk stars.  
All of the stars show enhanced Na/Fe and there is a 
mild trend with metallicity.
The dashed line at [Zn/Fe] = 0 indicates the solar meteoritic Zn/Fe ratio.}
\label{fig:Na}
\end{figure}

The observational picture for Na is complicated.
The majority of studies on sodium \citep{tomkin85,mcw95,chen00} report
that Na scales with Fe at all metallicities, although
\cite{edv93} found a mild Na enhancement at [Fe/H]~$\approx -1$ and
\cite{pilachow96} report a Na/Fe deficiency in a sample of 60 stars
with [Fe/H]~$< -1$.  Recently,
\cite{baum98} performed a non-LTE analysis of Na and report a trend
of decreasing Na with decreasing metallicity.  
We have measured Na
in our thick disk stars based on the observations of four Na~I
transitions: $\lambda 5682, 5688, 6154, 6160$.  None of the $gf$ values
are secure for these lines; the latter two are from theoretical work
by \cite{lamb78} and we adopt solar $gf$ values for $\lambda 5682,5688$.
Therefore, the analysis relative to our solar analysis is essential.
Figure~\ref{fig:Na} presents the solar-corrected [Na/Fe] values versus
metallicity for the 10 stars.  Every star exhibits mildly enhanced Na
with an average $\XFe{Na} = +0.087 \pm 0.014$.  
The overabundance is statistically significant but the 
systematic uncertainty (e.g.\ in the $gf$ values) is on the order
of the enhancement.
The Na/Fe ratio is remarkably insensitive to the atmospheric parameters
therefore the results are probably limited by the small number statistics
of measuring only three Na~I lines.
If we were to apply the results of the non-LTE analysis by \cite{baum98},
it is also possible that the observed Na enhancement would vanish.
In summary, we consider the Na/Fe results to be rather poorly constrained
yet consistent with no significant departure from the solar abundance.

\subsection{Iron-Peak Elements -- Sc, V, Cr, Mn, Co, Ni, Cu, Zn}

We now turn our attention to the elements with atomic number near Fe,
the so-called iron-peak elements.  
For several of these elements we observe no variations 
from solar abundance irrespective of the thick disk stellar metallicity. 
In particular,
the relative nickel and chromium 
abundances show no significant departure from solar
abundances for any of the thick disk stars.  There is a small
offset (+0.03~dex) from solar for the [Ni/Fe] and [Cr/Fe] values in the
standard analysis, but we observe 
a similar difference in the solar analysis such
that the corrected abundances are within 0.02~dex of solar
for every star but one.   The most metal-poor star (G84-37) exhibits
a low [Cr/Fe] value which is probably significant and is suggestive of
the underabundance observed for Cr in very metal-poor halo stars
(e.g.\ McWilliam et al.\ 1995). We note in passing that the Cr
abundance based on the Cr~II absorption lines suggest enhanced
[Cr/Fe], typically by $\approx +0.1$~dex.  The solar
measurements of Cr~II,
however, exhibit the same enhancement and therefore
we propose that the Cr~II $gf$ values of \cite{martin88}
may need to be revised upward by 0.1~dex.

The majority of iron-peak elements, however, 
exhibit significant departures from the solar
ratios and some have clear trends with the stellar metallicity.  As these
departures offer insights into the nucleosynthetic processes and
may distinguish the thick disk stars from other stellar populations, we
consider each element in greater detail.

\subsubsection{Scandium}
\label{sec:Sc}

\begin{figure}[ht]
\includegraphics[height=3.8in, width=2.8in,angle=-90]{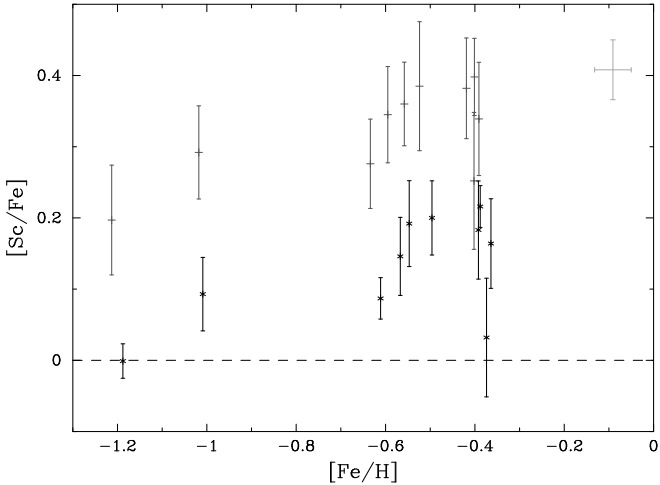}
\caption{Solar-corrected ($\times$) and standard (+)
[Sc/Fe] abundance ratios vs.\ [Fe/H] metallicity for the 10 thick disk stars.  
The ratios show an unusual dependence with metallicity and are enhanced
in all but the solar-corrected the most metal-poor star.
and no trend with metallicity.
The dashed line at [Sc/Fe] = 0 indicates the solar meteoritic Sc/Fe ratio.}
\label{fig:Sc}
\end{figure}

Previous studies of scandium have disagreed on the metallicity dependence
of [Sc/Fe].  \cite{zhao90} first suggested that Sc was enhanced 
by $\approx +0.25$ in metal-poor
stars based on their analysis of four Sc~II lines including several of the
lines that we have analyzed.  \cite{gratton91}, however, argued that this
enhancement was primarily due to inaccuracies in the Sc~II $gf$ values.
Our solar analysis agrees with this assessment; we find 
[Sc/H]$_\odot \approx +0.2$~dex using the $gf$ values from \cite{martin88}
and \cite{lawler89}.
Most recently, \cite{nissen00} published an analysis of Sc in over 100 G
and F dwarf stars with metallicities ranging from [Fe/H]~$\approx -1.4 - 0$.
Their study focused on five Sc~II lines 
($\lambda 5239, 5526, 5657, 6245, 6604$)
compared against a solar analysis.  Their results, based on an hfs
analysis from \cite{steffen85},  indicate an enhancement
of [Sc/Fe] with a trend that resembles the $\alpha$-enhancement of metal-poor
stars.  Compared to our hfs analysis (Table~\ref{tab:hfs}), however, the 
\cite{steffen85} compilation overestimates the hfs correction 
for these Sc~II lines,
which may account for a significant part if not all
of the reported trend \citep{promcw00}. 
Figure~\ref{fig:Sc} presents
[Sc/Fe] for our full sample as a function of metallicity for the
standard (+) and solar-corrected ($\times$) values; the latter reveal
a somewhat puzzling picture.  While the 8 stars at [Fe/H]~$\approx -0.5$
exhibit enhanced scandium, $<$[Sc/Fe]$> = +0.14 \pm 0.02$, the most
metal-poor stars show only a minor enhancement.  This difference could
be explained by unidentified blends, but it seems very unlikely given the
reasonably close agreement of more than 5 Sc~II lines in each star and 
because it would be
difficult to reconcile blends with the solar $\e{Sc}$ results.  An 
underestimate of $\approx 150$K in $T_{eff}$ for the stars
with [Fe/H]~$\approx -0.5$ could explain the observed enhancement,
but this too is unlikely given the fact that the
spectroscopic $T_{eff}$ values are already $\approx$100K 
higher than the \cite{carn94} photometric measurements.  
Yet another possibility is that
we have underestimated the hfs correction for these Sc~II lines.  The
equivalent widths are typically $< 60$~m\AA, however, and more importantly
an error in the hfs correction would be most severe for the Sun.
The net effect relative to the Sun would be to bring up the metal-poor
[Sc/Fe] values to $\approx +0.2$ while leaving the remaining [Sc/Fe]
values unchanged.  

Therefore, we contend that there is an overabundance of Sc 
at [Fe/H]~$\approx -0.5$ dex  and await future observations at 
[Fe/H]~$\approx -1$ to improve the statistical 
significance at that metallicity.
It is interesting to note that our observations qualitatively match the
[Sc/Fe] results from \cite{promcw00}.  In their reanalysis of \cite{nissen00}, 
there is evidence for an enhancement of [Sc/Fe] near [Fe/H]~$\approx -0.6$
yet no significant enhancement at [Fe/H]~$\approx -1$.
It is difficult to offer an explanation for the observed trend.
Perhaps the Sc production in supernovae has a metallicity dependent yield.
Or perhaps the enhancement at [Fe/H]~$\approx -0.5$ is a statistical anomaly
or is due to an overlooked source of systematic error. 

\begin{figure}[ht]
\includegraphics[height=3.8in, width=2.8in,angle=-90]{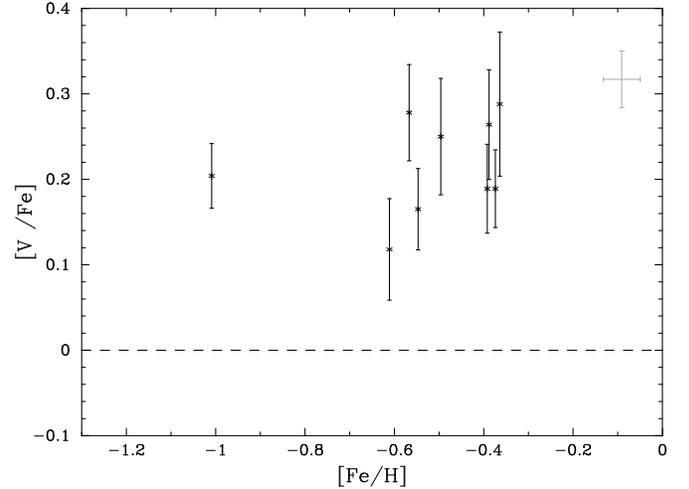}
\caption{Solar-corrected [V/Fe] abundance ratios vs.\ [Fe/H] 
metallicity for 9 of the 10 thick disk stars.  
The stars are all exhibit enhanced V and there is no obvious
trend with metallicity.
The dashed line at [V/Fe] = 0 indicates the solar meteoritic V/Fe ratio.}
\label{fig:V}
\end{figure}

\subsubsection{Vanadium}
\label{sec-V}

Even though there is an exhaustive list of useful V~I lines 
\citep{whaling85}, there have been only a few studies of vanadium to date.
\cite{gratton91} have observed V in a sample of $\approx 20$ stars with
a large range in metallicity.  Their analysis reveals no significant
departure from the solar V/Fe ratio at any metallicity.  As with many
of the other iron-peak elements, V suffers from significant hyperfine splitting
and we have been careful to account for it.  To our surprise, our 
solar analysis of vanadium based on over 15 clean V~I lines suggest
a small but significant offset from the meteoritic value:
$\e{V}_\odot = 3.90 \pm 0.014$.  This is contrary to the results of
\cite{whaling85} who derive $\e{V}_\odot = 3.99 \pm 0.01$, even though
we adopt identical $\log gf$ values and our solar $\ew$ measurements are
larger.  The discrepancy is eliminated, however, when we repeat the analysis
with the \cite{holw74}
solar atmosphere and the microturbulence value adopted by \cite{whaling85}.
In short, the V~I lines are very sensitive to the temperature of the 
stellar atmosphere and even the small differences between the Kurucz and
Holweger-M$\ddot{\rm u}$ller solar atmospheres lead to a 0.1~dex offset.
In terms of the abundance analysis of the thick disk stars relative to
the Sun, however, the relevant quantity is V/Fe which differs by $< 0.05$~dex
for the two models.

In Figure~\ref{fig:V}, we plot the solar-corrected [V/Fe] ratios versus
[Fe/H] metallicity for the 9 stars with V~I measurements.  
Every star exhibits an enhanced V/Fe ratio with no
apparent trend with metallicity: $\XFe{V} = +0.22 \pm 0.02$.  If anything,
one might have expected a deficiency for vanadium as predicted by
\cite{tmm95} in their chemical evolution model.
In fact, this marks the first conclusive evidence for enhanced V/Fe ratios
in any stellar population.
We have carefully
checked and rechecked our analysis for all possible systematic errors.
For the stars with [Fe/H]~$\approx -0.5$, the V/Fe ratio does depend
on $T_{eff}$ ($\Delta$[V/Fe]~$\approx +0.03$ for $\Delta T_{eff} = 50$K),
but it would require a very large temperature error to explain the
entire enhancement. Therefore,
we believe the V/Fe enhancement is a robust result
and a very puzzling one at that.  In $\S$~\ref{sec-discuss} we speculate
on possible explanations.

\begin{figure}[ht]
\includegraphics[height=3.8in, width=2.8in,angle=-90]{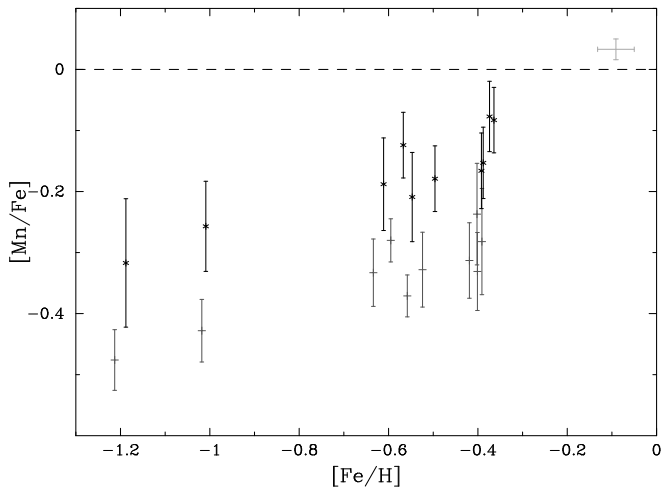}
\caption{Solar-corrected ($\times$) and standard (+)
[Mn/Fe] abundance ratios vs.\ [Fe/H] metallicity for the 10 thick disk stars.  
The stars are all deficient in Mn and there is a mild trend toward lower
[Mn/Fe] at lower metallicity.
The dashed line at [Mn/Fe] = 0 indicates the solar meteoritic Mn/Fe ratio.}
\label{fig:Mn}
\end{figure}

\subsubsection{Manganese}

Previous studies on manganese have indicated that it is underabundant
relative to Fe in metal-poor stars \citep{waller62,gratton89}.  
This trend has often been cited as evidence for
the 'odd-even effect' of $\alpha$-elements suggested by
\cite{helfer59}.  Because Mn is an odd-element it suffers from 
hyperfine splitting and even in the case of moderately saturated
absorption lines the hfs corrections can be large ($> 0.3$~dex).  
It is crucial, therefore, to carefully account for the hyperfine splitting
\citep{promcw00}.  In our analysis we 
adopted the hfs lines tabulated in Appendix~\ref{app-hfs} and   
relied on laboratory $gf$ values \citep{booth84a,martin88}.
In addition to the difficulties associated with hyperfine splitting,
Mn is special for the fact that its photometric solar
abundance is significantly discordant from the meteoritic abundance
\citep{booth84b}.  As noted in $\S$~\ref{sec-solar}, our solar analysis
also indicates that the solar photometric Mn value is significantly lower
than the meteoritic, $\e{Mn}_{phot} - \e{Mn}_{meteor} \approx -0.2$~dex.
We expect (as hypothesized by Booth et al.\ 1984b) that there is a zero-point 
error to the $\log gf$ values and, therefore, 
it is essential to consider a comparison relative to solar.
Figure~\ref{fig:Mn} plots the [Mn/Fe] values for our sample as a 
function of [Fe/H] for the standard (+) and solar-corrected  ($\times$)
analyses. In agreement with other
studies of Mn in metal-poor
stars \citep{gratton89}, we find sub-solar [Mn/Fe] values and 
evidence for a trend toward lower [Mn/Fe] values at lower metallicity.

\begin{figure}[ht]
\includegraphics[height=3.8in, width=2.8in,angle=-90]{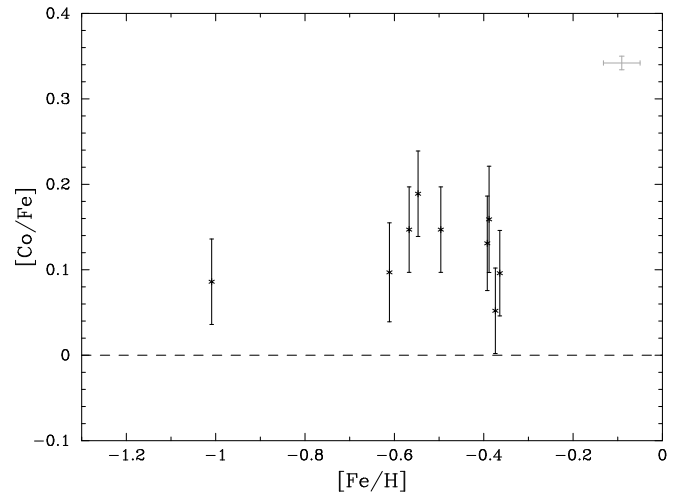}
\caption{Solar-corrected [Co/Fe] abundance ratios vs.\ [Fe/H] 
metallicity for the 10 thick disk stars.  
The stars are all exhibit enhanced Co and there is no obvious
trend with metallicity.
The dashed line at [Co/Fe] = 0 indicates the solar meteoritic Co/Fe ratio.}
\label{fig:Co}
\end{figure}

\subsubsection{Cobalt}
\label{sec-Co}

Cobalt is yet another odd-proton element in the iron-peak which suffers from 
hyperfine splitting.  For our stars the hfs corrections are small,
in particular
because the typical equivalent widths of the Co~I lines are small:
$\ew < 30$~m\AA. With the exception of the most metal-poor star
in our sample (G84-37), the [Co/Fe] measurements are based on $\approx 10$
absorption lines in good agreement ($\sigma<0.1$~dex).
Although the solar photospheric $\e{Co}$ values
are in good agreement with the solar 
meteoritic value ($\S$~\ref{sec-solar}), we plot the solar-corrected
[Co/Fe] values in Figure~\ref{fig:Co}
because the solar corrections significantly reduce the
observed scatter.  We find a clear enhancement of Co
relative to Fe with no significant evidence for a trend with metallicity:
$<$[Co/Fe]$> \; = +0.14 \pm 0.02$~dex.
The observed enhancement 
disagrees with the analysis of \cite{gratton91} who found that Co was actually
underabundant relative to Fe by $\approx 0.1$~dex over
the same metallicity range.  
As Tables~\ref{tab:erra} and \ref{tab:errb} indicate, the CoI/Fe ratio is very
{\it insensitive} to uncertainties in the atmospheric models, therefore
uncertainties in the equivalent width measurements
are the dominant source of error.
Nonetheless, we find good agreement among the individual
$\e{Co}$ measurements and have strong confidence in this result.  
This small enhancement hints at the significant overabundance of
Co/Fe observed in the extremely metal-poor halo stars \citep{mcw95}.

\begin{figure}[ht]
\includegraphics[height=3.8in, width=2.8in,angle=-90]{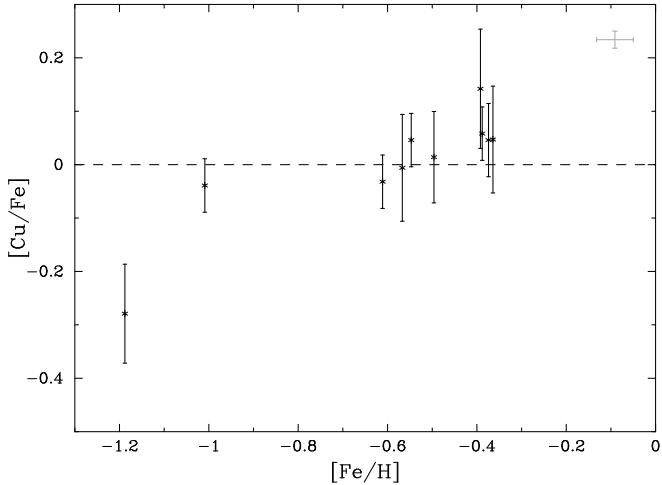}
\caption{Solar-corrected [Cu/Fe] abundance ratios vs.\ [Fe/H] 
metallicity for the 10 thick disk stars.  
With the exception of the most metal-poor star, all of the
stars exhibit solar Cu/Fe ratios and there is no significant
trend with metallicity.
The dashed line at [Cu/Fe] = 0 indicates the solar meteoritic Cu/Fe ratio.}
\label{fig:Cu}
\end{figure}

\subsubsection{Copper}
\label{sec-Cu}

In very metal-poor stars, \cite{sne88,sne91} found copper to be deficient 
relative to iron, reaching [Cu/Fe]~$\approx -1$~dex at [Fe/H]~$\approx -3$.
At metallicities comparable to our stars, however, their results show no
significant departures from the solar ratio.
Our observations include coverage of four Cu~I absorption lines:
$\lambda 5105, 5218, 5782, 8092$. 
The first three have laboratory $gf$ measurements 
\citep{kock68,hann83} but there is considerable
disagreement over the values ($\log gf$ for $\lambda 5782$ differs by
$\approx 0.5$~dex).   An analysis relative to solar
is essential and given that we will rely on solar $gf$ values 
we also include Cu~I $\lambda 8029$ in the analysis.
Unfortunately, the solar equivalent width for
$\lambda 5782$ is poorly constrained and we have excluded this line from 
the abundance analysis altogether.  
Copper suffers from significant hyperfine splitting and we 
have been careful to account for the effects in our analysis.
Figure~\ref{fig:Cu} presents the results from 
the solar-corrected analysis. 
While the uncertainties are large, the general picture is that the majority
of stars show nearly solar Cu abundances with the marked exception of the
most metal-poor star in our sample.  
There is an indication for a mild decrease in 
[Cu/Fe] with decreasing metallicity and the supersolar values 
at [Fe/H]~$\approx -0.4$ suggest a small
zero point error (0.05~dex) in our solar $gf$ values.

\begin{figure}[ht]
\includegraphics[height=3.8in, width=2.8in,angle=-90]{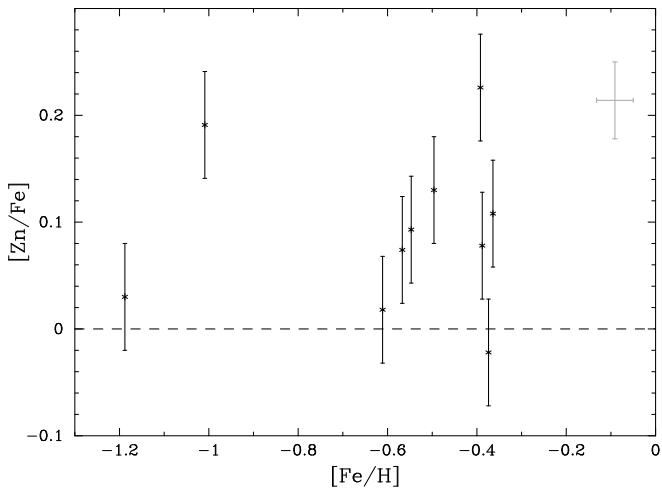}
\caption{Solar-corrected [Zn/Fe] abundance ratios vs.\ [Fe/H] 
metallicity for the 10 thick disk stars.  
The majority of stars show enhanced Zn/Fe and there is no trend with
metallicity.
The dashed line at [Zn/Fe] = 0 indicates the solar meteoritic Zn/Fe ratio.}
\label{fig:Zn}
\end{figure}

\subsubsection{Zinc}
\label{sec:Zn}

In their comprehensive study of Zn, \cite{sne88,sne91} 
found a nearly solar Zn/Fe ratio
over a large range of metallicity:
$\XFe{Zn}_{Sneden} = 0.04 \pm 0.02$.  We have analyzed
the same two Zn~I lines as \cite{sne91} and adopted the theoretical
$gf$ values \citep{biem80}.  Even though our solar analysis is in 
excellent agreement with the solar meteoritic value, we noticed that
the $\lambda 4722$ line yields systematically higher $\e{Zn}$ values
while the $\lambda 4810$ line gives systematically lower $\e{Zn}$.
We recommend that the oscillator strengths be adjusted by 0.04~dex as
follows: $\log gf_{\lambda 4722} = -0.35$, 
$\log gf_{\lambda 4810} = -0.21$. The 
agreement of the two
lines for the thick disk stars after the correction 
was generally $< 0.02$~dex.
Therefore the error bars plotted in Figure~\ref{fig:Zn}, which
represent the scatter in $\e{Zn}$ from the two Zn~I lines, underestimate
the true uncertainty in [Zn/Fe] which is dominated by the error in 
$T_{eff}$ ($\pm 0.03$ for $\Delta T = 50$K).
Figure~\ref{fig:Zn} gives the [Zn/Fe] values for the 10 thick disk stars
vs.\ [Fe/H].  Three of the stars exhibit Zn/Fe values consistent with the
solar ratio, yet the majority are enhanced relative to solar
with two stars showing enhancements of $\approx$~+0.2~dex.  
The mean enhancement is $<$[Zn/Fe]$> \, = 0.093 \pm 0.025$, which
is larger than the value reported by \cite{sne91}.  
While the enhancement may be unique to thick disk stars,
there are also indications of an overabundance of Zn in very metal-poor stars
\citep{jhnsn99}.  Given the tremendous impact of this ratio on studies in
the damped \lya systems, it will be very important to repeat a survey
similar to \cite{sne91} with more accurate Fe abundances and modern model
atmospheres.  It is worth noting that an overestimate in $T_{eff}$ 
of 100K (as might be the case for the majority of our stars), 
would imply an increase in the [Zn/Fe] abundances by 0.06$-$0.08~dex
implying a mean [Zn/Fe] enhancement of $+0.15 - 0.17$~dex.

\begin{figure}[ht]
\includegraphics[height=3.8in, width=2.8in,angle=-90]{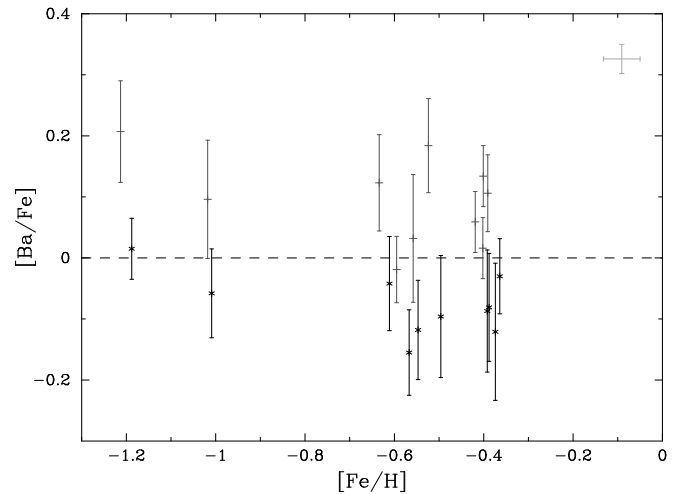}
\caption{Solar-corrected ($\times$) and standard (+) [Ba/Fe] abundance ratios 
vs.\ [Fe/H] metallicity for the 10 thick disk stars.  The observations
show a mild enhancement which we do not believe is statistically 
significant (see the text).
The dashed line at [Ba/Fe] = 0 indicates the solar meteoritic Ba/Fe ratio.}
\label{fig:Ba}
\end{figure}

\subsection{Heavy Elements -- Ba, Y, Eu}

Unfortunately, due to the lack of available absorption lines 
we have only been able to
obtain abundance measurements for three heavy elements: Ba, Y, and Eu.
These abundances are derived from very few lines
and are poorly constrained.  Nonetheless, they provide tentative insight into
the relative importance of the r and s-processes in these thick disk stars.

While barium can be synthesized through both the s and r-processes,
it is believed that the solar system composition
is $\approx 90 \%$ s-process \citep{kapp90} and one expects 
barium to be synthesized primarily via the s-process in metal-poor stars.
Our spectra include coverage of four Ba~II lines:
$\lambda 4554, 5853, 6141, 6496$.  The first absorption line is heavily
saturated for all of the thick disk stars so the analysis focuses only on the
latter three for which we adopt $gf$ values from \cite{mcw98}.
Because barium suffers from significant hyperfine splitting, we carefully
calculated hfs corrections with the hfs lines presented in Table~\ref{tab:hfs}
taken from \cite{mcw98}. For the hfs corrections we have adopted an
r-process isotopic composition and we note that the s-process composition
increase $\e{Ba}$ by less than 0.03~dex.
For the solar analysis, $\lambda 6141$ was too saturated to provide an 
accurate correction.  Furthermore, we estimate
the correction for $\lambda 6496$ to be very large, perhaps the result of
an unidentified blend.  In terms of the thick disk abundance analysis,
the solar-correction analysis leads to a decrease in the typical [Ba/Fe] value
by $0.17$~dex.  Figure~\ref{fig:Ba} plots the standard and solar-corrected
values which yield mean values $\XFe{Ba} \, = 0.093 \pm 0.024$ and
$\XFe{Ba} \, = -0.077 \pm 0.017$ respectively.  The offset between the two
is large and worrisome.  Our best interpretation
of the overall results is that the stars exhibit nearly solar Ba/Fe 
although a further investigation is warranted.
For the discussion in the following section, we will use the uncorrected
barium abundances as we fear the solar analysis introduces too large
an uncertainty.

\begin{figure}[ht]
\includegraphics[height=3.8in, width=2.8in,angle=-90]{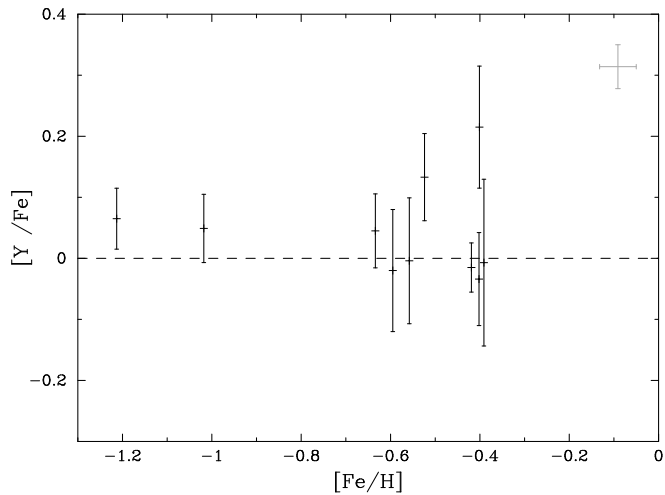}
\caption{Standard (uncorrected) [Y/Fe] abundance ratios 
vs.\ [Fe/H] metallicity for the 10 thick disk stars.  All of the observations
are consistent with the solar ratio.
The dashed line at [Y/Fe] = 0 indicates the solar meteoritic Y/Fe ratio.}
\label{fig:Y}
\end{figure}

Previous analyses on yttrium \citep{zhao91,gratton94} have demonstrated a mild
metallicity dependence of [Y/Fe] with [Fe/H] where metal-poor stars show
mildly deficient Y.  These studies focused on Y~II lines which are plentiful
and for which a reasonably large database of $gf$ values exist
\citep{hann82}.
While we observed $\approx 10$ Y~II lines for each star, 
almost every line suffers
from significant line blending.  Therefore the final Y/Fe
abundance measurements are based on only $1-3$ clean Y~II lines.  Because
the solar-corrected analysis further reduces the number of lines
considered and does not give significantly different results from 
the standard analysis, we present
the uncorrected [Y/Fe] values in Figure~\ref{fig:Y}.
All of the Y/Fe values are consistent with the solar ratio and we observe
no trend with metallicity.  Note that the high [Y/Fe] measurement from
G97-15 was derived from a single Y~II line which is partially blended.

\begin{figure}[ht]
\includegraphics[height=3.8in, width=2.8in,angle=-90]{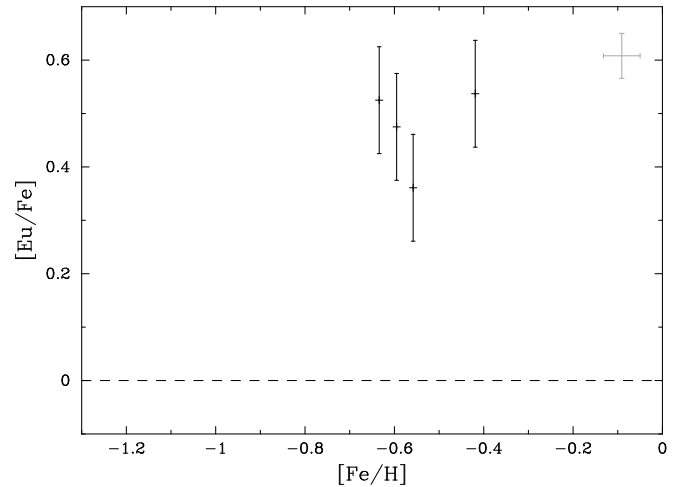}
\caption{Standard (uncorrected) [Eu/Fe] abundance ratios 
vs.\ [Fe/H] metallicity for the 4 stars with wavelength coverage of the
Eu~II $\lambda 6645$. All of the observations are significantly enhanced
above the solar ratio.
The dashed line at [Eu/Fe] = 0 indicates the solar meteoritic Eu/Fe ratio.}
\label{fig:Eu}
\end{figure}

Europium is predominantly synthesized through the
r-process.  Because the r-process takes place almost
exclusively in Type~II SN, Eu should trace other
elements formed primarily through Type~II SN (i.e.\ the $\alpha$-elements).
Unfortunately, the observational challenge for a europium abundance
analysis in the thick disk stars is severe.
While we have coverage of Eu~II $\lambda 6437$, it was too
weak to reliably measure.  Furthermore, the Eu~II $\lambda 6645$ line lies
at the red edge of our setup and could be measured only in those stars with
geocentric velocity $v < 60 \mkms$. Finally, the Eu~II
$\lambda 6645$ line is very weak and given that it is found near the end
of a spectral order, its equivalent width is particularly uncertain:
$\sigma(\ew) > 3$m\AA. We find [Eu/H]~$\approx +0.1$ based on a solar
analysis of $\lambda 6645$ but consider this value to be unreliable. 
Therefore we present the standard results in Figure~\ref{fig:Eu} and
warn that they may be systematically high by $\approx 0.1$~dex.  
We observe Eu to be significantly enhanced in these stars, 
$\XFe{Eu} \, = 0.47 \pm 0.05$.  
Unfortunately, there are too few measurements from our sample to
meaningfully comment on any trend in [Eu/Fe].  

\subsection{Summary} 

Table~\ref{tab:absum} provides a summary of the abundance measurements
for the thick disk sample.  Column 1 lists element X, column 2 details
the number of stars ($N$) with measurements of X, column 3 gives the mean
[X/Fe] ratio $\bar x$, column 4 is the error in the mean 
$\sigma(\bar x)$, column 5 marks whether 
the abundances were corrected by the solar analysis $(\odot$), column 6
indicates if we believe a trend of [X/Fe] with [Fe/H] metallicity is likely,
and column 7 provides additional comments.
Forgoing a formal error analysis to future papers, 
the error in the mean merely represents
the reduced standard deviation: $\sigma(\bar x) = \sigma / \sqrt{N-1}$.
This value is more representative than the formal statistical
error given the many systematic uncertainties involved in measuring the
abundance of a given star.   We warn, however, that systematic errors which
affect all of the measurements -- $gf$ offsets, errors in the Kurucz
atmospheres, etc.\ -- may exceed the $\sigma(\bar x)$ values.

\begin{table}[ht]
\begin{center}
\caption{ {\sc ABUNDANCE SUMMARY} \label{tab:absum}}
\begin{tabular}{lccccc}
\tableline
\tableline
X& $N$ & $\bar x$ & $\sigma(\bar x)$ & Trend & App\tablenotemark{a} \\
\tableline
O  &  10 & $ 0.488$ & 0.056 & y &            \\
Na &  10 & $ 0.101$ & 0.017 & n & $\odot$    \\
Mg &  10 & $ 0.371$ & 0.017 & n &            \\
Al &   9 & $ 0.334$ & 0.028 & n & $\odot$    \\
Si &  10 & $ 0.172$ & 0.021 & ? & $\odot$    \\
S  &   9 & $ 0.111$ & 0.075 & ? & $\odot$    \\
Ca &  10 & $ 0.197$ & 0.020 & y & $\odot$    \\
Sc &  10 & $ 0.131$ & 0.025 & ? & $\odot$    \\
Ti &  10 & $ 0.290$ & 0.021 & n & $\odot$    \\
V  &   9 & $ 0.216$ & 0.020 & n & $\odot$    \\
Cr &  10 & $ 0.000$ & 0.011 & n & $\odot$    \\
Mn &  10 & $-0.175$ & 0.025 & y & $\odot$    \\
Co &   9 & $ 0.123$ & 0.015 & n & $\odot$    \\
Ni &  10 & $ 0.021$ & 0.007 & n & $\odot$    \\
Cu &  10 & $ 0.000$ & 0.037 & y & $\odot$    \\
Zn &  10 & $ 0.093$ & 0.025 & n & $\odot$    \\
Y  &  10 & $ 0.043$ & 0.026 & n &            \\
Ba &  10 & $ 0.094$ & 0.024 & n &            \\
Eu &   4 & $ 0.475$ & 0.046 & n &            \\
\tableline
\end{tabular}
\end{center}
\tablenotetext{a}{The $\odot$ marks those elements for which a solar-correction was applied}
\end{table}

\begin{figure*}
\begin{center}
\includegraphics[height=9.0in, width=7.2in]{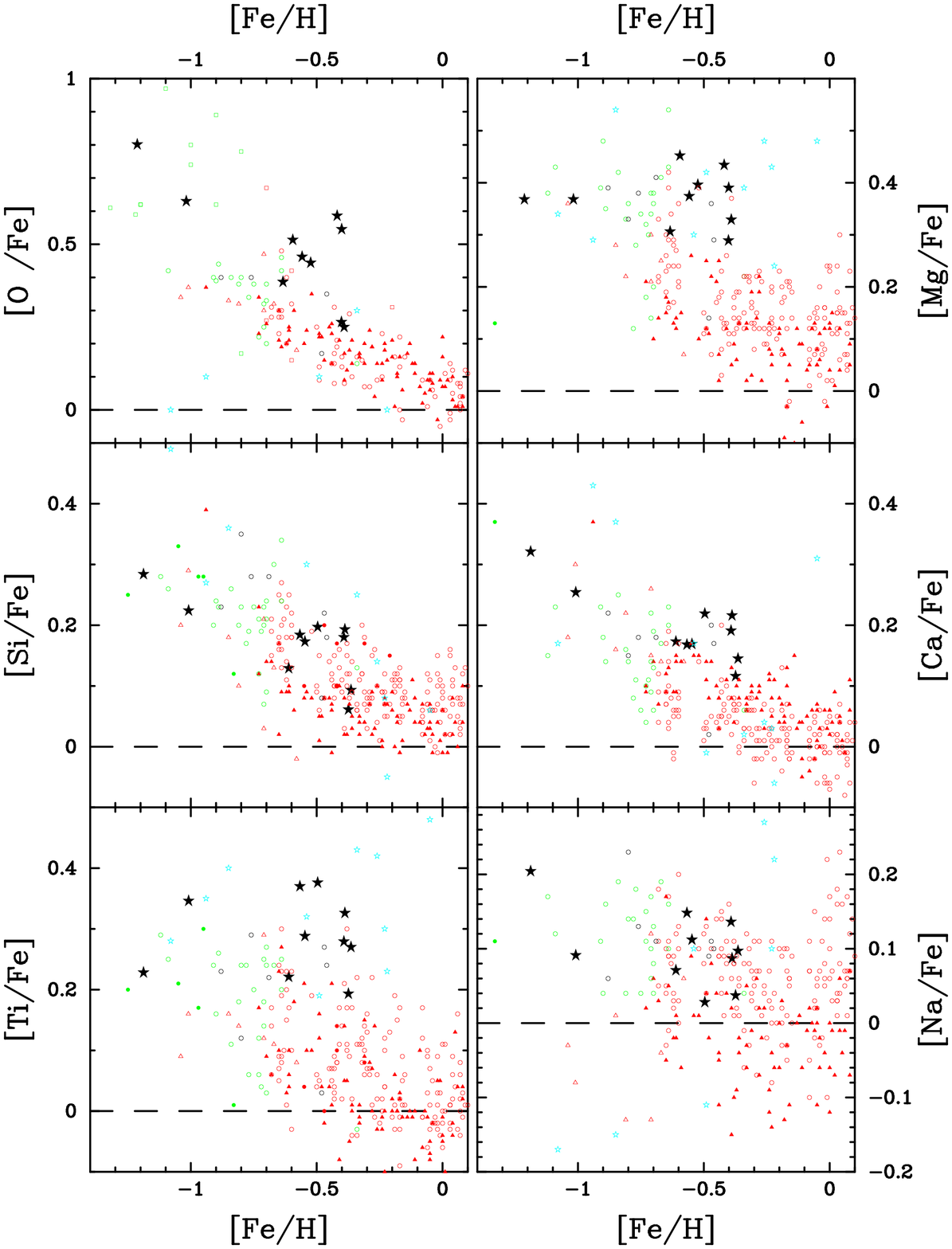}
\caption{Abundance patterns for the thick disk stars (large black stars)
for 18 of the elements analyzed in this paper.  For comparison, we plot 
the abundance patterns for halo stars (green points), thin disk
stars (red points), other thick disk measurements (small black stars),
and a small sample of bulge stars (light blue points).  The different 
symbols refer to the various literature sources indicated in the text.
There are three principal points to take from the figure: (1) for many of
the elements, the thick disk abundances are distinct from the thin disk;
(2) the thick disk abundances tend toward the halo star values at 
[Fe/H]~$\approx -1$; (3) the thick disk abundances are in good agreement
with the bulge values for essentially every element.}
\label{fig:abptta}
\end{center}
\end{figure*}
\begin{figure*}
\begin{center}
\includegraphics[height=9.0in, width=7.2in]{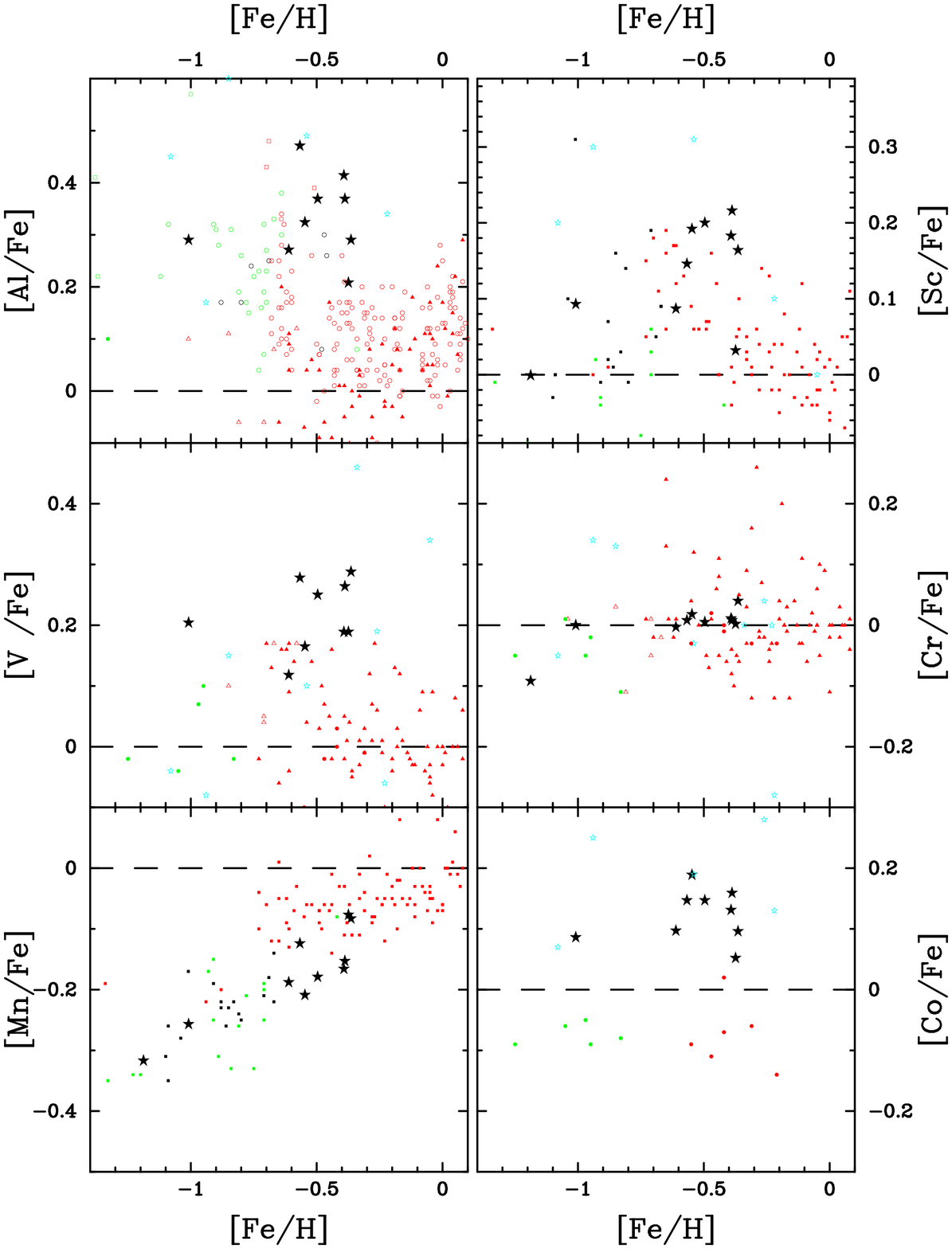}
\caption{Abundance patterns for the thick disk stars (large black stars)
for 18 of the elements analyzed in this paper.  For comparison, we plot 
the abundance patterns for halo stars (green points), thin disk
stars (red points), other thick disk measurements (small black stars),
and a small sample of bulge stars (light blue points).  The different 
symbols refer to the various literature sources indicated in the text.
There are three principal points to take from the figure: (1) for many of
the elements, the thick disk abundances are distinct from the thin disk;
(2) the thick disk abundances tend toward the halo star values at 
[Fe/H]~$\approx -1$; (3) the thick disk abundances are in good agreement
with the bulge values for essentially every element.}
\label{fig:abpttb}
\end{center}
\end{figure*}
\begin{figure*}
\begin{center}
\includegraphics[height=9.0in, width=7.2in]{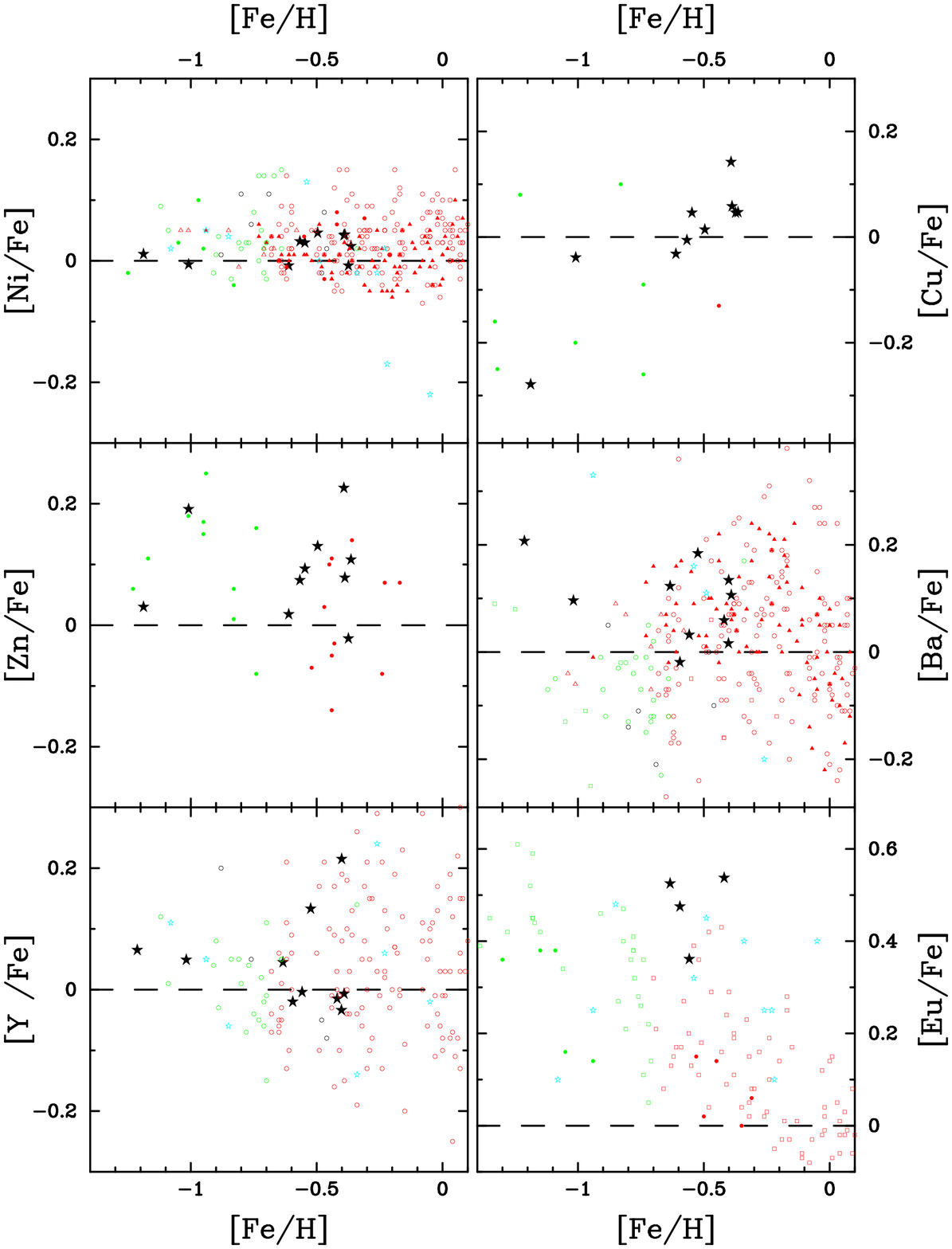}
\caption{Abundance patterns for the thick disk stars (large black stars)
for 18 of the elements analyzed in this paper.  For comparison, we plot 
the abundance patterns for halo stars (green points), thin disk
stars (red points), other thick disk measurements (small black stars),
and a small sample of bulge stars (light blue points).  The different 
symbols refer to the various literature sources indicated in the text.
There are three principal points to take from the figure: (1) for many of
the elements, the thick disk abundances are distinct from the thin disk;
(2) the thick disk abundances tend toward the halo star values at 
[Fe/H]~$\approx -1$; (3) the thick disk abundances are in good agreement
with the bulge values for essentially every element.}
\label{fig:abpttc}
\end{center}
\end{figure*}

\section{DISCUSSION}
\label{sec-discuss}

\subsection{Comparisons with Other Stellar Populations}

In the previous section, we presented an element-by-element account of the
abundances for the thick disk stars offering comparisons between
our results and previous work. In this section we will contrast the elemental
abundances against other stellar populations, namely the metal-poor halo
stars, the thin disk stars, and a small
sample of bulge stars.  In the following subsections, we comment on the
implications for the formation history of the thick disk and Galaxy as well
as the nucleosynthesis of these elements in the early universe.

Figures~\ref{fig:abptta}-\ref{fig:abpttc} present abundances for 18 elements 
with our sample of thick disk stars plotted as large black stars.
We do not include S in this subsection because of the very large 
uncertainties associated with these measurements in all stellar populations.
In each sub-panel, we overplot abundance measurements taken from a number of
literature sources which describe the metal-poor halo (green),
the thin disk (red), the thick disk (black), and the bulge (light blue).  
Abundance trends are sensitive to the stellar atmosphere,
spectrum synthesis algorithm, hyperfine splitting corrections, etc.\
adopted by each group and therefore systematic uncertainties exist in 
comparing various data sets.   It is important at the least to
keep this uncertainty
in mind and if possible to correct for identifiable systematic offsets.
The plot symbols distinguish 
the various data sets: (i) the \cite{edv93} data are plotted as open
circles.
We have noted ($\S$~\ref{subsec-pparm}) a systematic
offset between their temperature scale and ours of $\Delta T_{eff} \sim 100$~K.
We correct for this offset by applying the values in Table~9 of \cite{edv93}
which gives the sensitivity of various element ratios on $T_{eff}$.  
We have identified thick disk stars in the
\cite{edv93} sample according to the same kinematic criteria imposed for
our sample ($Z_{max} > 600$~pc; $-100 < \tilde V < -20$~\kms;
$-1.1 < {\rm [M/H]} < -0.4$).  We also identified a few halo stars
which have $\tilde V < -100 \mkms$ and large eccentricity. The remaining
stars are plotted as thin disk stars if [Fe/H]~$> -0.7$ and as halo
stars otherwise.  We warn, however, that many of these
stars at [Fe/H]$~\approx -0.7$ could be thick disk members.
(ii) The triangles represent the chemical abundances from
\cite{chen00}. 
We believe \cite{chen00} adopted a similar temperature
scale to ours so we make no correction to this data set.
\cite{chen00} identified a small subsample of thick disk
stars according to kinematic and metallicity criteria; these are plotted
as open red 
triangles in the figure.  As noted in the introduction, 
these stars have implied ages younger than that typically assumed for
the thick disk and therefore we contend that they are actually thin disk stars;
(iii) The solid circles depict the values taken from \cite{gratton88},
\cite{gratton91}, \cite{sne91}, and \cite{gratton94}.
Because these authors did not
characterize these field stars according to specific stellar populations,
we have plotted them under the assumption that those stars with 
[Fe/H]~$< -0.7$ are halo stars and the remaining are thin disk stars.
For the few stars near [Fe/H]~$= -0.7$ it is possible that there is 
some contamination from thick disk stars; 
(iv) The Sc and Mn values (solid squares) are taken
from \cite{promcw00} who reanalyzed the abundances reported by
\cite{nissen00}.  As \cite{promcw00} emphasize, their values may not 
represent the true Mn and Sc abundances, but they should be more accurate
than the \cite{nissen00} values which are based on an incorrect hfs treatment.
For these measurements, we have adopted the stellar population identifications
given by \cite{nissen00};
(v) For the O abundances, we have restricted our
comparisons to those measurements based on the O~I lines at 7770~\rAA
as there are concerns over systematic differences between the various 
approaches to measuring oxygen \citep{tomkin92,abia89};
(vi) The bulge abundances are taken from \cite{mcw94}; and
(vii) The open squares represent values taken from the
remaining literature sources \citep{brown92,mcw92,shetr96}.
For all of the data sets we have normalized the measurements to the
updated \cite{grvss96} solar meteoritic abundance scale as necessary.

It is difficult to digest all of the
abundance trends presented in Figures~\ref{fig:abptta}-\ref{fig:abpttc}, yet
several comparisons stand out.  
In terms of the $\alpha$-elements, the
thick disk stars exhibit significantly larger overabundances 
than the thin disk stars.  The O and Mg results agree
with the overabundances claimed by
previous thick disk studies \citep{grtt00,fuhr98}
while the Ca and Ti ratios lend even further support that the thick disk is 
chemically distinct from the thin disk.  
The Si/Fe ratios are the least enhanced over the thin disk, yet
other than the two lowest Si/Fe values even these values are higher
than the average thin disk star at [Fe/H]~$\approx -0.5$.  
Therefore,
the $\alpha$-elements offer convincing evidence that the thick disk population
is chemically discrete from the thin disk.
This point contradicts the interpretation of
\cite{chen00} who comment that the thick disk abundances
extend smoothly from the thin disk.
As noted above, however,
the \cite{chen00} sample of thick disk stars are younger
than the fiducial age of the thick disk casting their conclusions into
serious doubt. This point not withstanding,
carefully examining Figures~\ref{fig:abptta}-\ref{fig:abpttc}
one notes that the thick disk stars from 
\cite{chen00} all have metallicity [Fe/H]~$< -0.6$ where a comparison
with the thin disk is difficult owing to the paucity of measurements of
thin disk stars with [Fe/H]~$< -0.6$.
In contrast with the thin disk, the halo star values are in much better
agreement with the thick disk stars.
The Mg and Ti abundances show similar enhancements for the two populations 
while the remaining $\alpha$-elements 
approach the halo star values in the most metal-poor thick disk stars.
Finally, we find an impressive match between the bulge and thick disk
$\alpha$-element abundances\footnote{
There is the notable exception of oxygen, yet
McWilliam \& Rich (priv.\ comm.) now contend that their original O/Fe
ratios are in error.}.  The Ca, Mg, and Ti trends are all in 
agreement, while the thick disk Si abundances show somewhat
smaller enhancements. 
To summarize our results on the $\alpha$-elements, of the four
stellar populations the thick disk and bulge star abundance patterns
most resemble one another, the most metal-poor thick disk stars 
closely match the halo values, and all of these populations show 
$\alpha$-enhancements which are 
significantly larger than the values observed for thin disk stars.  

As with the $\alpha$-elements, the light elements -- sodium and aluminum --
are enhanced in the thick
disk stars.  The Na/Fe results are in good accord with the \cite{edv93}
analysis yet contradict the results from \cite{chen00}.
\cite{chen00} noted this discrepancy between their results and
\cite{edv93}, but did not determine the cause.  
With respect to the halo and bulge, the thick disk values
are somewhat larger than the typical halo values (either solar or sub-solar)
and the Na/Fe values from the metal-poor bulge stars 
(based on only 1 or 2 Na~I lines). 
Given the large uncertainties associated with measuring Na in each of
these stellar populations, we do not feel confident in drawing any
conclusions aside from the observation that none of the populations show 
significant Na/Fe departures from the solar ratio.
The Al measurements, on the other hand, offer more compelling comparison.
The thick disk Al abundances are enhanced over the thin disk
observations by both \cite{chen00} and \cite{edv93}
and are in reasonable agreement with the halo star observations.
Finally the majority of Al measurements from the bulge stars match the
thick disk, although there are several bulge measurements with 
[Al/Fe]~$> +0.6$~dex.

With respect to the iron-peak measurements of the thick disk,
the Ni, Cu, and Cr abundances generally track Fe, yet the remaining 
elements all show significant departures from the solar abundance.  
In several ways, the thick disk stars significantly distinguish themselves from
the halo and thin disk stellar populations.
First, unlike the thin disk the
V, Co, and Zn abundances are all enhanced in the thick disk.
Second, all but 
two of the thick disk stars exhibit significantly
lower Mn/Fe ratios than the thin disk.  Whereas the thin disk and halo
appear to exhibit plateaus in [Mn/Fe] at $-0.05$ and $-0.3$~dex respectively,
the thick disk values show a trend with metallicity which connects the
other two populations.
In the case of Zn, V, and Co, the thick disk ratios
are even enhanced over the halo abundances for halo stars with 
[Fe/H]~$\geq -1.8$ (the extremely metal-poor halo stars also have 
overabundances of Co and Zn; McWilliam et al.\ 1995, Johnson 1999).
As discussed in $\S$~\ref{sec:Sc}, 
the thick disk Sc/Fe results are very peculiar.
The stars with [Fe/H]~$\approx -0.5$~dex show a marked Sc/Fe overabundance
while the two metal-poor stars have nearly solar Sc/Fe ratios.
These results are in surprisingly good agreement with the Sc/Fe results from
\cite{promcw00}.  The Sc abundances are very unusual and 
require further investigation before one can meaningfully comment on 
their implications.
Given the significant number of distinctions between the thick disk and the
halo/thin disk stellar populations, the agreement between the
iron-peak abundance patterns of the thick
disk and bulge is stunning. In particular, the metal-poor
bulge stars show enhancements in V and Co at levels consistent
with the thick disk results and there is even an indication of enhanced Sc.
It must be noted, however, that the iron-peak
bulge values are primarily based on 1-2 lines and that \cite{mcw94}
did not take particular care to account for line blending.  The latter
point, in particular, gives pause because with the exception of Mn the
iron-peak abundances we are considering are enhanced.  
We anxiously await confirmation of these trends by \cite{mcw00}
who have recently obtained a sample of bulge stars observed with HIRES
on Keck~I at significantly higher resolution than their previous study.

Finally, we can compare the various stellar populations with
respect to the heavy element abundance trends.
This is somewhat ambitious, however, given the
large uncertainties of our Ba, Y, and Eu measurements.  Nevertheless,
the Ba and Y trends are in reasonable agreement for all of the stellar
populations in this metallicity range; in general, they are all consistent with
the solar composition.  Europium, however, presents a different story.
In agreement with the the $\alpha$-element observations and the notion
that Eu is formed predominantly in Type~II SN, the thick disk
Eu/Fe ratios are well in excess of the thin disk measurements and
comparable to the halo and bulge values. 

\subsection{Implications for the Formation of the Thick Disk}

In the previous subsection, we drew comparisons between the abundance
patterns of the thick disk stars with the halo, thin disk, and bulge
stellar components.  For the majority of elements --
the $\alpha$-elements, Co, V,
Zn, Al, Mn, Eu -- the thick disk stars show abundance patterns
which clearly distinguish them from the thin disk.  
These results provide compelling confirmation that 
the chemical history of the two stellar components is distinct as first
suggested by the Mg and O analysis of \cite{grtt00} and \cite{fuhr98}.
In turn, our analysis implies the thick disk is a physically
distinct stellar component from the thin disk with its own specific
formation history.

\cite{grtt00} and \cite{fuhr98} have further argued that the discrete
separation of the O/Fe and Mg/Fe ratios
between the two stellar populations implies a significant time delay between
the formation of the thick disk and the onset of star formation in the thin
disk.  This assertion appears well supported by our results. The
larger $\alpha$/Fe enhancement demonstrated in the thick disk indicates 
the thin disk stars formed from gas more significantly polluted by 
Type~Ia SN requiring a significant delay between the formation
epochs.  As described below, however, this assertion is now complicated by
the tentative evidence for Type~Ia contributions in the thick disk abundance
patterns.  Nevertheless, the fact that the thick disk abundance patterns
do not smoothly transition to the thin disk does require a significant delay
in the star formation rate following the formation of the thick disk stars.
The overabundance of the $\alpha$-elements in the thick disk as well
as the underabundance of Mn/Fe indicate that the thick disk formed prior
to the thin disk on the grounds that there was less pollution from the 
long-lived Type~Ia SN.  In addition, the observed halo-like Ba/Eu ratio
(Figure~\ref{fig:BaEu}) indicates r-process dominated
enrichment which supports a great age for these stars.  

In the abundance studies by \cite{grtt00} and
\cite{fuhr98} there is no indication for a trend of [O/Fe] or [Mg/Fe]
with metallicity.  \cite{grtt00} therefore asserted that 
the formation of the thick disk occurred on time-scales shorter than the
Type~Ia SN, i.e.\ $t_{form} < 1$~Gyr.  
If confirmed, the tentative trends of Ca, Si, and O with metallicity in our
sample of thick disk stars contradict this assertion.
Traditionally, the decrease of an $\alpha$-element enhancement with
increasing [Fe/H] metallicity is attributed to the onset of Type~Ia SN.
This explanation is well motivated by our theoretical understanding of 
Type~Ia and Type~II SN. If it explains the Si, Ca, and O trends for the
thick disk, it will be difficult
to reconcile our observations with the notion of a brief thick disk
formation time.  Instead the observed 
$\alpha$-element trends may indicate that the formation time of the thick disk
exceeds the Type~Ia SN timescale, i.e.\ $t_{form} \gtrsim 1$~Gyr.  
To reconcile the nearly constant Mg/Fe trend under this scenario,
however, one may require fine tuning in the form of enhanced formation of
very massive Type~II SN, an enhanced production factor for Mg in metal-poor
Type~Ia SN, or increased Mg production in more metal-rich Type~II SN.
Another possibility altogether is that the $\alpha$-element trends are
the sole result of Type~II SN with an evolving IMF; specifically, the
importance of the moderate mass Type~II SN decreases as [Fe/H] increases.
In this case, $t_{form} \ll 1$~Gyr, but it is difficult to understand the
O/Fe trend with [Fe/H].  In the following, we will adopt the hypothesis
that the O,Si, and Ca trends imply $t_{form} \gtrsim 1$~Gyr but warn that
further investigation into the nucleosynthetic history of the thick disk
is crucial.

To date, several scenarios have been put forward to 
explain the formation of the
thick disk.  In particular, these include dissipative collapse models
\citep{larson76,jones83}, 
dynamical heating \citep{noguchi98}, and merger scenarios
\citep{qnn93,huang97,sell98}.  Our observations place important constraints
on these models.  Consider first the formation of the thick disk
during the dissipative collapse of a gaseous halo.  In order to maintain
the chemical (and kinematic) distinction between the thick and thin disks,
the thick disk must form prior to the thin disk,
{\it in situ} from a gaseous disk with a velocity
dispersion consistent with the thick disk stars $(\sigma_W \approx 40 \mkms$).
As suggested by \cite{jones83}, this may be the 
natural evolution of a gaseous halo which 
has undergone some initial dissipation.  
The dissipational time-scale, however, is significantly shorter than
1~Gyr, i.e. $t_{form} \ll 1$~Gyr in the dissipational collapse scenario.
One could invoke additional energy sources such as primordial magnetic
fields \citep{jdk99} or an enhanced supernovae rate and demand that they
maintain the 40~\kms velocity dispersion of the gas disk for 1~Gyr
while allowing the dissipational collapse and formation of molecular
clouds and stars, but this is unlikely.
On the grounds that the formation time 
of the thick disk might well exceed 1~Gyr,
we contend it is unlikely that the thick disk 
formed {\it in situ} as envisioned in an ELS-like scenario.  
Therefore, the thick disk must have been preceded by a stellar thin
disk which was heated -- either gradually over the course of
$\approx 1$~Gyr or suddenly -- to the current thick disk. 
The clumpy star-forming region model envisioned by \cite{noguchi98} provides
a gradual heating mechanism ($t_{form} \approx 1$~Gyr)
for the formation of the thick disk.  In this
model, $10^9 \msol$ clumps form during the dissipational collapse of the
Galaxy which initiate star formation, spiral to the galaxy center
via dynamical friction, scatter the stars in the initial thin disk to form
a thick disk, and eventually merge to form the bulge.  
The model does not describe the formation of the
current thin disk;  presumably it will form after the clumps have merged
through the accretion of a new reservoir of gas, but 
this may require some fine tuning to insure that the resulting
thin disk is chemically distinct from the thick disk.  This complication aside,
the model accounts for the majority of our observations.
Finally, consider merging scenarios where the initial thin disk is heated
via the accretion of one or more satellites.  Unfortunately, it is difficult
to assess this model as the various numerical studies conflict on the
effects of a merger event (e.g.\ heating efficiency; Quinn et al.\ 1993;
Huang \& Carlberg 1997; Sellwood et al.\ 1998).  
Nonetheless, this scenario should naturally
allow for the formation of the initial thin disk over the course of 1~Gyr
and the merger event(s) would erase all trace of the initial thin disk
providing the discrepancy between the resulting thick disk and the future
thin disk.  Similar to the clumpy region model, however, the formation
of the thin disk after the merger event has not been considered in these
numerical simulations.  Furthermore, they disagree on the robustness of the
thick disk to future accretion events.  Hopefully, future numerical simulations
will resolve the current conflicts and pursue the formation of both the
thick and thin disks.

What does a comparison of the thick disk with the halo chemical abundances
reveal?  It is informative that with the exception of Co and V
(and maybe Zn) the thick disk abundances for all of the elements
match or approach the abundances
of halo stars with [Fe/H]~$\approx -1.3$.  This implies that the
metal-poor thick disk stars
formed from similar material as that for the most metal-rich halo stars.  
This naturally follows from the notion that the
halo stars formed first, i.e., before the majority of gas had dissipated
to form a disk (thin or thick).  
Subsequently, those stars which formed
first in the disk exhibit abundances similar to the metal-rich halo stars while
the later disk stars show higher metallicity and evolved abundance patterns.
This explanation does not account for the discordant Co
and V trends.  Perhaps these abundances are the result of a 
specific type of 
nucleosynthesis event (e.g.\ supernova) whose time-scale exceeds
the formation of the halo.  Alternatively, the overproduction of these
elements may require 
specific physical conditions (i.e.\ surface density, binary 
fraction, metallicity) which did not exist in the halo.  Therefore,
by identifying the mechanisms responsible for the Co and V enhancements, 
one might gain further insight into the time-sequence of the formation of the 
Galactic halo and thick disk.

The similarities of the thick disk and bulge chemical abundance patterns are
striking.  With the exception of the O/Fe (which can be disregarded)
and possibly the Si/Fe ratios,
{\it there is no significant difference in the abundance
patterns of any element}.  The obvious interpretation of this agreement
is that the two stellar components formed at essentially the same time and
from the same gas reservoir.  
It is also interesting to note that the presence of thick disk in external
galaxies has been linked to systems with significant bulges \citep{bur79}.
The most straightforward explanation is that
a single merger event sparked the formation of both stellar components, 
although this is not the only possibility. 
The two components may have formed together through a lengthy dynamical
process as described by \cite{noguchi98}. 
It is also worth noting that the dissipational collapse model of 
\cite{jones83} also predicts a close association between the metal-poor
bulge stars and the thick disk.
Nevertheless, the claim that the two components
shared a similar gas reservoir and formation epoch is largely
independent of the exact physical processes involved in their formation.  
We eagerly look forward
to future analyses of the bulge \citep{mcw00}, in particular the metal-poor
tail.  If these observations confirm the current picture (in particular
the Co and V enhancements), the two stellar components will be 
irrefutably wed and all viable Galactic formation scenarios will need to
account for them simultaneously.

\subsection{Implications for Nucleosynthesis}

Irrespective of comparisons with other stellar populations, the chemical
abundances of the thick disk stars provide unique constraints on the 
processes of nucleosynthesis in the early universe.  Consider first the
$\alpha$-elements which exhibit an overabundance relative to Fe for
all of the thick disk stars; the enhancements indicate that the gas from which
these stars formed was primarily enriched by Type~II SN.  Furthermore, by 
comparing the various abundance trends of the
$\alpha$-elements one gains insight into the yields
from different mass and/or metallicity Type~II SN.
As noted in the previous subsection,
the most striking comparison is that 
the Mg and Ti abundances show no dependence on [Fe/H] metallicity whereas
the Ca, Si, and O ratios appear to decrease steadily with increasing 
[Fe/H] for the thick disk stars.  While this statement is largely
dependent on the observations of the two metal-poor stars in our sample and
therefore suffers from small number statistics, 
we speculate on the implications.
Differences between the trends of Mg and Si or 
Ca may be expected on theoretical grounds because these elements 
are believed to be synthesized in different mass 
Type~II SN \citep{ww95}.  In particular, the SN models suggest that Mg
is produced primarily in the highest mass Type~II SN whereas the 
production of Si and Ca is dominated by moderate mass SN.  The observed 
trends, however, are difficult to explain under the assumption of a 
constant IMF.  If anything, one might expect significant evolution in Mg
as it arises in the more massive Type~II SN.
Therefore, either nucleosynthesis occured in Type~II SN with an 
evolving IMF or Type~Ia SN are playing a 
significant role in the observed abundance trends. 
If the latter explanation is adopted (as assumed in the previous subsection),
then it becomes a challenge to explain the constancy of the Mg/Fe ratio.
Perhaps at [Fe/H]~$\approx -0.5$ there is an increase in the number of
very massive Type~II SN or maybe the 
the first generation of Type~Ia SN overproduces Mg.
It is also difficult to reconcile the difference in the 
Mg and O abundance trends 
as both elements are expected to be
produced primarily in massive Type~II SN and therefore should track one
another reasonably well.   It is possible the O trend is
the result of a systematic error (i.e.\ non-LTE effects) in
measuring O from the O~I triplet at $\lambda \approx 7775$\AA.
If this is not the case, the observations pose a meaningful challenge to
the current models of Type~II SN.  Finally,
while the similarity between the Mg and Ti trends may be a coincidence,
it does suggest the possibility that the two elements are produced in
similar nucleosynthetic sites.  At the very least, while searching for the
processes which yield enhanced Ti one may wish to first focus on those
mechanisms which produce Mg.

The behavior of the light elements in the thick disk stars
differs considerably.  The Na/Fe
ratios follow the $\alpha$-element enhancements -- albeit with a more mild
overabundance -- in a fashion possibly
consistent with some Type~II SN contribution.
The enhancement is small, however, and could be the result of a systematic
error in the Na analysis. In contrast, aluminum is significantly enhanced
over the solar ratio which clearly points to a significant production of
Al in massive stars.  Like Mg and Ti, Al
exhibits no obvious trend with [Fe/H] metallicity in our sample.  Again,
we can speculate that Al is formed in similar sites as Mg and Ti.

With respect to nucleosynthesis, the
abundance trends of several of the iron-peak elements may offer the
most startling results.  In particular,
the thick disk stars show enhanced Co and V where if anything
one predicts a deficiency for these 'odd' nuclei.  Furthermore, the Zn
enhancement -- while mild -- appears to contradict the previous 
empirical belief that Zn/Fe is solar at all metallicities.
The Co enhancement brings to mind the overabundance observed
for extremely metal-poor stars \citep{mcw95}.
Cobalt is believed to be synthesized during the $\alpha$-rich freeze out
fueled by a neutrino-driven wind \citep{woo92}.  An enhancement of Co
might then be expected to be correlated with other elements slightly
more massive than Fe.  Similar to the extremely metal-poor stars, however,
we do not observed enhanced Ni or Cu, yet we note that the Zn enhancement
(also seen in the extremely metal-poor stars; Johnson 1999) could be
related to the Co pattern.  It will be important to focus on the Co/Zn
ratio in future studies.  The vanadium overabundance is
not predicted by any current theoretical model of nucleosynthesis.
According to \cite{ww95}, the dominant vanadium isotope $^{51}{\rm V}$
is primarily produced during the $\alpha$-rich freeze out, 
yet the leading theories predict only a fraction of 
the observed solar vanadium abundance \citep{woo86}.  
In fact, to the best of our knowledge this marks the first significant
evidence for enhanced V in any stellar population.
The V overabundance poses an excellent challenge for nucleosynthesis research
on an element which until now has been largely ignored.
Finally, note that the unusual Sc/Fe trends agree surprisingly well 
with the reanalysis of \cite{nissen00} by \cite{promcw00}.  
The trend is very 
peculiar with the [Fe/H]~$\approx -1$ stars exhibiting essentially no 
Sc/Fe enhancement, most stars at [Fe/H]~$\approx -0.5$
enhanced above solar, 
and the nearly solar metallicity stars show solar Sc/Fe.  
As with Co and V, scandium is believed to be produced primarily 
through the $\alpha$-rich freeze out process.  Given the very unusual
behavior of Sc/Fe, a combination of metallicity dependent yields and
various supernovae appears likely.  Perhaps the overabundance at
[Fe/H]~$\approx -0.5$ is due to the onset of Type~Ia SN whose relative Sc/Fe
production decreases with increasing metallicity. 
Altogether, the Co, V, Sc, and Zn enhancements point toward a further
investigation of the $\alpha$-rich freeze out nucleosynthesis.  In turn,
one may learn about the IMF of the thick disk stars.
Finally, we note that the Mn/Fe values for the thick disk are generally
lower than the thin disk values.  We consider this evidence for the
overproduction of Mn in Type~Ia SN \citep{gratton89}.  
This assertion is further supported by
the fact that the stars with the highest Mn/Fe ratios show the lowest
Si/Fe, Ca/Fe, and O/Fe values and argue against
a metallicity dependent yield for Mn.

\begin{figure}[ht]
\includegraphics[height=3.8in, width=2.8in,angle=-90]{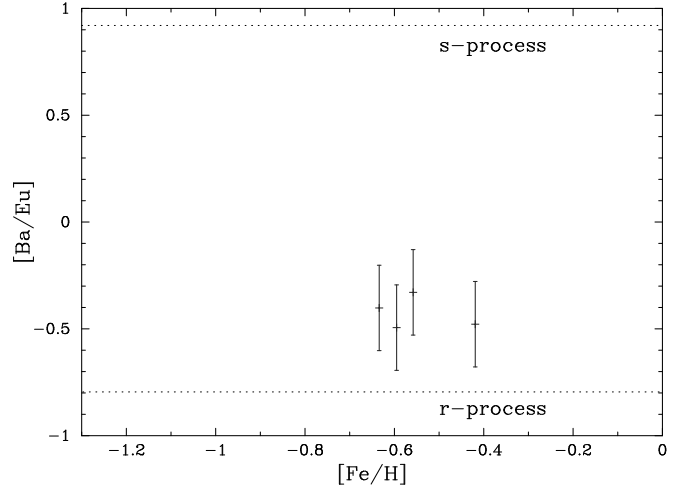}
\caption{[Ba/Eu] ratios vs.\ [Fe/H] metallicity for the 4 stars with
measured Eu abundance.  Overplotted in the figure are the fiducial values
for the s and r-process Ba/Eu ratios.  The results suggest a mix of the
two processes with the r-process being dominant.
}
\label{fig:BaEu}
\end{figure}

We conclude our discussion of nucleosynthetic implications with 
a few comments on the heavy element results.  
Figure~\ref{fig:BaEu}
presents the [Ba/Eu] ratio against [Fe/H] for the four stars with wavelength
coverage of the Eu~II $\lambda 6645$ line.  Overplotted on the figure
are two lines indicating the fiducial values for the solar s and r-process
values of the Ba/Eu ratio.  The thick disk ratios lie 
in between the two fiducial values suggesting that both processes are important
in the nucleosynthesis of the heavy elements, although the r-process appears
dominant.  As noted in the previous section, this is consistent with the very
large age believed for the thick disk stars.  Lastly, the Ba/Y ratio is 
approximately solar in line with the halo composition \citep{gratton94}.

\subsection{Comparisons with the Damped \lya Systems}
\label{sec:dla}

This final section describes the implications of our results on 
interpretations of the abundance patterns of the damped \lya systems.
The damped \lya systems are absorption line
systems identified along the sightlines to distant quasars and have 
neutral hydrogen column densities $\N{HI} \geq 2 \sci{20} \cm{-2}$.  
Owing to their very large $\N{HI}$ values, 
these systems dominate the neutral hydrogen content of the
universe at all epochs \citep{wol95,rao00}.  
Furthermore, the very large column densities
imply overdensities suggestive ($\delta \rho / \rho \gg 100$) 
of protogalaxies.  Finally, the comoving baryonic mass density in gas
inferred from the damped systems at $z = 2 - 3$ coincides with the current
comoving baryonic mass density in stars today \citep{wol95,storr00}.
For these reasons
the damped \lya systems are widely believed to be the progenitors of modern
galaxies. 
The majority of studies on the chemical abundances of these protogalaxies
has focused on damped systems with absorption redshift $z_{abs} > 1.7$ 
(i.e.\ $t > 10.5$~Gyr for a universe 15~Gyr old) where the \lya absorption
line is observable with optical spectrographs.  Therefore,
with the exception of Mn (whose transitions lie at large rest wavelength
and are most easily observed in lower $z_{abs}$ systems), our discussion
will focus on these very old systems.

A comparison of the damped \lya systems with the Galactic thick disk is
motivated by: (i) the damped systems are believed to be the progenitors
of modern galaxies like the Milky Way; 
(ii) the thick disk is believed to have formed at an epoch
consistent with $z_{abs}$ of the damped systems; 
(iii) the Galactic thick disk kinematics are surprisingly consistent
with a model introduced by \cite{pro97a} to explain the absorption line
profiles of the damped \lya systems;
and (iv) the metal-rich damped systems contain enough baryons at the thick
disk metallicity to account for the stellar mass of the Galactic thick disk
\citep{wol98}.
While these similarities are present, we should note that the thick disk 
component may not be a generic component of disk galaxies and that 
the majority of damped \lya systems exhibit significantly lower metallicity
than the Galactic thick disk.  Therefore, the two
systems may not have a one-to-one correspondence.  
These points not withstanding,
our observations of Zn in the thick disk have
immediate impact on the damped \lya abundance studies. 
For the damped \lya systems, Zn currently plays the most pivotal role in
interpreting abundance patterns in the early universe.  The key point is
that measurements of the damped \lya systems are based on observations of
Fe, Si, Zn, Cr, Ni, etc.\ in the {\it gas-phase} (analogous to abundance
measurements made of the Galactic ISM; e.g.\ Savage \& Sembach 1996) where 
elements like Fe, Ni, and Cr can be significantly depleted
onto dust grains.  Concerns over the effects of dust depletion in the
damped systems are well motivated by \cite{ptt94} who demonstrated 
an overabundance of Zn/Cr relative to solar of $\approx 0.4$~dex.  This
enhancement suggests dust depletion because Zn, unlike Cr, is not heavily
depleted onto dust grains in the ISM.
The implication, therefore, is that 
in order to assess even the [Fe/H] metallicity
of a damped \lya system one must first account for the depletion of Fe onto
dust grains.  An alternate approach -- the one typically implemented
in damped \lya research -- is to utilize Zn as a surrogate for Fe because:
(1) Zn is not expected to be significantly depleted onto dust grains, and
(2) Zn was found to track Fe at essentially all metallicities in stars
\citep{sne91}.   It is on the second point that our observations play a 
meaningful role.

In the majority of studies on the damped \lya systems, researchers have
assumed that [Zn/Fe]~$\approx 0$ irrespective of [Fe/H].  While empirically
this assertion has the support of stellar abundance analysis \citep{sne91},
it is difficult (if not impossible) to motivate theoretically.  Zinc is
an iron-peak element, but it is not expected to be synthesized in
a similar fashion to Fe.  The leading theory of Zn nucleosynthesis 
contends that Zn forms in the neutrino-driven winds of Type~II SN
\citep{hff96}.  Under this scenario, Zn may be expected to behave like an
$\alpha$-element and therefore exhibit an enhancement relative to Fe
in metal-poor stars.

Our observations suggest such an enhancement ([Zn/Fe]~$\sim +0.1$), 
albeit at a level
below the majority of $\alpha$-elements and -- more importantly --
below the Zn/Fe enhancement observed
in the damped \lya systems.  The thick disk abundances, however,
do reflect a more complicated origin for Zn than the one readily adopted
in quasar absorption line studies.  Furthermore, as noted in $\S$~\ref{sec:Zn},
the Zn/Fe ratio is quite sensitive to the effective temperature adopted
for the stellar atmosphere.  
In fact, we may have overestimated $T_{eff}$ in the
majority of stars ($\S$~\ref{subsec-pparm}), such that the true $\XFe{Zn}$
value in the thick disk stars would be closer to $+0.15$~dex.  
Even this enhancement
would not fully account for the [Zn/Fe]~$\approx +0.4$
enhancement observed for the damped \lya systems \citep{pro99},
but it does emphasize that a dust explanation alone is not accurate for
the damped \lya observations.  Furthermore, 
note that the typical [Fe/H] of the thick
disk stars is significantly higher than that of the damped \lya systems
([Fe/H]~$\approx -1.5$). Recent studies of   
the extremely metal-poor stars indicate a 
[Zn/Fe] enhancement of $\approx +0.2$ to +0.3~dex 
at [Fe/H]$~< -2.5$ \citep{jhnsn99}.
An extensive survey for Zn in stars 
with [Fe/H]~$\approx -1.5$ is clearly well motivated. 

\begin{figure*}
\includegraphics[height=9.3in, width=7.2in]{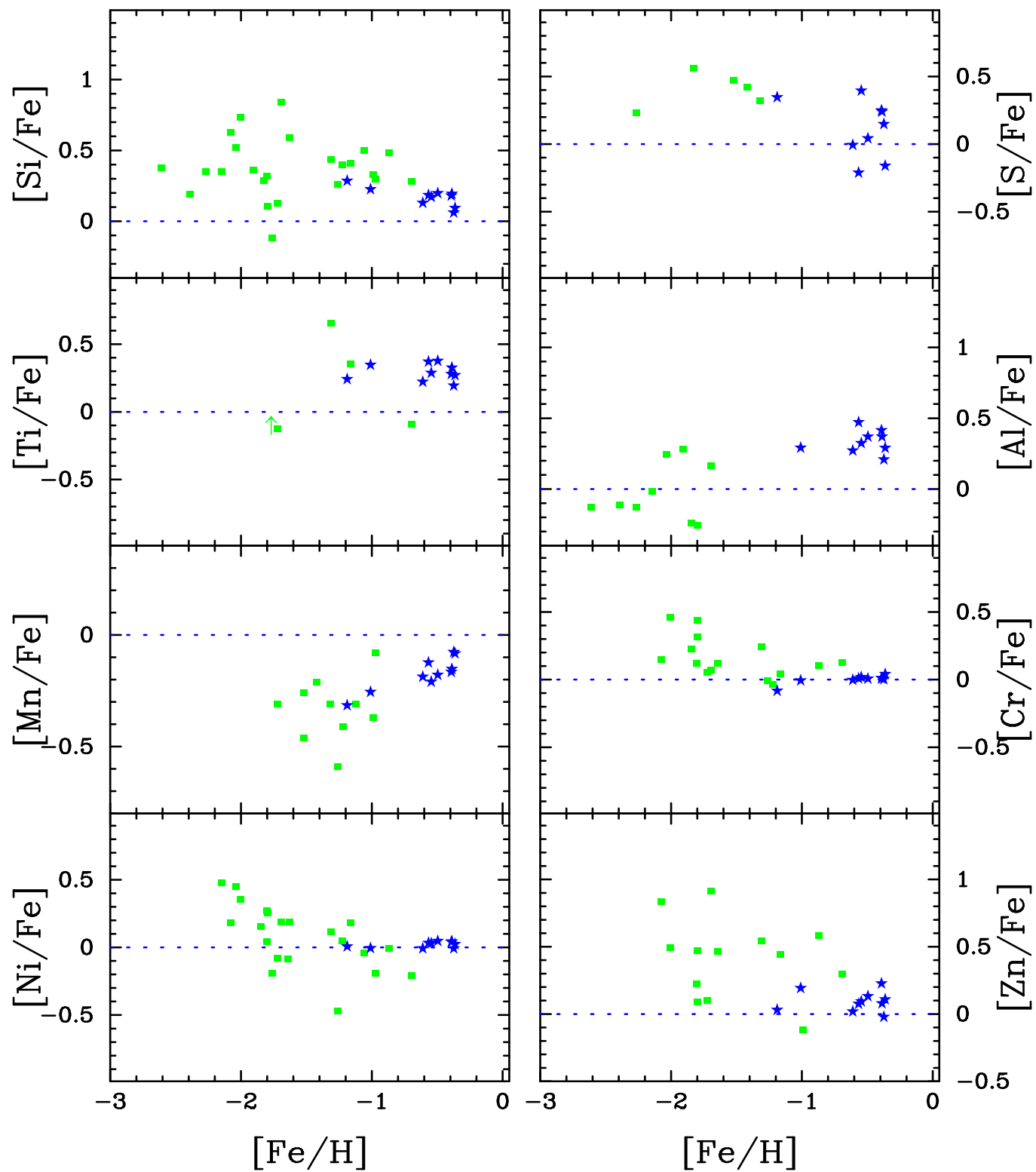}
\begin{center}
\caption{A comparison of the abundance patterns for our sample of thick
disk star with those for a sample of damped \lya systems taken from 
\cite{lu96,lu97,pro99}.}
\label{fig:dla}
\end{center}
\end{figure*}

The prospect of nucleosynthetically
enhanced Zn/Fe may actually be a welcome sight to the interpretations
of the damped \lya abundance patterns.  The current controversy regarding
the abundance patterns is the following
(see Prochaska \& Wolfe 1999 for a more detailed discussion). 
Figure~\ref{fig:dla}
plots the abundance patterns for a compilation of damped \lya systems
\citep{lu96,pro99}
with a range of metallicity\footnote{The Cr and Ni observations at the
lowest metallicities in the damped systems are at the detection limit
and are likely to have been biased to larger [Ni/Fe] and [Cr/Fe] values.}.
When a stellar abundance
researcher examines the plot, s/he sees a classic example of halo
abundances with enhanced $\alpha$-elements, deficient Mn and
unpeculiar Ni, Cr and Al\footnote{Note that field stars with 
metallicities similar to the damped \lya systems do 
not show significantly enhanced Al/Fe; Shetron 1996, Gratton \& Sneden 1988}.  
The overabundance of Zn relative to Fe would be puzzling yet the overall
implication of Type~II SN enriched gas would be clear.  In fact, one might
go so far as to claim that the figure depicts a natural evolutionary
sequence in the abundances from the damped systems to the thick disk stars. 
In contrast, if an expert in the interstellar medium
studies the figure, s/he would identify a warm, halo gas dust-depletion
pattern to explain the enhanced Si, Zn, and S.  This interpretation 
is consistent
with the Ni, Cr, and Al abundances, too, but fails to account for the Mn trend
or the two systems with enhanced Ti 
as neither underabundant Mn nor enhanced Ti is observed in 
dust depleted gas in the ISM, SMC, or LMC \citep{sav96,welty97,welty99}.  
The difficulty in interpreting the damped \lya abundance 
patterns, therefore, has been that two explanations exist
which account for the majority of the observations but not all.  Because
the two interpretations are degenerate in the observed elements, it
may even be possible to allow for both Type~II SN $\alpha$-enrichment
and a dust depletion pattern (although the S/Fe ratio may set an upper limit
to the combination; Prochaska \& Wolfe 1999).  

The implications of a Type~II SN enrichment pattern for the damped \lya
abundances are reasonable.  These systems have low metallicity and
presumably young ages and one would expect them to have been enriched
primarily by Type~II SN.  The difficulty remains, however,
in explaining the large Zn/Fe enhancements.  On the other hand, if one
adopts the ISM perspective then the underlying nucleosynthetic pattern must
more closely resemble a Type~Ia SN yield \citep{vld98}.
The implications for this scenario are troubling.  First, we know that a
significant number of stellar systems exhibit enhanced $\alpha$-elements
\citep[e.g.][]{mcw97,trager00}.  If the damped \lya systems do not contain
the gas which gave rise to these systems, then where is that gas?
We reemphasize that the damped systems dominate the neutral hydrogen gas
content of the universe at all observed epochs ($z = 0 - 4.5$), i.e.,
that gas which
is most likely responsible for star formation.  Second, if the gas in the
damped systems has been significantly enriched by Type~Ia SN, then the stars
which polluted these systems must have formed $\approx 1$~Gyr prior to 
the observed epoch.  For a damped system at $z_{abs} \approx 3$, this
implies a formation epoch which quickly approaches the Big Bang!  For
these reasons it is difficult to accept a pure dust depletion
explanation for Zn/Fe.

Now consider that Zn/Fe
may be enhanced in metal-poor stars and therefore presumably in the
metal-poor damped \lya systems.  In this case, a scenario with both
dust depletion and Type~II SN enhancement is valid, if not favored.
While one must introduce dust to account for some of the observed
Zn/Fe overabundance, there could still be an underlying Type~II SN pattern.
The exact level of $\alpha$-enhancement would depend on the dust depletion
pattern one assumes, particularly for Si.  Dust-corrected enhancements
of [Si/Fe]~$\approx +0.2$~dex are viable and even enhancements of 
+0.3~dex are conceivable.  Unfortunately, there will always be uncertainties
in interpreting Si owing to the exact depletion pattern one adopts.
Oxygen would be an ideal prospect for removing this degeneracy if not
for the difficulties in measuring O in the damped \lya systems \citep{pro00}.

\begin{figure}[ht]
\includegraphics[height=3.8in, width=2.8in,angle=-90]{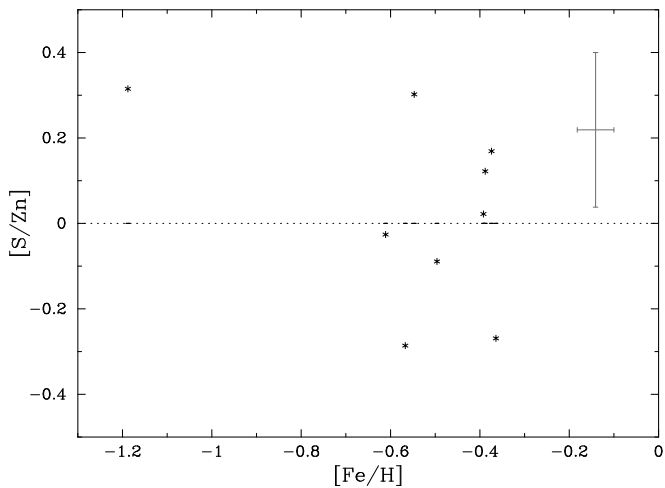}
\caption{Solar-corrected [S/Zn] ratios vs.\ [Fe/H] metallicity for the
9 thick disk stars with S measurements.  All of the data points scatter
around the solar meteoritic ratio.}
\label{fig:SZn}
\end{figure}

As a final point regarding the damped \lya systems, consider the S/Zn ratio
observed for the thick disk stars.  In several recent studies on the damped
systems, researchers have suggested that the S/Zn ratio may provide the
best indication of nucleosynthesis in these systems \citep{cent00}.  
The argument follows
from the fact that neither S nor Zn are significantly affected by dust
depletion and S is expected to be a Type~II SN tracer while 
Zn is a surrogate for Fe.  In Figure~\ref{fig:SZn}, we plot the [S/Zn]
measurements for the 9 thick disk stars with a sulfur measurement. 
Recall that the sulfur abundances are very sensitive to the assumed 
$T_{eff}$ values and are based on weak, single line measurements, i.e., 
the $\e{S}$ values are very uncertain.  This uncertainty is somewhat
tempered by the fact that the Zn~I lines have a reasonably 
significant dependence on $T_{eff}$ such that S/Zn is less affected by errors
in $T_{eff}$ than S/Fe.  
Examining Figure~\ref{fig:SZn} one notes that
the [S/Zn] values suggest no significant enhancement
in these stars ([S/Zn] = $0.03 \pm 0.08$) even though the
$\alpha$-elements Ca, Ti, Mg, O, and Si are all enhanced relative to Fe.  This
result is due to the fact that we find only a mild S/Fe enhancement in
the thick disk stars (in contradiction with the observations of
Francois 1988) and the observed Zn/Fe enhancement.  Of course,
it would be most informative to repeat this analysis at metallicities
more representative of the damped \lya systems.  Nonetheless,
the results suggest that [S/Zn] is not a reliable indicator of 
nucleosynthesis.  In fact, our observations may even help to explain
the puzzling
sub-solar S/Zn ratios observed by \cite{cent00} in a few damped systems.

\subsection{Concluding Summary}

We have presented a detailed chemical abundance analysis of 10 kinematically
selected thick disk stars with metallicity ranging from $-1.2$ to $-0.4$~dex.
The majority of [X/Fe] ratios for the elements studied exhibit significant
($> 0.1$~dex) departures from the solar abundance.  These include
(i) overabundances for all of the $\alpha$-elements except S;
(ii) enhancements above solar for the light elements Na and Al;
(iii) overabundances of the iron-peak elements Sc, Co, Zn, and V and
deficient Mn; (iv) significantly enhanced Eu.
For S, Ni, Cr, Cu, Ba, and Y we find essentially solar abundances relative
to Fe.

We compared our results with abundance studies of the halo, bulge, and thin
disk taken from the literature.  These comparisons reveal that in the 
majority of cases the thick disk stars exhibit X/Fe values distinct from the
thin disk.  In particular, O, Ca, Mg, Ti, Eu, Al, V, and Co all 
show enhancements
in the thick disk well in excess of those found in the majority of thin
disk stars with comparable [Fe/H] metallicity.  We argure, therefore, that
the thick disk has a distinct chemical history from the thin disk.
In general, the thick disk abundances either match or at low [Fe/H]
tend toward the patterns observed for halo stars with [Fe/H]$~> -1.5$.
Most impressive, however, is the excellent agreement between the thick disk
and metal-poor bulge star
abundance patterns.  With the possible exception of Si,
the chemical histories of these two stellar populations 
appear remarkably similar.

Unlike previous elemental abundance studies of the thick disk
\citep{grtt00,fuhr98}, we find tentative evidence that several 
$\alpha$-elements (O,Si,Ca) show trends of declining enhancements with
increasing [Fe/H].  If these trends are interpreted as evidence for 
the onset of Type~Ia SN (we caution that 
this is not the only viable explanation), then 
the thick disk formed over the course of $\gtrsim 1$~Gyr.
Because this conclusion has significant impact on formation scenarios for
the thick disk, future efforts to confirm the $\alpha$-element 
trends and investigate other explanations are essential.

We discussed the implications of our observations on the leading formation
models of the thick disk.  The differences between thick and thin disk
abundance patterns imply that the thick disk formed prior to the thin disk
and that there exists a significant delay between its formation and that
of the thin disk.
We argued that if the formation time of the
thick disk is $\sim 1$~Gyr then most dissipational collapse models are
ruled-out.  On the other hand, the clumpy region model of 
\cite{noguchi98} and the merger scenarios \citep[e.g.][]{qnn93} are viable
scenarios, although none of these studies considered the formation of the thin
disk in a self-consistent manner.  

The excellent agreement between the thick disk and the metal-poor bulge
abundance patterns provides a strong argument that the two populations
shared a common gas reservoir and formation epoch.  A conncetion between
the two populations has previously been suggested through imaging studies
of edge-on spirals \citep{bur79} and can be explained by several of the
formation scenarios \citep{jones83,noguchi98}.  Our observations 
tighten the association between the two populations, hinting they
are intimate from the onset of their formation. 

In the final sections, we discussed implications of our analysis on 
nucleosynthesis in the early universe and interpretations of the damped
\lya abundance patterns.  In terms of nucleosynthesis, the conflicting
$\alpha$-element trends of Mg, O, Si, Ti, and Ca present a challenge to
our current understanding of nucleosynthetic production in
Type~Ia and Type~II SN.  
We also noted that the Sc, V, Co, and Zn
overabundances hint at an overproduction through an enhanced 
$\alpha$-rich freeze-out process.  For the damped \lya systems, we 
described the implications of enhanced Zn/Fe in the thick disk stars.  In
particular, this enhancement allows for interpretations of the damped \lya
abundance patterns which include a combination of dust depletion and 
Type~II SN enrichment patterns.  Finally, we noted that the S/Zn ratio
is not a good nucleosynthetic indicator.

Our future observational efforts will include obtaining 
additional spectra of kinematically selected thick disk stars.  In particular, 
we will further investigate the trends of O, Si, and Ca by increasing the
sample of thick disk stars with [Fe/H]~$\approx -1$ and by adjusting our
setups to include the O~I [6300] forbidden transition.
We also intend to
observe thick disk stars at greater radial distances to examine radial
dependencies of the abundance trends.  To minimize potential systematic
errors of comparing our results against abundance studies taken from the
literature, we will observe a significant sample of halo stars with
[Fe/H]~$\approx -1$, thin disk stars with [Fe/H]~$\approx -0.5$, and metal-poor
bulge stars.  The results presented in this paper demonstrate that these
stellar populations offer remarkable insight into the Galactic formation
history.

\acknowledgments

We wish to acknowledge
J. Fulbright, J. Johnson, M. Albada, B. Weiner, and R. Bernstein
for insightful discussion and comments.
We thank T. Barlow for providing the HIRES reduction package.
We acknowledge the very helpful Keck support staff for their efforts
in performing these observations.   We thank J. Omeara and
D. Tytler for taking observations of G92-19.
Finally, we thank R. Kurucz for help in
using his codes and databases.
JXP acknowledges support from a Carnegie postdoctoral fellowship. 
BC acknowledges support from an NSF grant to UNC: AST-9619381.
AM acknowledges support from NSF grant AST-9618623.
AMW was partially supported by 
NASA grant NAGW-2119 and NSF grant AST 86-9420443.  

\appendix

\section{Stellar Gravity}
\label{app:grav}

In those cases where there are Hipparcos parallax measurements,
we calculated the stellar gravity according to the following
equations. 

\begin{equation}
\log (g/g_\odot) = \log (L/L_\odot) + \log (M/M_\odot) +
4 \log (T/T_\odot)
\end{equation}

\noindent by assuming the following mass relation for these 
main sequence stars,

\begin{equation}
\log (M/M_\odot) = 0.48 - 0.105 M_{bol}\cmma
\end{equation}

and

\begin{equation}
\log (L/L_\odot) = - (M_{bol} - 4.64) / 2.5 \perd
\end{equation}

\noindent  To estimate the bolometric correction ($BC \equiv M_V - M_{bol}$), 
we interpolated
between the tabulated values of \cite{alonso95}.  Given the uncertainties
in the parallax and $BC$ values, the error in $\log (g/g_\odot)$ from
this approach is $\approx 0.1$~dex.  

\section{Hyperfine Splitting}
\label{app-hfs}

The coupling of the nuclear angular momentum {\bf I} with the angular
momentum of the outer electrons {\bf J} is known as the hyperfine interaction.
These interactions lead to the splitting of absorption lines by typically
$1 - 10$m\rAA and can have the effect of de-saturating strong features.  
As such, it is important to account for hyperfine splitting in order
to accurately measure elemental abundances. 

Analogous to spin-orbit coupling, one can define the vector sum 
of the angular momenta as {\bf F} which has quantum numbers ranging from
$|I-J|$ to $|I+J|$.  The number of energy states is $2J+1$ when
$I \geq J$ and $2I+1$ for $J \leq I$.  The allowed transitions must satisfy
the Wigner-Eckart theorem, namely $\Delta F = 0, \pm 1$ with
$0 \to 0$ forbidden.  While one could calculate the wavelength splittings
from first principals, it is usually more accurate to determine them
empirically.  The energy splitting for a given energy level is typically
represented as:

\begin{equation}
\Delta E = \ohf A K + B \frac{ \frac{3}{4} K (K+1) - J(J+1) \, I(I+1)}{
2I(2I-1)(2J-1)}
\end{equation}
where
\begin{equation}
K \equiv F(F+1) - I(I+1) - J(J+1) \perd
\end{equation}
By determining the $A$ and $B$ constants empirically from very high
resolution laboratory experiments for both the upper and lower energy
levels, one can calculate the wavelength shifts of hyperfine lines to
the precision required for stellar spectroscopy.  Kurucz (1999) has compiled
the most accurate $A$ and $B$ constants from the literature and calculated
the hyperfine lines for nearly all of the absorption lines identified in
our spectra.  With the exception of Ba (for which we adopt the values 
provided by McWilliam 1998), we take the values presented by Kurucz.

To calculate the relative line strength $\eta$ of each hyperfine line, one
can adopt the Russell-Saunders terms \citep{condon35}.
In the case where J = J$'$ the following equations apply:

\begin{eqnarray}
\eta &= \frac{(2F + 1) (F(F+1) - I(I+1) + J(J+1))^2}{4 F (F+1)} 
	\quad\quad\quad [F = F'] \\
     &= \frac{(F - I + J) (F + I - J) (I+J+F+1)(I+J+1-F)}{4 F} \quad [F>F']\\
     &= \frac{(F - I + J+1) (F+I-J+1) (I+J+F+2)(I+J-F)}{4 (F+1)} \quad [F<F'] 
\end{eqnarray}

\noindent If J is greater than J$'$ then,

\begin{eqnarray}
\eta &= \frac{(2F + 1) (F+J-I) (F+I-J+1) (F+I+J+1) (I+J-F)}{4 F(F+1)}
	\quad [F=F'] \\
     &= \frac{(F+I-J+1) (I+J-F) (F+I-J+2)(I+J-1-F)}{4 (F+1)} \quad [F<F']\\
     &= \frac{(F-I+J-1) (F+J-I) (F+I+J+1)(I+J+F)}{4 F} \quad [F>F']
\end{eqnarray}

\noindent Finally, if J is less than J$'$:

\begin{eqnarray}
\eta &= \frac{(2F + 1) (F+J'-I) (L+I-J'+1) (F+I+J'+1) (I+J'-F)}{4 F(F+1))}
	\quad [F=F'] \\
     &= \frac{(F'+I-J'+1) (I+J'-F') (F'+I-J'+2)(I+J'-1-F')}{4 (F'+1)} 
	\quad [F<F']\\
     &= \frac{(F'-I+J'-1) (F'+J'-I) (F'+I+J'+1)(I+J'+F')}{4 F'} \quad [F>F']
\end{eqnarray}

\noindent We then normalize the relative line strengths to give the total
$\log gf$ value listed in Table~\ref{tab:ew}.  We present the actual values
that we have adopted in Table~\ref{tab:hfs} in the event that the Kurucz
table is updated after publication.

\begin{table}[ht]
\begin{center}
\caption{ {\sc HFS TABLES} \label{tab:hfs}}
\begin{tabular}{lccc}
\tableline
\tableline
Ion & $\lambda$ (\AA) & EP (eV) &$\log gf$\\
\tableline
Sc II &   4670.400 &  1.357 & $ -1.182  $ \\
Sc II &   4670.403 &  1.357 & $ -1.353  $ \\
Sc II &   4670.405 &  1.357 & $ -1.560  $ \\
Sc II &   4670.408 &  1.357 & $ -1.831  $ \\
Sc II &   4670.409 &  1.357 & $ -1.922  $ \\
Sc II &   4670.410 &  1.357 & $ -2.249  $ \\
\tableline
\tablecomments{The complete version of this table is in the 
electronic edition of the Journal.  The printed edition contains only
a sample}
\end{tabular}
\end{center}
\end{table}

\end{document}